\documentclass[aps,reprint,twocolumn,pra,groupedaddress,floatfix]{revtex4-1}
\usepackage{amsmath,amssymb,amsfonts}
\usepackage{mathrsfs}
\usepackage{stackengine}
\usepackage{soul}
\usepackage{graphicx}
\usepackage{color}
\begin{document}
\title{$N$-channel comb filtering and lasing in $\mathcal{PT}$-symmetric superstructures}
\author{S. Vignesh Raja}
\email{vickyneeshraja@gmail.com}
\author{A. Govindarajan}
\email{govin.nld@gmail.com}
\author{A. Mahalingam}
\email{drmaha@annauniv.edu}
\author{M. Lakshmanan}
\email{lakshman.cnld@gmail.com}
\affiliation{$^{*,\ddagger}$Department of Physics, Anna University, Chennai - 600 025, India}
\affiliation{$^{\dagger,\mathsection}$Centre for Nonlinear Dynamics, School of Physics, Bharathidasan University, Tiruchirappalli - 620 024, India}
\begin{abstract}

A comb spectrum generating device based on  Bragg grating superstructures  with gain and loss is suggested in this paper. It includes a comprehensive analysis of the device formulation, generation and manipulation of the comb spectrum with a number of degrees of freedom such as duty cycle, sampling period and gain-loss parameter. For applications such as RF traversal filters and tunable multi-wavelength laser sources, the reflected intensities of the comb resulting from the superstructures should have uniform intensities, and this is guaranteed by optimizing the physical length of the device, gain and loss in the unbroken $\mathcal{PT}$-symmetric regime. Alternatively, it can be accomplished by reducing the duty cycle ratio of the superstructure to extremely small values in the broken $\mathcal{PT}$-symmetric regime. Such a customization will degrade the reflectivity of the conventional grating superstructures, while it gives rise to narrow spectral lines with high reflectivity in the proposed system. Remarkably, combs with an inverted envelope are generated for larger values of gain and loss.
\end{abstract}
\maketitle
\section{Introduction}

 Optical frequency comb (OFC) is a widely used notion to denote a spectral source featuring a series of isolated spectral lines, which are uniformly spaced \cite{Pilozzi2017}. OFCs are readily built from a diverse range of all-optical building blocks, including mode locked fiber lasers \cite{jones2000carrier,fortier201920,jin2006absolute}, ring resonators \cite{del2007optical,Herr2014}, sampled or superstructured fiber Bragg gratings (SFBGs) \cite{Dong2006,Lee2003,Lee2004,Lee2004a,Li2003,Loh1999,Navruz2008,Zhang2019,Zou2006}, $\mathcal{PT}$-symmetric topological structures \cite{Pilozzi2017} and so forth. OFCs have broad spectral span (with or without uniform amplitudes) \cite{Li2003}, and due to this inherent property, the footprints of OFCs are found in both classical and quantum optical applications such as telecommunication systems \cite{Dong2006,Lee2003,Lee2004,Lee2004a,Li2003,Loh1999,Navruz2008,Zhang2019,Zou2006}, ultrafast spectroscopy \cite{Gohle2005}, generation of attosecond pulses \cite{baltuvska2003attosecond}, quantum computing \cite{Pfister2020},  and optical frequency metrology \cite{udem2002optical}. Precise control over the spectral characteristics of OFCs is one of the challenging aspects of scientific investigation \cite{jayaraman1993theory,Li2003,Zou2006,fortier201920} besides the scalability and on-chip integration of these OFC sources \cite{baltuvska2003attosecond}. For an outstanding contribution to the generation of OFC and its application to laser-based precision spectroscopy \cite{holzwarth2000optical}, Hall and Hansch were awarded Nobel prize back in 2005 \cite{hall2006nobel,hansch2006nobel}. Since then, investigations on the generation, control and applications of OFCs in various fields remain as one of the active areas of research \cite{fortier201920}.  
 
 Research interest on modern light wave communication systems is primarily targeted at effective utilization of the available channel bandwidth via judicious design of multi channel devices that would simultaneously handle many independent frequencies  \cite{Dong2006,Lee2003,Lee2004,Lee2004a,Li2003,Loh1999,Navruz2008,Zhang2019,Zou2006}. In particular, OFCs based on SFBGs are fascinating for the reasons that they are much compact, less complex and they can be employed to handle multi-channel functionalities like multi-wavelength lasers \cite{jayaraman1993theory,ishii1993multiple, Sourani2019}, broadband dispersion compensators \cite{Li2003,Lee2003,Lee2004a}, transverse-load sensing devices \cite{Shu2003}, space-tunable multi-channel notch  filters \cite{Li2008}, etc. Previously, frequency comb spectrum in the presence of gain and loss was realized in a topological cascaded laser \cite{Pilozzi2017} and its realization in simple optical devices like fiber Bragg grating (FBG) still remains to be unexplored, which is a  subject of the present investigation. Hence, it is important to critically analyze some of the fundamental concepts in designing a sampled grating structure alias superstructure in order to realize an OFC.

SFBG is a class of Bragg periodic structures, which is customized by not allowing any spatial variation in the refractive index (RI) of the core in between two adjacent samples \cite{Shu2003}. The samples are the regions of the core whose RI is permanently modulated by exposing them to intense UV exposure \cite{Li2003,Lee2003}. These samples are commonly referred as seed gratings since they form the basic building units of the structure \cite{Li2003}. Even though these basic building blocks are nothing but the conventional FBG structures (uniform or nonuniform), the SFBG exhibits some unique characteristics both in its structure and spectrum \cite{Eggleton1994,DeSterke1997,Eggleton1996,DeSterke1996}. Physically, the term sampling length ($s_L$) refers to the length of one sample and each sample is separated from its predecessor by a distance of  $s_\Delta$ (region of constant RI which is unexposed to UV). Hence, the period of the sampled grating is given by $s_\Lambda = s_L +s_\Delta$ \cite{Zhang2019,Zou2006,Navruz2008} as shown in Fig. \ref{fig1}. The RI modulation of the individual sample is low, while the modulation of the grating depth of the overall structure is large \cite{DeSterke1995,DeSterke1996}. Also, the grating period ($\Lambda$) of the each FBG is very low ($<$ 1 $\mu$m) when compared to the sampling period (of the order of mm) \cite{erdogan1997fiber,DeSterke1995} which results in a broad spectral span of the SFBG in contrast to the conventional FBG spectrum \cite{erdogan1997fiber,DeSterke1995,DeSterke1997} . 

The SFBG spectrum consists of many spectral lines (discrete) of high reflectivity and narrow width \cite{Zou2006,Lee2004} with each spectral line dedicated to one particular wavelength. The origin of such distinct spectrum can well be understood from the photonic band gap structure of a SFBG which possesses additional Bragg resonances or multiple band gaps that differentiate it from a conventional FBG \cite{DeSterke1995,DeSterke1997,Eggleton1996}. Thus higher order reflection modes become inevitable attributes of the SFBG spectrum \cite{Zhang2019,DeSterke1995}. The periodic nature of the inter-coupling parameter is yet another remarkable property of  SFBGs \cite{DeSterke1997,DeSterke1995}. These features can be deemed as a precursor for realizing the  multi-wavelength applications such as filters and others mentioned previously \cite{Li2003}.

Theoretical and experimental investigations on SFBGs are predominantly focused on improving the spectral characteristics such as increasing number of usable optical channels within the available bandwidths  by reducing the channel spacing \cite{Navruz2008,Zhang2019,Zou2006}, and nonuniformity in the reflectivity (transmittivity) of the multi-channels \cite {ibsen1998sinc}. These desirable attributes can be altered in accordance with the application of interest thanks to the availability of a wide range of sampling functions and flexible design methodologies in fabricating them \cite{ibsen1998sinc,Lee2003,Li2003}. Broadly, these sampling windows can be categorized into amplitude sampling \cite{ibsen1998sinc}, phase only sampling \cite{Li2003,Lee2003}, and hybrid sampling functions \cite{Navruz2008}. The uniform sampling technique is the simple and most straight forward approach in realizing SFBGs. The channels in the output spectrum of a uniformly sampled SFBG is likely to be nonuniform in their amplitude and their control remains a challenge. The SFBG mainly suffers from a decrease in the reflectivity with an increase in the number of channels in general \cite{Li2003,ibsen1998sinc,Zou2006,Lee2003}. In the perspective of SFBGs without gain and loss, phase sampling techniques are widely incorporated to overcome this issue \cite{Lee2003,Lee2004,Navruz2008}.

Having concisely discussed the general concepts in the physical realization of SFBGs and the spectrum exhibited by them, we would like to note that all the above mentioned types of  SFBG structures can be revisited again by researchers from the perspective of $\mathcal{PT}$-symmetric structures. It is worthwhile to note that invoking the notion of $\mathcal{PT}$-symmetry in a conventional FBG itself is of high scientific interest as it leads to novel non-Hermitian optical systems. We here take a step further ahead to establish $\mathcal{PT}$-symmetry in a  FBG superstructure which to our knowledge  has no relevant scientific articles in literature to deal with.
Realizing a $\mathcal{PT}$-symmetric SFBG (PTSFBG) simply requires the modulation of RI of the seed grating (samples) to obey the $\mathcal{PT}$-symmetric condition $n(z) = n^*(-z)$ \cite{kottos2010optical,el2007theory,ruter2010observation,sarma2014modulation,lin2011unidirectional,phang2013ultrafast,miri2012bragg,huang2014type,govindarajan2018tailoring}. It is well-known that the real and imaginary parts of the RI profile, respectively, need to be even and odd functions of propagation distance ($z$) for a $\mathcal{PT}$-symmetric FBG (PTFBG) \cite{huang2014type,lin2011unidirectional,miri2012bragg}. Prior to this work, linear spectrum of homogeneous \cite{miri2012bragg,lin2011unidirectional,kulishov2005nonreciprocal} and inhomogeneous $\mathcal{PT}$-symmetric gratings \cite{huang2014type,lupu2016tailoring,raja2020tailoring,raja2020phase} were investigated and the literature seems to lack any comprehensive investigation on the linear spectrum based on sampled PTFBG. Nevertheless, the demonstration of frequency comb in a supersymmetric  (SUSY) DFB structure  by Longhi \cite{longhi2015supersymmetric} could be translated into the context of PTFBGs.

 In this paper, we consider a SFBG with gain and loss  under the expectation that the concept of reversal of direction of incidence \cite{kulishov2005nonreciprocal,longhi2010optical,phang2015versatile} can possibly overcome the problem of reduction in reflectivity with an increase in the number of channels in contrast to the conventional SFBGs. Recently, discrete comb lasing modes were demonstrated in a topological $\mathcal{PT}$-symmetric structure \cite{Pilozzi2017}. It should be noted that the broken $\mathcal{PT}$-symmetric spectrum of any PTFBG feature a  lasing behavior as reported by many authors \cite{raja2020tailoring,phang2014impact,huang2014type,raja2020phase,longhi2010pt}. The natural question arises from these investigations is that whether it is possible to realize discrete and identical lasing modes with the aid of a SFBG operating in the broken $\mathcal{PT}$-symmetric regime.  We believe that the inherent nature of PTFBGs
 to offer multi-functionalities in different $\mathcal{PT}$-symmetric regimes and the
 degrees of freedom it offers for the  optimization of the desired spectral characteristics looks promising to engineer applications like  filters and tunable laser sources in a SFBG structure.
 
 With these motivations, we organize the article as follows. Section \ref{Sec:2} describes the theoretical modeling of the PTSFBG in addition to the mathematical description of the system based on the transfer matrix method. In Sec. \ref{Sec:3}, the  comb filtering application of the system and its optimization with the grating parameters are presented in the unbroken $\mathcal{PT}$-symmetric regime with a special emphasis on the right light incidence direction. The reflection-less wave transport phenomenon at the unitary transmission point is illustrated in Sec. \ref{Sec:4}. Section  \ref{Sec:5} illustrates the discrete multi channel lasing spectrum of the system in the broken $\mathcal{PT}$-symmetric regime.  The inferences from the previous sections are summarized in Sec. \ref{Sec:6}.

\section{Mathematical model}
\label{Sec:2}

\begin{figure*}
	\centering	\includegraphics[width=0.85\linewidth]{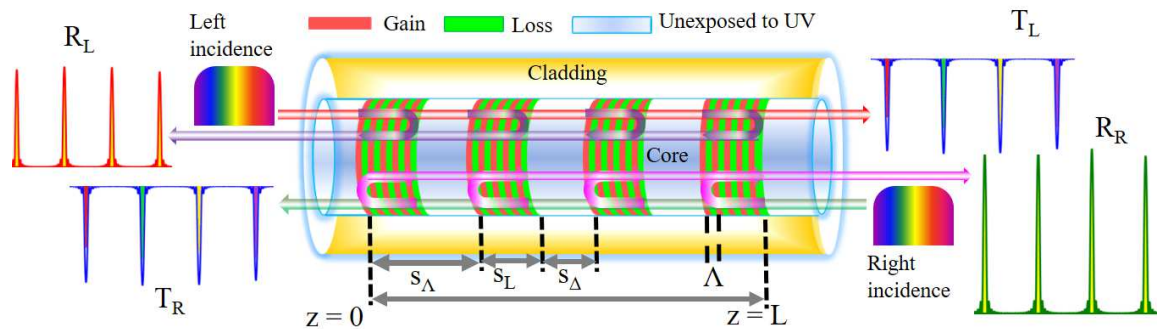}\\
	\caption{Schematic of a PTSFBG of length $L$ made of  uniform PTFBG samples of length $s_L$ and sampling period $s_\Lambda$. Here we have taken number of samples ($N_s$) to be four. Each sample is separated from the next sample by a region of $s_\Delta = s_\Lambda - s_L$. 
	Each sample consists of many unit cells of grating period $\Lambda$. The $\mathcal{PT}$-symmetric RI condition is achieved in each unit cell by having alternate regions of gain (red) and loss (green) \cite{phang2013ultrafast,phang2014impact,raja2020tailoring}} 
\label{fig1}
\end{figure*}
\begin{figure}[t]
\centering	\includegraphics[width=1\linewidth]{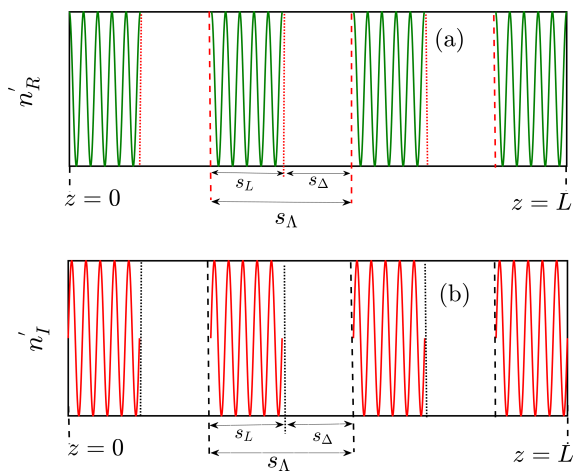}
\caption{Schematics of the modulation in (a) real ($n^{'}_{R}$) and (b) magnitude of imaginary ($n^{'}_{I}$) parts of RI profile. The solid lines represent modulated RI of the sample over a distance of $s_L$. The blank spaces in each sample period indicate that the core RI is unexposed to UV over a region of $s_\Delta$. }
	\label{fig2}
\end{figure}
  The description of the device is as follows: The structure is built up of (multiple) uniform PTFBG samples separated from each other by regions of core unexposed to UV. The  RI distribution [$n(z)$] of the sample that includes the effect of $\mathcal{PT}$-symmetry is given by 
\begin{equation}
n(z)=n_{0}+n_{1R}\cos\left(\frac{2\pi}{\Lambda}z\right)+in_{1I}\sin\left(\frac{2\pi}{\Lambda}z\right),
\label{Eq:norm1}
\end{equation}

where $n_0$ stands for the constant RI of the core. The grating's modulation strength is a complex entity whose real and imaginary parts are given by $n_{1R}$ and $n_{1I}$, respectively. The notation $\Lambda$ in Eq. (\ref{Eq:norm1}) refers to the grating period of each one of the gratings and it is indicated pictorially in Fig. \ref{fig1}. Each sample has a uniform length of value $s_L$ and it is followed by a region which is unexposed to UV (of length $s_\Delta$). This behavior occurs cyclically in each sampling period ($s_\Lambda$). It is important to point out that the system locally satisfies the $\mathcal{PT}$-symmetric condition in each unit cell as well in the sampling length. 

One unit cell is formed by having a real [$n_{R}^{'}$ $=$ $n_{1R}$ $\cos(2 \pi z/\Lambda)$] and imaginary [$ n_{I}^{'}$ $=$ $n_{1I}$ $\sin(2 \pi z/\Lambda)$] modulation of the refractive index. The modulations  $n_{R}^{'}$ and  $n_{I}^{'}$  are depicted in Figs. \ref{fig2}(a) and \ref{fig2}(b). Each sample (uniform PTFBG) features a number of alternating regions of gain and loss.  

It is useful to introduce an important parameter, namely duty cycle, which must be judiciously varied to achieve the desired spectrum \cite{erdogan1997fiber}. Mathematically, the duty cycle is defined as the ratio of the length of the sample to the sampling period and it reads as
\begin{equation}
d=\cfrac{s_L}{s_\Lambda}.
\label{Eq:norm2}
\end{equation}
  
The PTSFBG proposed here is simply a uniform PTFBG in which the grating elements are stripped off in a periodic fashion. The resulting spectrum of the device can be found by  the coupled mode theory (CMT) formalism. The transfer matrix method (TMM) stands out to be a first choice for theoretical physicists to analyze any complex FBG structure as it offers high accuracy and consumes less computation time over other techniques like the Gel-Fand–Levitan–Marchenko inverse scattering method, standard thin-film techniques or the Rouard theory of waveguides \cite{erdogan1997fiber}, \cite{kashyap2009fiber}. Each of these techniques has its own demerits and they are listed as below: First, the accuracy of the simulated results is limited by the rounding off errors in the computation in thin film based approaches.  Also, it cannot fully characterize both the phase and amplitude responses of the complex type of gratings like nonuniform FBGs or superstructures. If the number of grating periods or the length of grating itself is large, the number of matrices also gets increased in Rouard method \cite{erdogan1997fiber}. Thus, the computation becomes more complex and consumes more time. Nonetheless, the TMM is capable of addressing all these issues. Many types of physically realizable gratings have been fully characterized by employing this technique \cite{kashyap2009fiber}. This is possible because of the fact that the TMM approach allows computation of the output field of a short section of the grating in a single iteration \cite{kashyap2009fiber}. In the subsequent iterations, the resulting matrix that represents the output fields from the previous section is taken as the input matrix for a given section and this process is repeated for $n$-number of cycles until the whole FBG is computed \cite{raja2020tailoring}. Another important reason to choose  TMM  over other methods (for modeling SFBG) is that the direct analytical solutions are tedious to calculate if the number of samples is large \cite{erdogan1997fiber,kashyap2009fiber}. The direct integration of coupled mode equations may not work easily, if the sample contains abrupt phase jumps in its RI profile \cite{Li2003,erdogan1997fiber}. Here, the mathematical relation between the number of samples ($N_s$), sampling period ($s_\Lambda$) and the length of the whole device ($L$) can be given by,
\begin{gather}
N_s=L/s_\Lambda.
\end{gather}
Note that the total number of samples ($N_s$) can be varied as per the requirement by manipulating the sampling period, by fixing the length of the device (unless specified). Mathematically, let the matrices corresponding to these samples be assumed as $\mathcal{L}_{\mathcal{S}}$, where $\mathcal{S} = 1, 2, \dots, N_s$. As an example, the PTSFBG with four samples ($N_s = 4$) is shown in Fig. \ref{fig1}. To model this PTSFBG structure, the following routine is adopted:

All the samples in the above discussion are taken to be \emph{identical} in the present investigation. Therefore, the matrices that represent these samples are also identical ($\mathcal{L}_1 = \mathcal{L}_2 = \mathcal{L}_3 = \mathcal{L}_4$). The regions unexposed to UV within each sample period $s_\Lambda$ are simply modeled by a phase matrix $\mathcal{L}_{\Delta}$ whose matrix elements are given by \cite{erdogan1997fiber}

\begin{gather}
\nonumber \\\mathcal{L}_{\Delta}=\left[\begin{array}{cc}
\exp\left(\cfrac{2\pi i n_0s_\Delta}{\lambda}\right) & 0\\
0 & \exp\left(\cfrac{-2\pi i n_0s_\Delta}{\lambda}\right)
\end{array}\right],
\label{Eq:norm6}
\end{gather}
where $s_\Delta$ and $\lambda$ stand for the length of the region unexposed to UV and the operating wavelength, respectively. 

Let $u_0$ and $v_0$ represent the input fields of the PTSFBG. Similarly, the output fields are given by $u_{out}$ and $v_{out}$.   
Since the PTSFBG is built up of repeated units of samples followed by the core region unexposed to UV, the corresponding phase matrix ($\mathcal{L}_\Delta$) should be inserted in between the matrices representing the sample. Hence, the total electric field propagating through the device is given by
\begin{widetext}
\begin{gather}
\left[\begin{array}{c}
u_{out}\\
v_{out}
\end{array}\right]= \mathcal{L}_1 \times \mathcal{L}_\Delta \times \mathcal{L}_2 \times \mathcal{L}_\Delta \times  \mathcal{L}_{3} \times \mathcal{L}_\Delta \times \mathcal{L}_4
\left[\begin{array}{c}
u_0\\
v_0
\end{array}\right] = \left[\begin{array}{cc}
\mathcal{L}_{11}& \mathcal{L}_{12} \\
\mathcal{L}_{21}  & \mathcal{L}_{22} 
\end{array}\right] \left[\begin{array}{c}
u_0\\
v_0.
\end{array}\right] 
\label{Eq:Norm9}
\end{gather}

To find the matrices $\mathcal{L}_1$, $\mathcal{L}_2$, $\mathcal{L}_3$, and $\mathcal{L}_4$, each sample (of length $s_L$) is divided into $n_s$ number of \emph{piece-wise uniform} sections (of length $\Delta z$). Thus, a sample is assumed to be a  functional block formed by cascading $n_s$ number of sections of uniform PTFBG ($s_L = n_s \times \Delta z$). 	It is well known that cascading (physically) of different sections leads to multiplication of their respective transfer matrices (mathematically). Let $\mathcal{N}_j$ ($j = 1, 2, \dots, n_s$) represent the matrix that corresponds to the  $n^{th}$ section of the sample. Therefore, 
\begin{gather}
\mathcal{L}_1 = \mathcal{N}_1 \times \mathcal{N}_2 \times \mathcal{N}_3 \dots \mathcal{N}_{n_s}
\end{gather}

 	The standard forms of $\mathcal{N}_j$ for a uniform PTFBG have been already dealt by us in detail in our previous work \cite{raja2020phase} and its final form is given by Eq. (\ref{Eq:Norm8})
 	
 	\begin{gather}
  \mathcal{N}_j=\left[\begin{matrix}
 	\cosh(\hat{\sigma}_j {\Delta z}  )+i\left(\cfrac{\delta_j}{\hat{\sigma}_j}\right)\sinh(\hat{\sigma}_j {\Delta z}) & i\left(\cfrac{\kappa_j+g_j}{\hat{\sigma}_j}\right)\sinh(\hat{\sigma}_j \Delta z)
 	\\-i\left(\cfrac{\kappa_j-g_j}{\hat{\sigma}_j}\right)\sinh(\hat{\sigma}_j {\Delta z}) &\cosh(\hat{\sigma}_j {\Delta z}  )-i\left(\cfrac{\delta_j}{\hat{\sigma}_j}\right)\sinh(\hat{\sigma}_j {\Delta z})
 	\end{matrix}\right],
 	\label{Eq:Norm8}
 	\end{gather}
 	where $j= 1, 2 \dots n_s$. In Eq. (\ref{Eq:Norm8}),  $\hat{\sigma}_j$, $k_j$, $g_j$, and $\delta_j$ represent, respectively, the propagation constant, coupling, gain-loss and detuning coefficients of the piece-wise uniform $n^{th}$ section of the sample. For a uniform PTSFBG, these coefficients are same in all the sections and therefore,
 	\begin{gather}
 	\nonumber \hat{\sigma}_j=\left(\kappa_j^2-g_j^2-\delta_j^2\right)^{1/2}, \qquad \kappa_j = \kappa = \pi n_{1R}/\lambda,\qquad g_j = g = \pi n_{1I}/\lambda, \\ \delta_j = \delta = 2\pi n_{0} \left (\cfrac{1}{\lambda}-\cfrac{1}{\lambda_b}\right),  \quad \lambda_b = 2 n_{0} \Lambda.
 	\end{gather}
 	
Note that the model can be extended to $N_s$ number of samples by increasing the value of $\mathcal{S}$. In such a case, Eq. (\ref{Eq:Norm9}) can be rewritten as

\begin{gather}
\left[\begin{array}{c}
u_{out}\\
v_{out}
\end{array}\right]= \mathcal{L}_1 \times \mathcal{L}_\Delta \times \dots \times \mathcal{L}_{N_{s}-1} \times \mathcal{L}_\Delta \times \mathcal{L}_{N_{s}}
\left[\begin{array}{c}
u_0\\
v_0
\end{array}\right] = \left[\begin{array}{cc}
\mathcal{L}_{11}& \mathcal{L}_{12} \\
\mathcal{L}_{21}  & \mathcal{L}_{22} 
\end{array}\right] \left[\begin{array}{c}
u_0\\
v_0
\end{array}\right]. 
\label{Eq:Norm9a}
\end{gather}

 It is well known that the reflection and transmission coefficients are nothing but the squared magnitudes of the amplitudes. These coefficients can be directly computed by applying the FBG boundary conditions, $u_0 = 1$ and $v_{out} = 0$ in Eq. (\ref{Eq:Norm9}) and they read as \cite{lin2011unidirectional,raja2020tailoring}
\begin{gather}
 R_{L}=|- \mathcal{L}_{21}/\mathcal{L}_{22}|^2, \quad R_{R} = |\mathcal{L}_{12}/\mathcal{L}_{22}|^2, \quad T=|(|\mathcal{L}_{11}\mathcal{L}_{22}-\mathcal{L}_{12}\mathcal{L}_{21}|)/\mathcal{L}_{22}|^{2}. 
\label{Eq: mul3}
\end{gather}

	\end{widetext}
	
The nature of $\mathcal{PT}$-symmetry can be described based on the coupling and gain-loss coefficients as,
\begin{gather}
g\begin{cases}<k, & \text{for: unbroken $\mathcal{PT}$-symmetric regime}\\=k, & \text{at: exceptional point} \\>k, & \text{for: broken $\mathcal{PT}$-symmetric regime}.\end{cases}
\end{gather}
The Bragg wavelength of the sample is assumed to be $1550$ nm and constant core index ($n_0$) is taken as $1.45$ throughout the paper. 


\section{Unbroken $\mathcal{PT}$-symmetric regime}
\label{Sec:3}
It should be remembered that the output fields of uniform, chirped and other PTFBG devices consist of a single spectrum centered at the Bragg wavelength \cite{lin2011unidirectional,raja2020tailoring}. In contrast to these structures, the optical field (reflected and transmitted light) emerging out from a PTSFBG is a distinct one and commonly referred to as a comb spectrum  for two reasons. First, the generated spectrum is characterized by periodic maxima (minima) or a serious of sharp spectral lines (resembling the teeth of a comb) \cite{jayaraman1993theory,Li2003}. The other factor is that all these spectral lines are equidistant from each other in the wavelength domain and share a common phase evolution as a result of which the frequency dynamics of every mode in the comb spectrum is deterministic in nature \cite{fortier201920}. With this brief description, we directly look into the comb spectrum of a PTSFBG by following the routine proposed in Sec. \ref{Sec:2}. Among the various spectral features (such as delay, dispersion, and so on), this article deals only with the variations in the reflection and transmission characteristics of a comb spectrum (with respect to the changes in the PTSFBG parameters) which are directly computed from Eq. (\ref{Eq: mul3})
\subsection{Influence of $\mathcal{PT}$-symmetry ($n_{1I}$)}

\begin{figure}
	\centering	\includegraphics[width=0.5\linewidth]{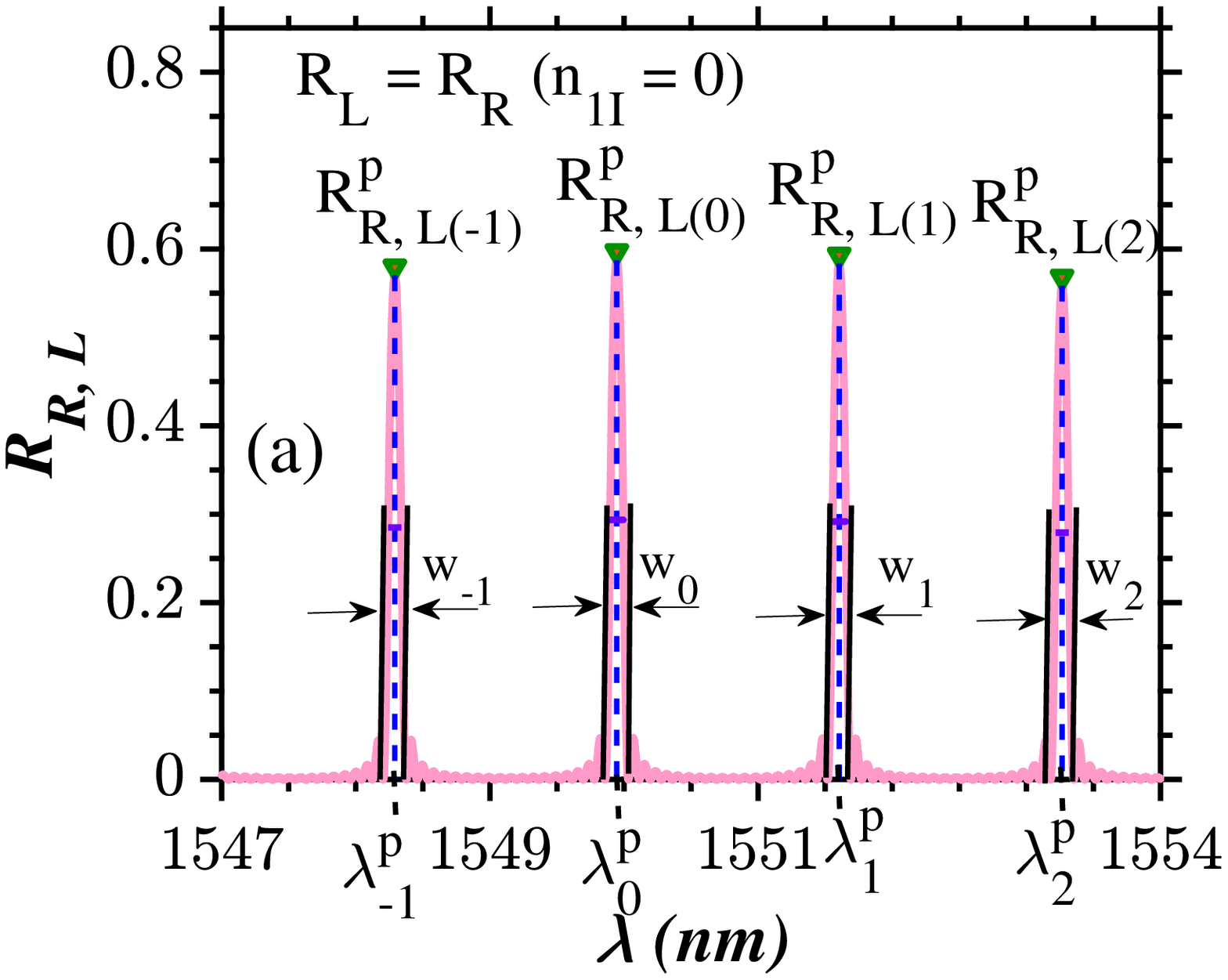}\includegraphics[width=0.5\linewidth]{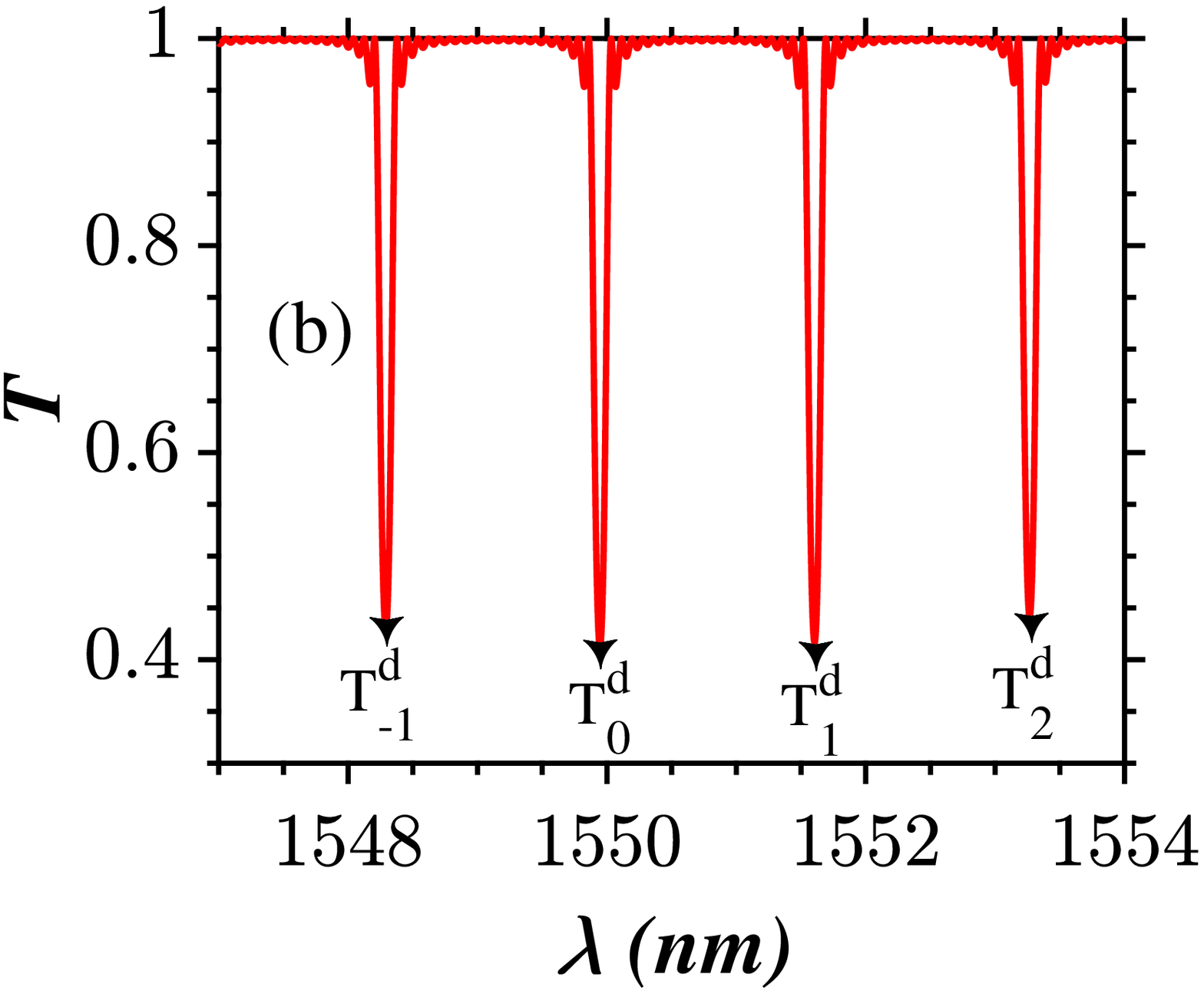}
		\caption{(a) Reflection and (b) transmission spectrum of a conventional SFBG ($n_{1I}=0$) having grating parameters $L = 10$ mm, $n_{1R}=5\times 10^{-4}$, $d=0.1$, and $s_\Lambda=500$ $\mu$m. The reflectivity peaks are denoted by the notation $R_{R(m)}^{p}$ and $R_{L(m)}^{p}$, where $L$, $R$ in the subscript indicate left and right incidences,  respectively. The transmission dips are represented as $T_m^d$  and the full width at half maximum [FWHM] is denoted as $w_m$. $m$ indicates the order of the mode and it takes the values $-1$ ,$0$, $1$, $2$ for sampling period of $500$ $\mu$m in the wavelength range $1547$  -- $1554$ nm and the corresponding values of $\lambda^p_m$ are given by 1548.3, 1549.9, 1551.6, and 1553.3 nm, respectively.}
	\label{fig4}
\end{figure}

\begin{figure}
	\centering	\includegraphics[width=0.5\linewidth]{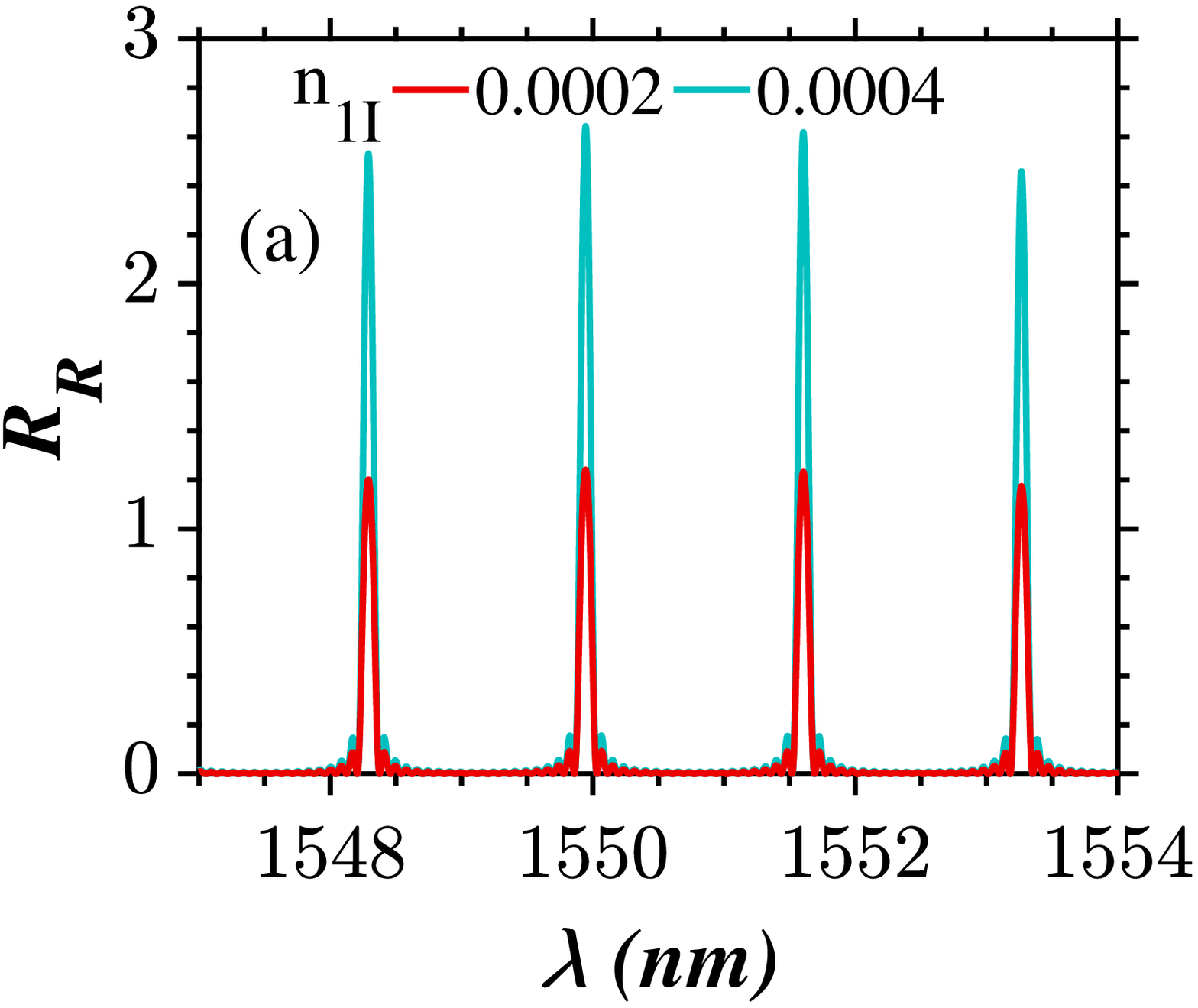}\includegraphics[width=0.5\linewidth]{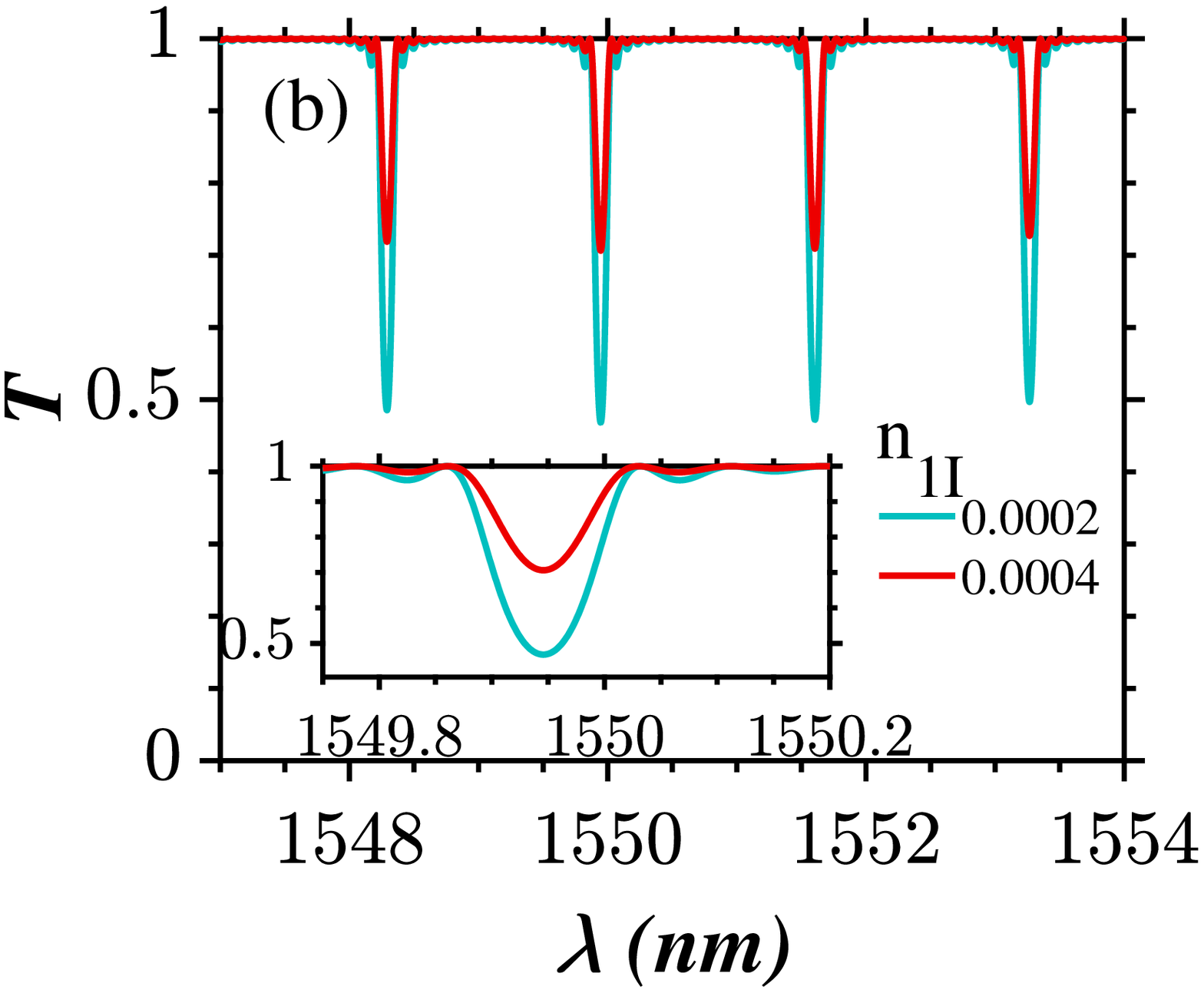}\\\includegraphics[width=0.5\linewidth]{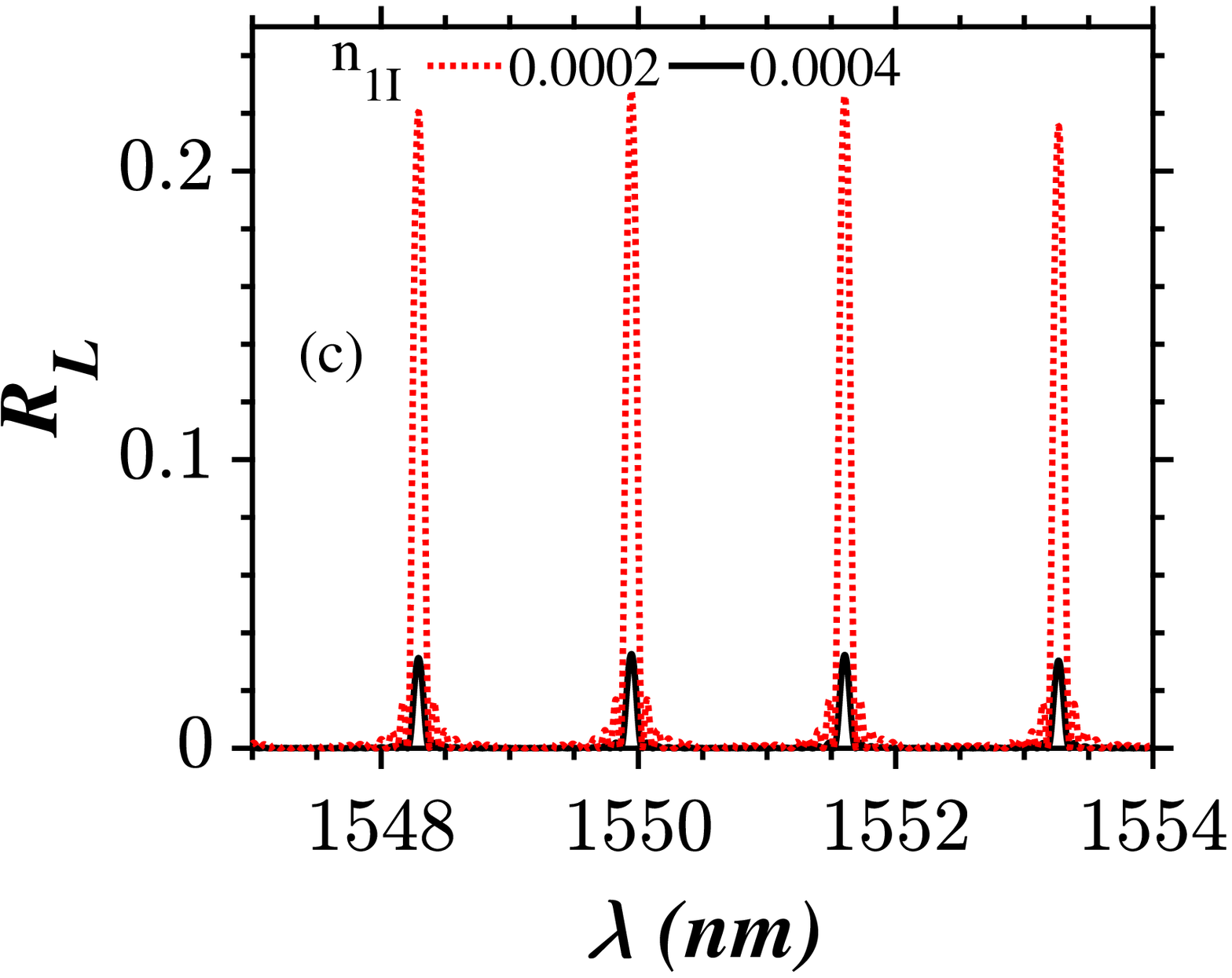}\includegraphics[width=0.5\linewidth]{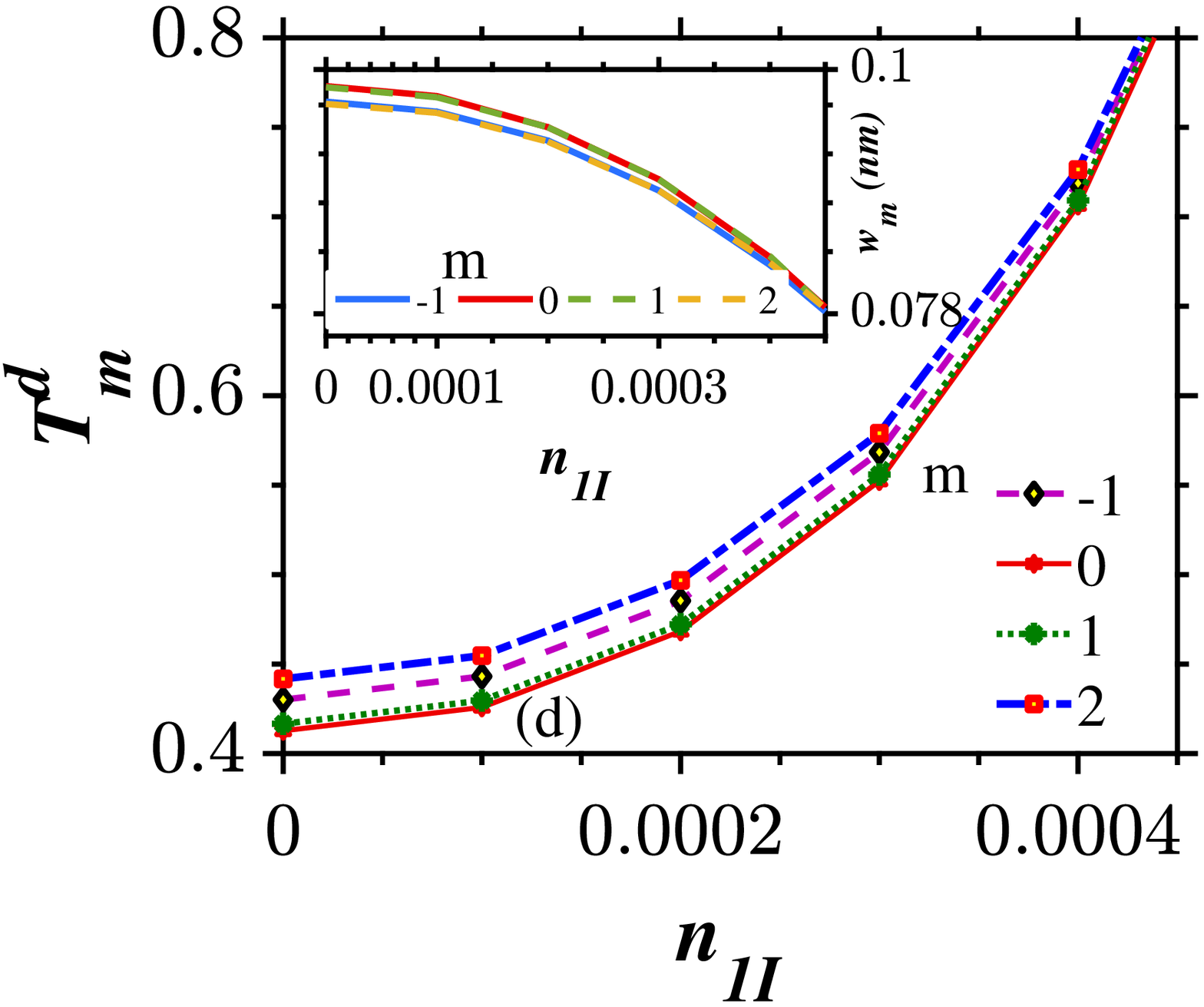}\\\includegraphics[width=0.5\linewidth]{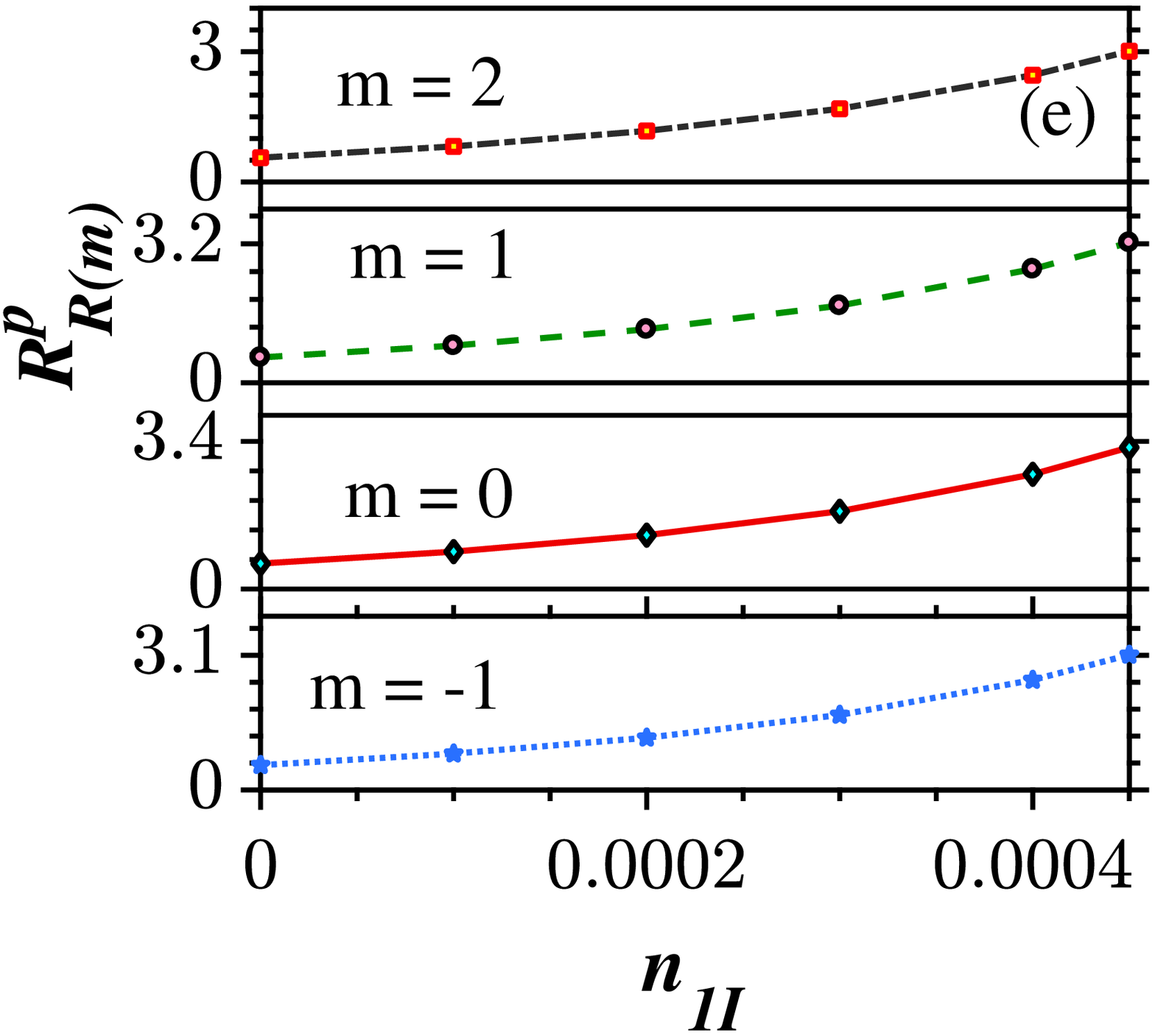}\includegraphics[width=0.5\linewidth]{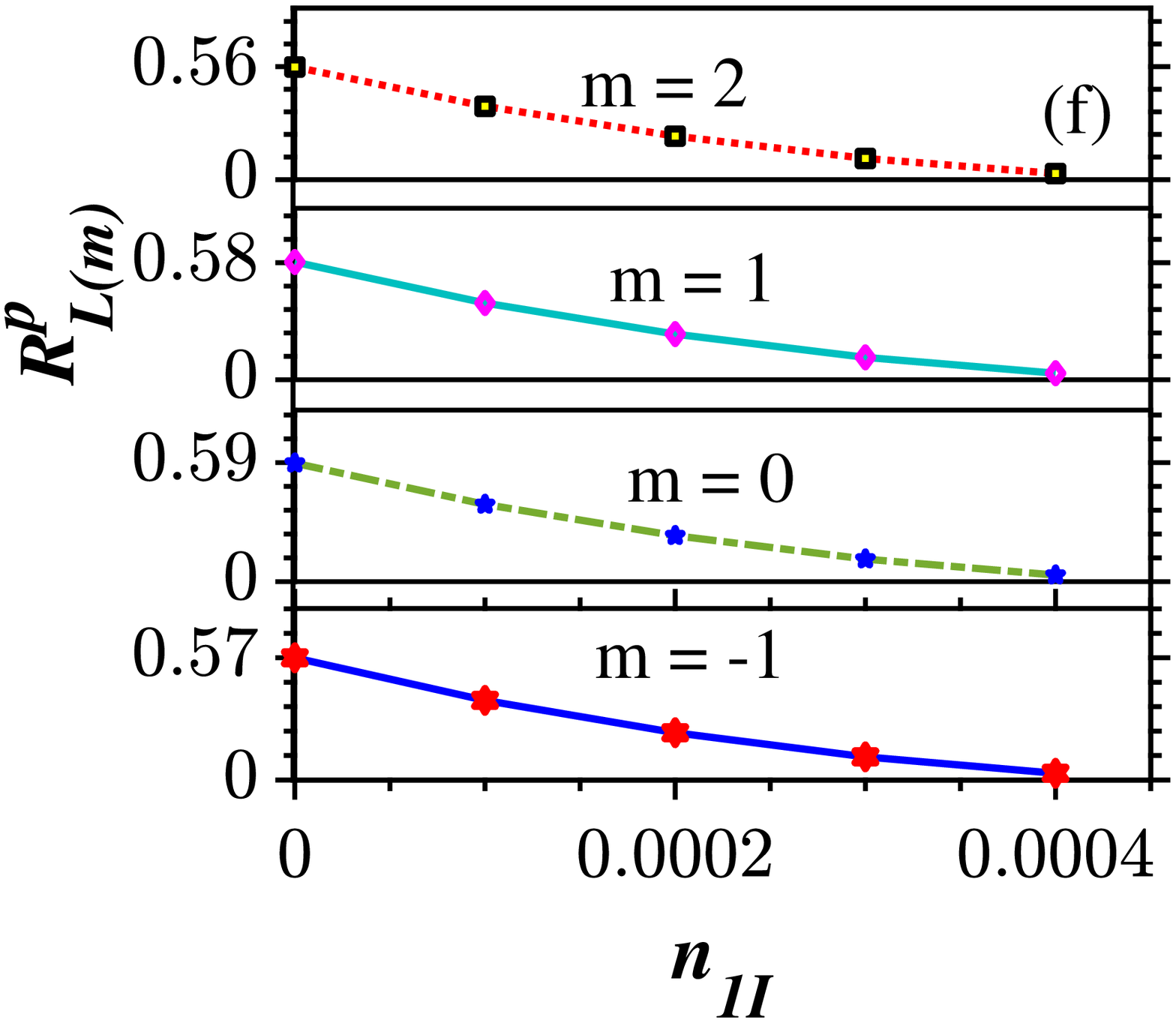}
	\caption{(a, c) Reflection and (b) transmission spectrum of a PTSFBG  having the same device parameters as in Fig. \ref{fig4}. The inset in (b) shows the transmission spectrum pertaining to a single channel. (d) Continuous variation of transmission dips ($T_d^m)$  and the inset depicts variation in FWHM ($w_m$)  against variation in $n_{1I}$. (e) and (f) Continuous variation of reflectivity peaks [$R_{R(m)}^{p}$ and $R_{L(m)}^{p}$] against variation in $n_{1I}$.}
	\label{fig5}
\end{figure}

For simplicity, we first consider the spectrum with four modes ($N = 4$) in the wavelength range $\lambda = 1547$ to $1554$ nm by assuming a sampling period of $s_\Lambda$ = 500 $\mu$m and $d = 0.1$ as shown in Fig. \ref{fig4}. The mode which occurs close to the Bragg wavelength features highest reflectivity (in general) and the detuning parameter corresponding to this mode is nearly zero ($\delta \approx 0$) and hence the mode is designated as zeroth order mode ($m = 0$).  On either side of zeroth order mode, higher order modes occur at discrete wavelengths corresponding to positive ($\delta>0$) and negative detuning regimes ($\delta<0$) and hence the modes in Fig. \ref{fig4} are designated as $m = -1$, 0 , 1, and 2 rather than $m = 0$, 1, 2, and 3.  The linear spectrum of a conventional sampled FBG is illustrated in Figs. \ref{fig4}(a) and \ref{fig4}(b) which confirm that the reflection and transmission from each channel always obey the condition $R+T=1$, which suggests that the Hamiltonian of the system is conserved in the absence of $\mathcal{PT}$-symmetry. But, with the inclusion of gain and loss, the reflectivity of each channel (indicated by subscript $m$) increases provided that the light launching direction is the right ($R_R$) direction as shown in Figs. \ref{fig5}(a) and \ref{fig5}(e). Nevertheless, when the incident direction is reversed, reflectivity ($R_L$) of all the channels gets reduced as seen in Figs. \ref{fig5}(c) and \ref{fig5}(f) when compared to the reflection spectrum shown in Fig. \ref{fig4}(a). The dips in the transmittivity ($T^d_m$) of the individual channels are clearly influenced by the presence of $\mathcal{PT}$-symmetry as shown in Figs. \ref{fig5}(b) and \ref{fig5}(d). Moreover, the transmittivity at the side lobes of the individual channels gets reduced provided that $n_{1I}$ is sufficiently large, say $n_{1I}=4\times10^{-4}$. Also, significant reduction in the FWHM of a single channel is observed which is shown in the inset of Figs. \ref{fig5}(b) and \ref{fig5}(d).

\subsection{Variations in the sampling period ($s_\Lambda$)}
\begin{figure}
	\centering	\includegraphics[width=0.5\linewidth]{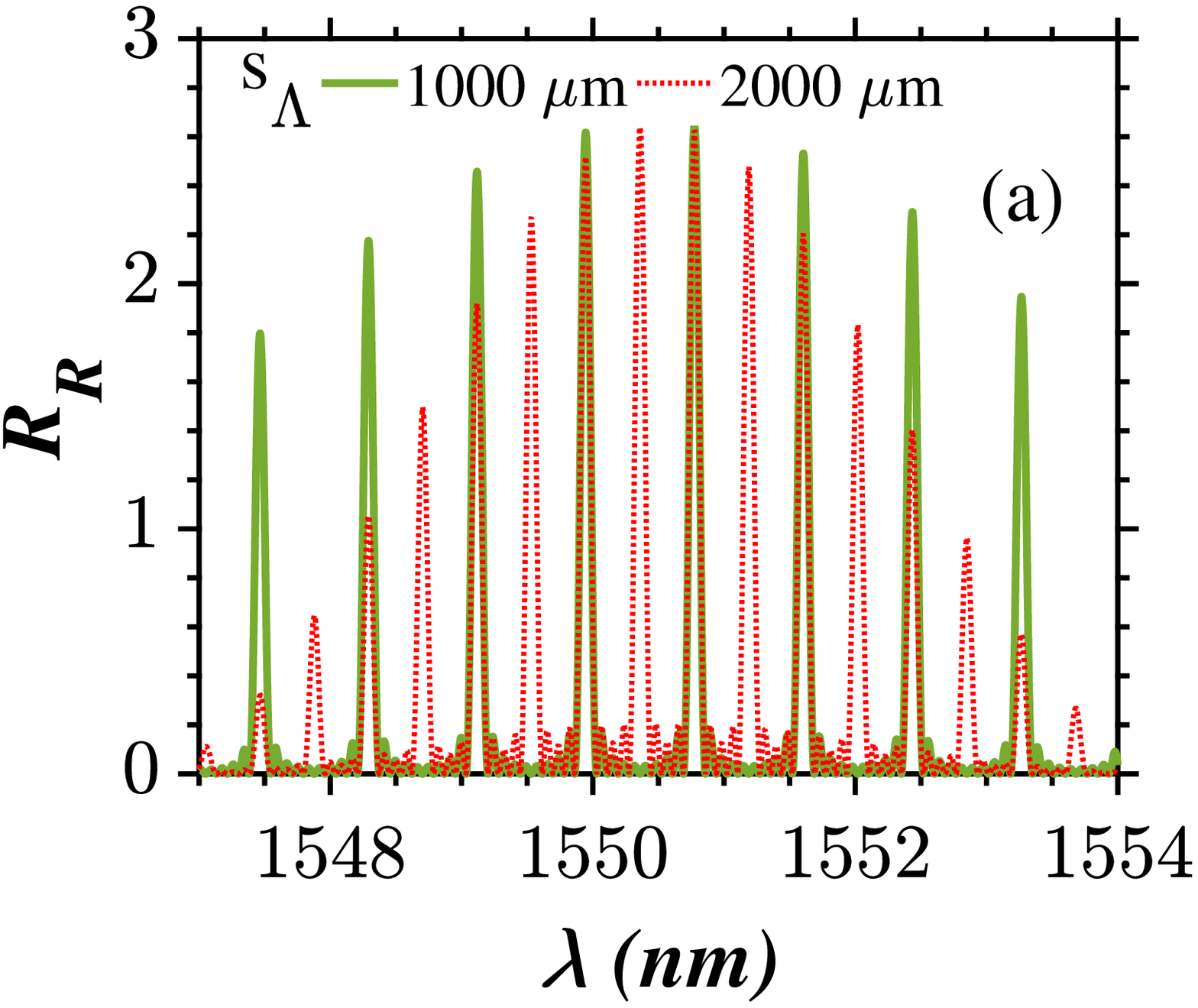}\includegraphics[width=0.5\linewidth]{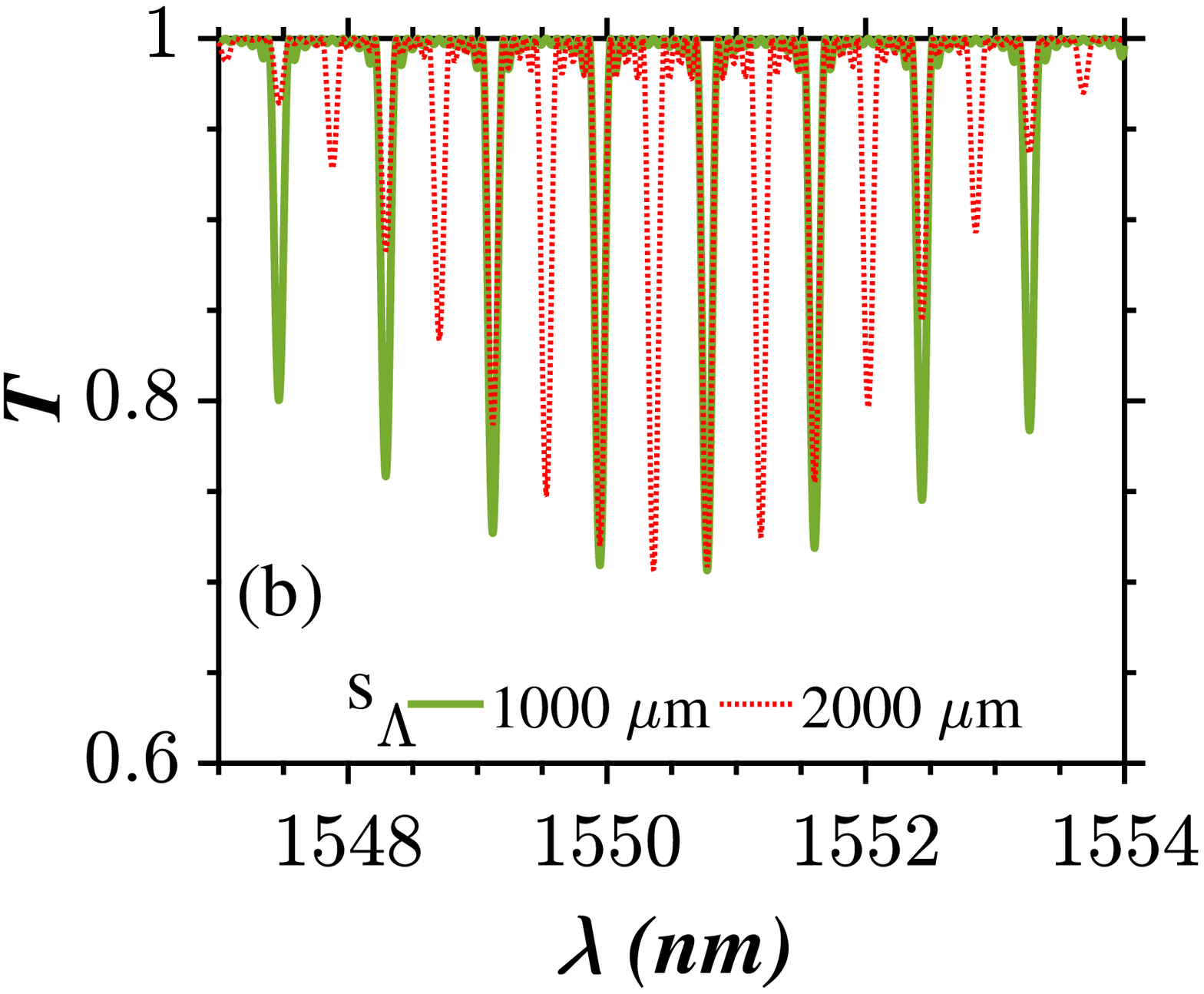}\\\includegraphics[width=0.5\linewidth]{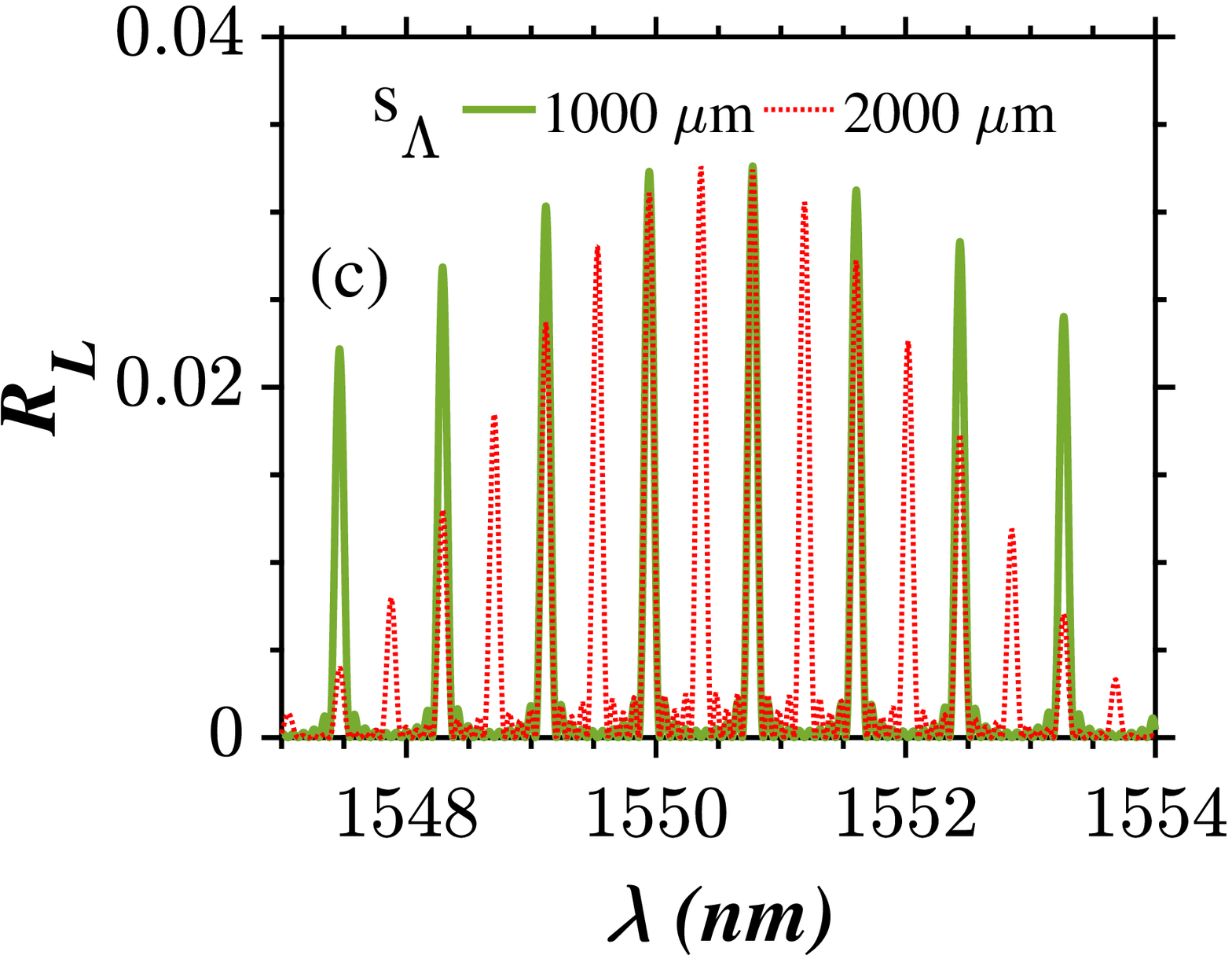}\includegraphics[width=0.5\linewidth]{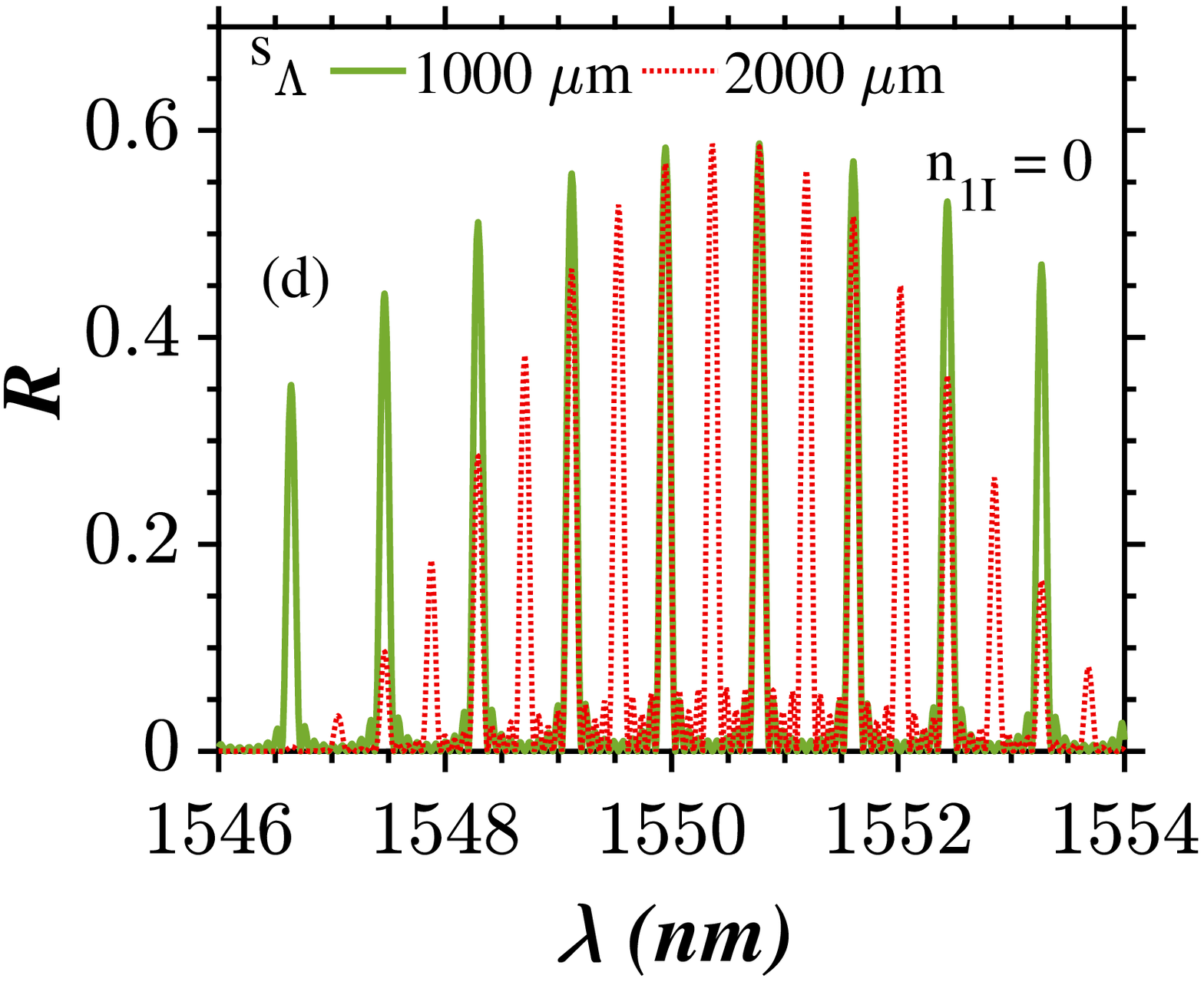}\\\includegraphics[width=0.85\linewidth]{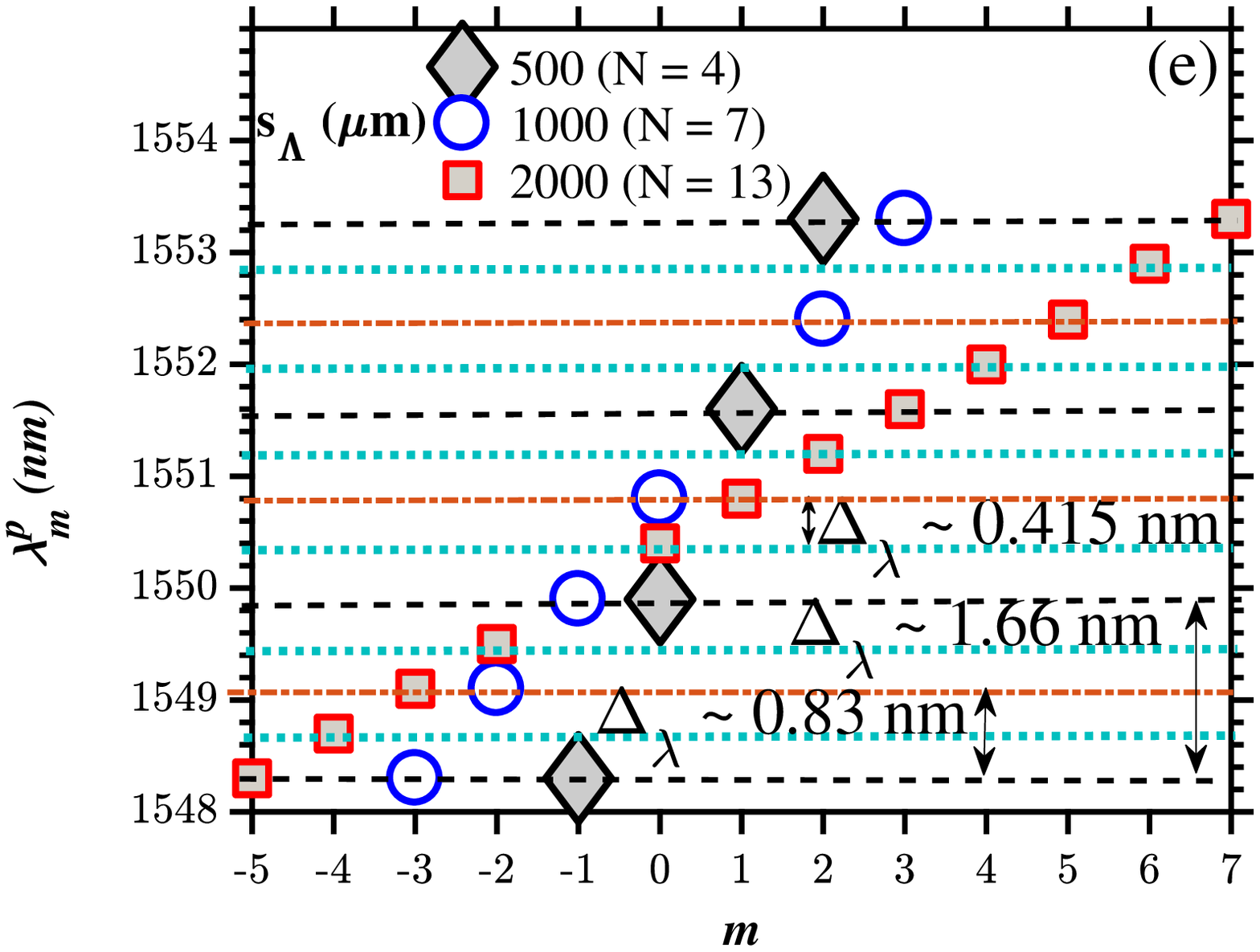}
	\caption{(a) -- (c) Influence of sampling period on unbroken PTSFBG spectrum with same device parameters as in Fig. \ref{fig5}. Also, $n_{1I}$ is kept constant at 0.0004. (d) Illustrates the reflection spectrum in a conventional SFBG ($R_L$ = $R_R$ $=$ $R$) for different sampling periods. (e) Variation in the number of modes ($N$), center wavelength ($\lambda_{m}^{p}$) and separation between adjacent modes ($\Delta_\lambda$)  against $s_\Lambda$. Here the parameter $m$ which indicates the order of the modes takes discrete values $m = -3$, $-2$, $-1$, $0$, $1$, $2$, $3$ for $s_\Lambda = 1000$ $\mu$m. The values of $\lambda_m^p$ for these modes in the same order are given by 1548.3, 1549.1, 1549.9, 1550.8, 1551.6, 1552.4, and 1553.3 nm, respectively. Similarly, $m = -5$ to $7$ (in steps of unity) for $s_\Lambda = 2000$ $\mu$m and the corresponding values of $\lambda^p_m$ are found to be 1548.3, 1548.7,    1549.1,    1549.5,    1549.9,    1550.4,    1550.8,    1551.2,    1551.6,    1552, 1552.4,    1552.9,    1553.3, and 1553.7 nm.}
	\label{fig6}
\end{figure}
The sampling period ($s_\Lambda$) is a crucial parameter in the construction of sampled PTFBG structure, since it dictates the number of usable channels within the available spectral span. Also, the channel spacing between any two adjacent channels is controlled by the sampling period. Among the two spectra shown in Fig. \ref{fig6}(a), the first one [$s_\Lambda$ = 1000 $\mu$m (green and solid lines)] is characterized by less number of channels and thus features more inter channel separation width than the second one [$s_\Lambda$ = 2000 $\mu$m (red and dotted lines)]. Also, this is true for the reflection spectrum for left incidence as shown in Fig. \ref{fig6}(c). By the inherent property of the sampled FBG and the nature of sample (uniform), all these channels are equally spaced in the spatial domain as shown in Fig. \ref{fig6}(e). From these figures, it can be concluded that the number of channels is directly proportional to the sampling period ($s_\Lambda$), whereas the channel separation is inversely proportional to $s_\Lambda$ and it satisfies the relation \cite{Lee2003,Li2003,Li2008,jayaraman1993theory},
\begin{gather}
\Delta_\lambda =\cfrac{\lambda_{b}^{2}}{2n_0 s_\Lambda}
\label{Eq:norm18}
\end{gather}
 From Eq. (\ref{Eq:norm18}), it is very obvious that designing a SFBG with larger $s_\Lambda$ will result in a minimal channel separation width and thereby leading to a increased number of channels within the desired spectral range. But, it should be recalled that the sampling period cannot be arbitrarily large in the perspective of conventional FBGs \cite{Navruz2008,Li2003,Li2008} as it leads to a decrease in the reflectivity of the channels away from the Bragg wavelength as shown in Fig. \ref{fig6}(d). Increasing the index of the core ($n_0$) may appear as a good choice to compensate this reduction in reflectivity from Eq. (\ref{Eq:norm18}) but it suffers from fabrication difficulties. Instead, SFBGs with gain and loss can be employed to increase the reflectivity as shown in Fig. \ref{fig6}(a) as long as the input field is launched from the rear end of the device which is a unique outcome of the $\mathcal{PT}$-symmetry. It is important to point out that the gain and loss parameters amplify the reflectivity pertaining to all the individual wavelengths of the comb spectrum (rather than equalizing the reflectivity of all channels to that of the zeroth order) so that even the channels on the edges of the designed filter will have reflectivity larger than unity as shown in Fig. \ref{fig6}(a).  In opto-electronic approach, Erbium doped fiber amplifiers (EDFA) are used as signal boosters and they do perform the same functionality. Yet, fewer modes corresponding to $s_\Lambda$ = 2000 $\mu$m [Fig. \ref{fig6}(a)] still have weak reflectivity on both sides which means that the parameter $g$ must be increased accordingly for larger sampling periods.
 \subsection{Impact of  device length ($L$)  on uniformity of combs}
 \begin{figure}
 	\centering	\includegraphics[width=0.5\linewidth]{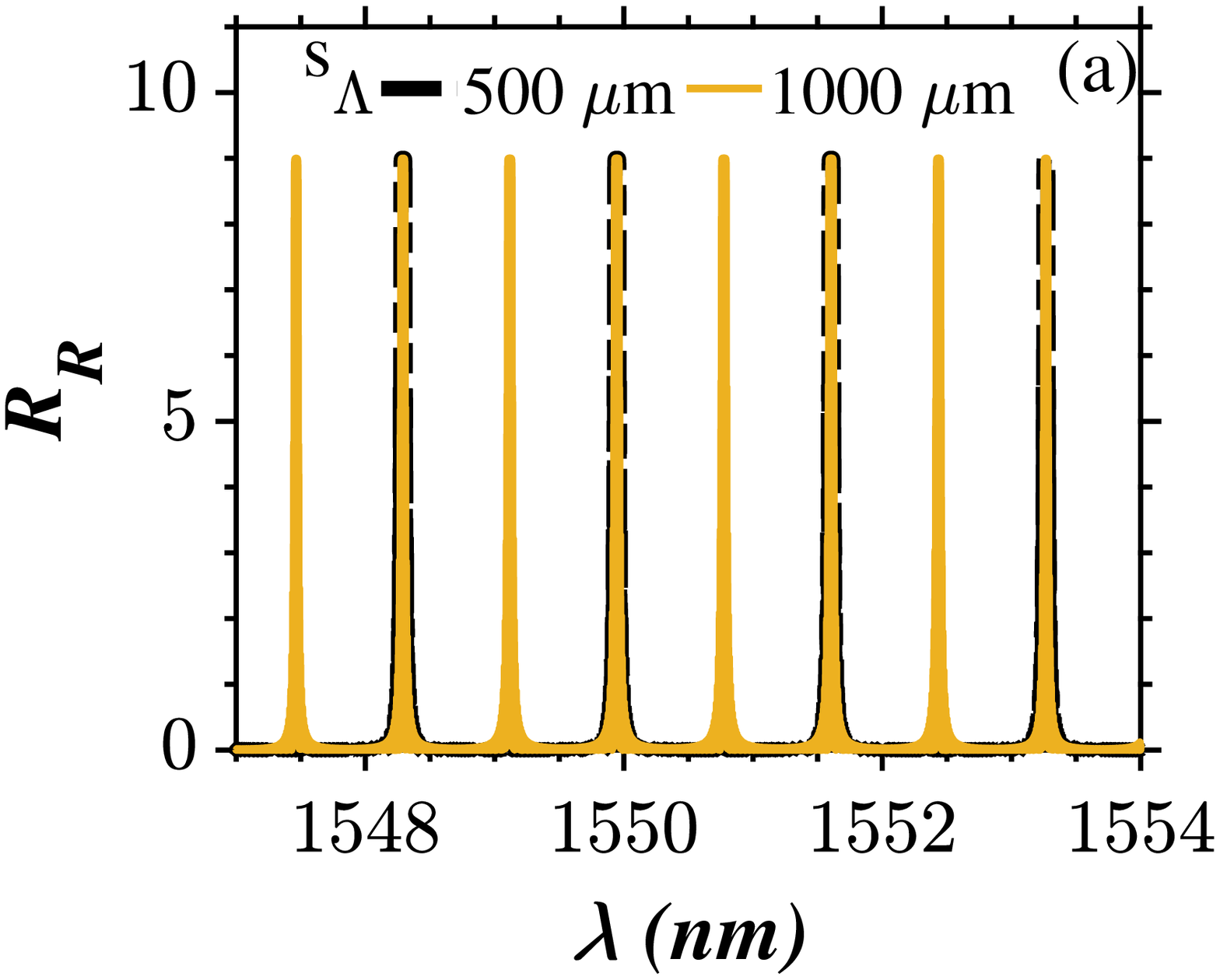}\centering	\includegraphics[width=0.5\linewidth]{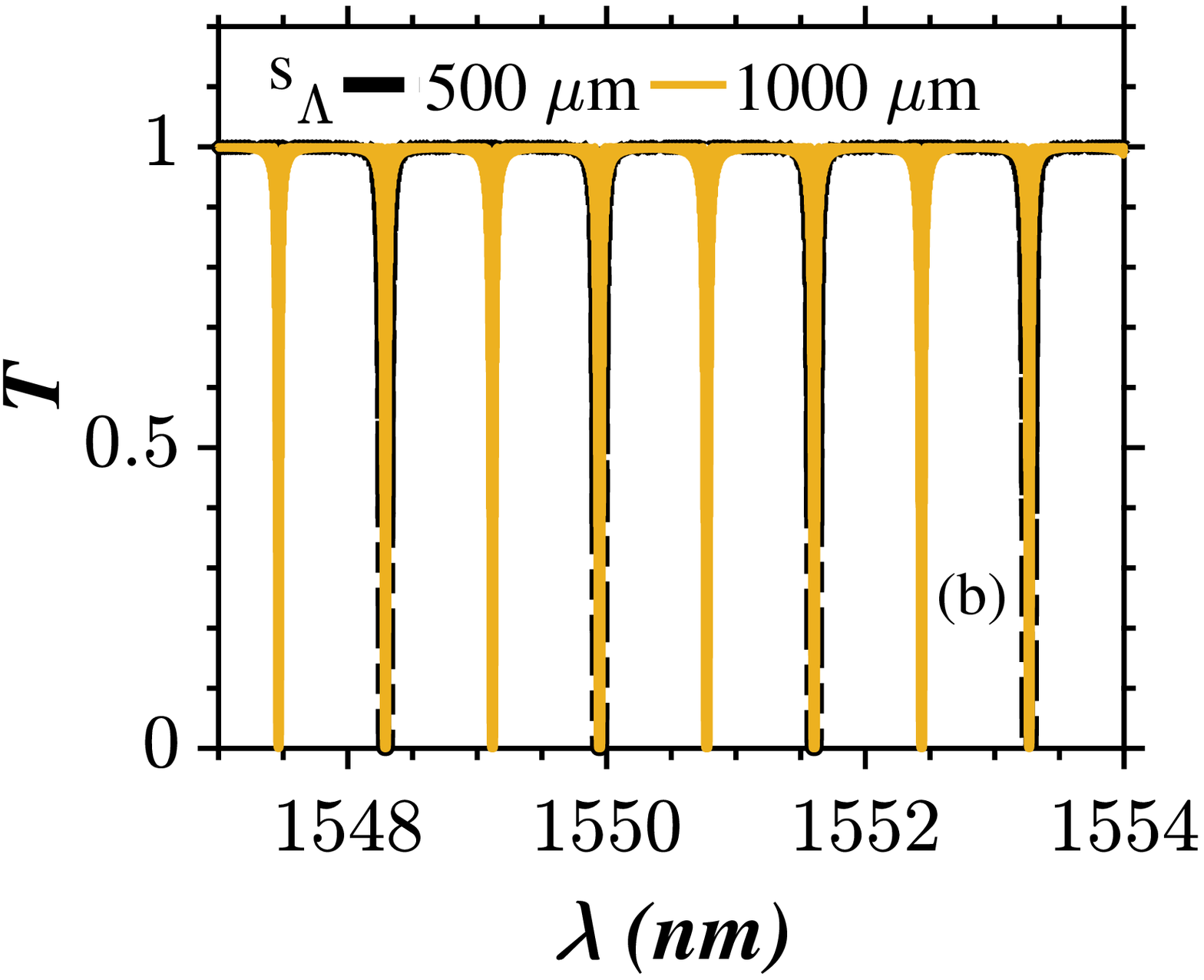}\\\centering	\includegraphics[width=0.5\linewidth]{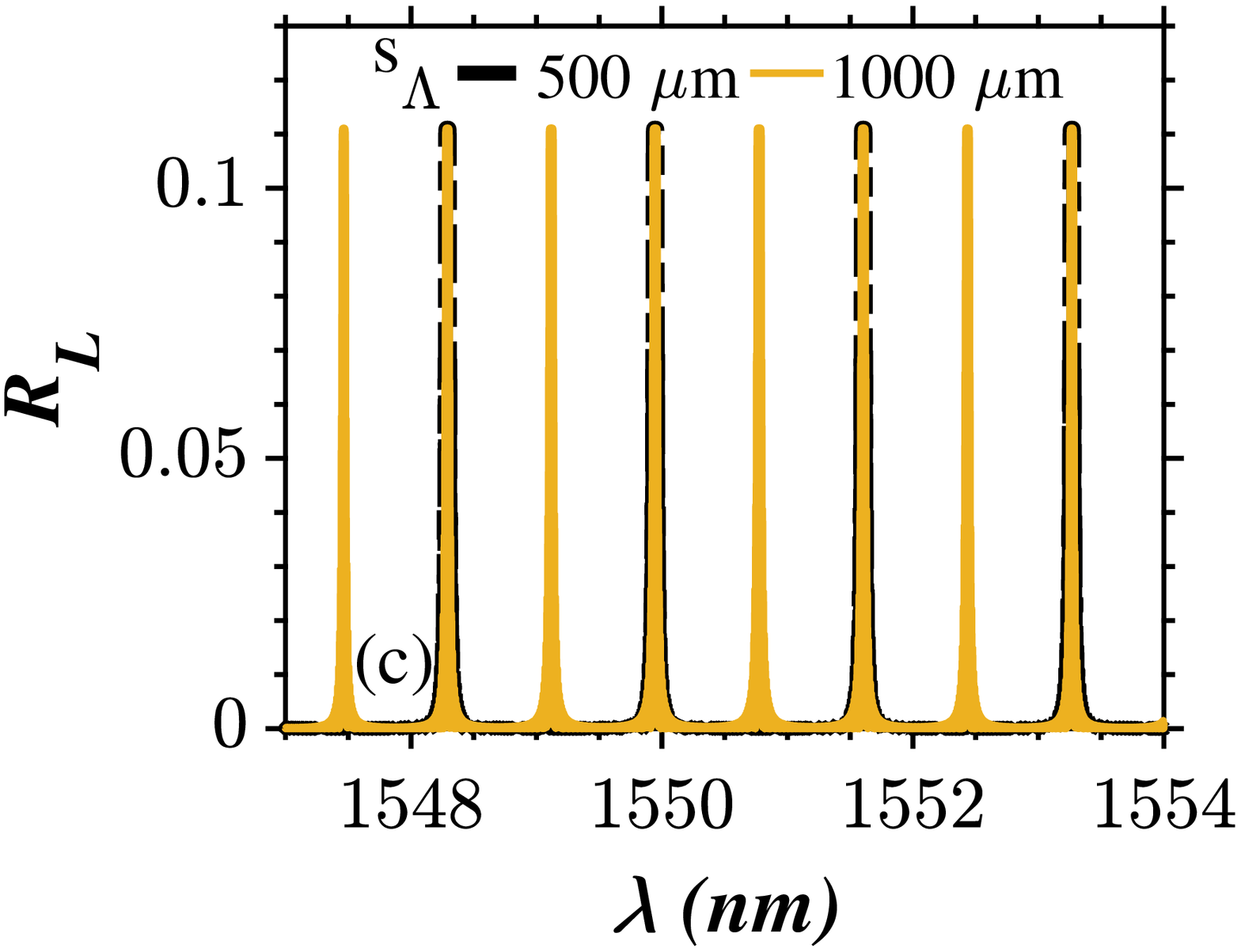}\includegraphics[width=0.5\linewidth]{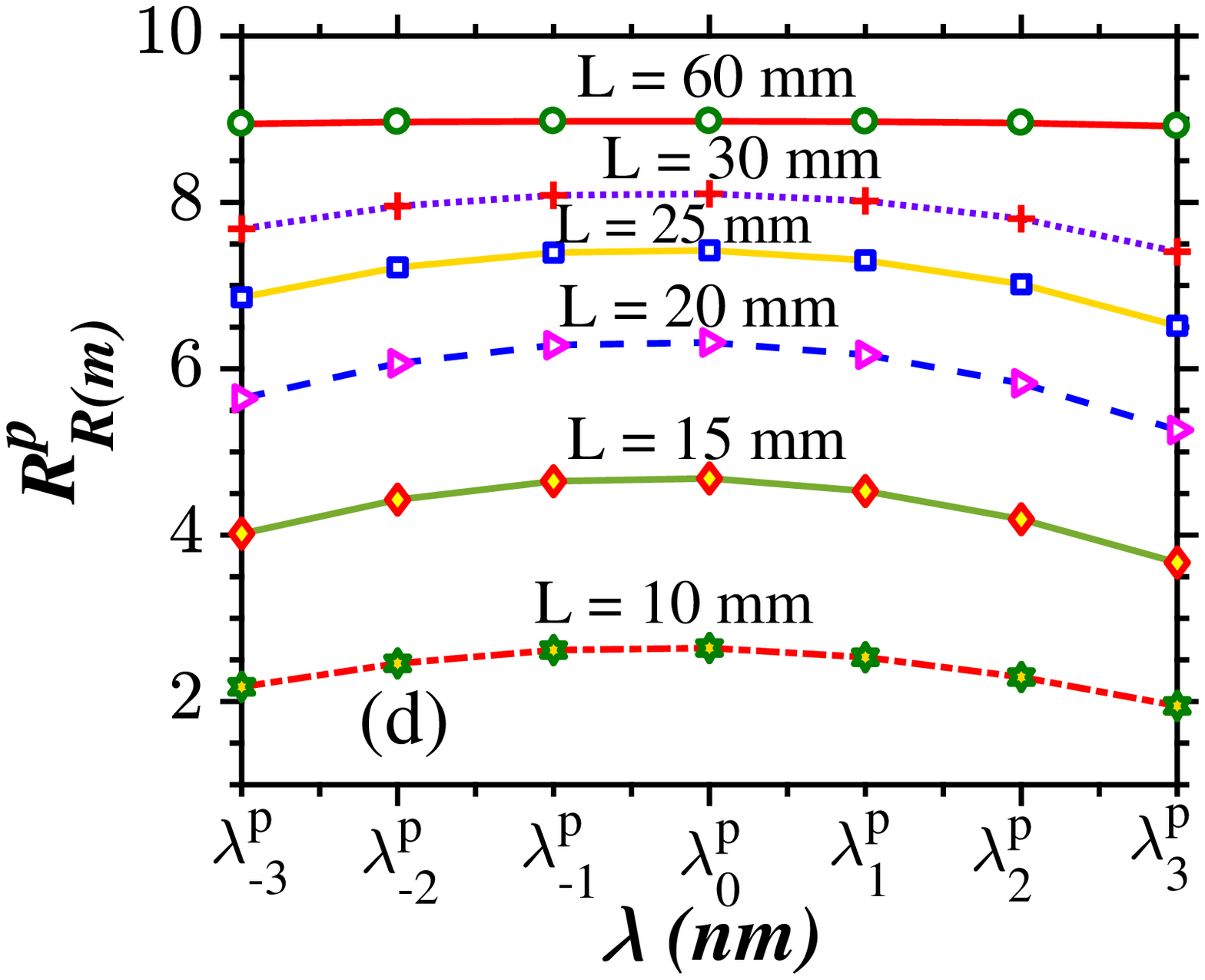}
 	\caption{(a) -- (c) Linear spectrum of a PTSFBG of length $L = 100$ mm for two different sampling periods $s_\Lambda = 500$ $\mu$m and 1000 $\mu$m and other device parameters are the same as in Fig. \ref{fig6}(a). (d) Flattening of the envelope and generation of spectrum with uniform amplitudes by varying the length ($L$) given that the values of $m$ and $\lambda_m^p$ are the same as given in Fig. \ref{fig6} for a sampling period of $s_\Lambda$ = 1000 $\mu$m.}
 	\label{fig7}
 \end{figure}
It can be inferred from Figs. \ref{fig7}(a) -- \ref{fig7}(c) that the reflectivity and transmittivity of all the individual channels in the selected spectral span can be made almost uniform by increasing the physical length of the device ($L$). This behavior is observed to be true for all the values of the sampling period. Also, the reflectivity of each channel is dramatically increased with an increase in the physical length of the device ($L$) for both left and right light incidences. As a special case, the flattening of the envelope of the spectrum  for right incidence is shown in Fig. \ref{fig7}(d) which confirms that the systems with longer physical lengths are optimal for generation of spectrum with uniform reflectivity peaks which is a highly desired feature in the perspective of PTSFBG spectra. 
 
\subsection{Variations in the duty cycle ($d$)}
\begin{figure}
	\centering	\includegraphics[width=0.5\linewidth]{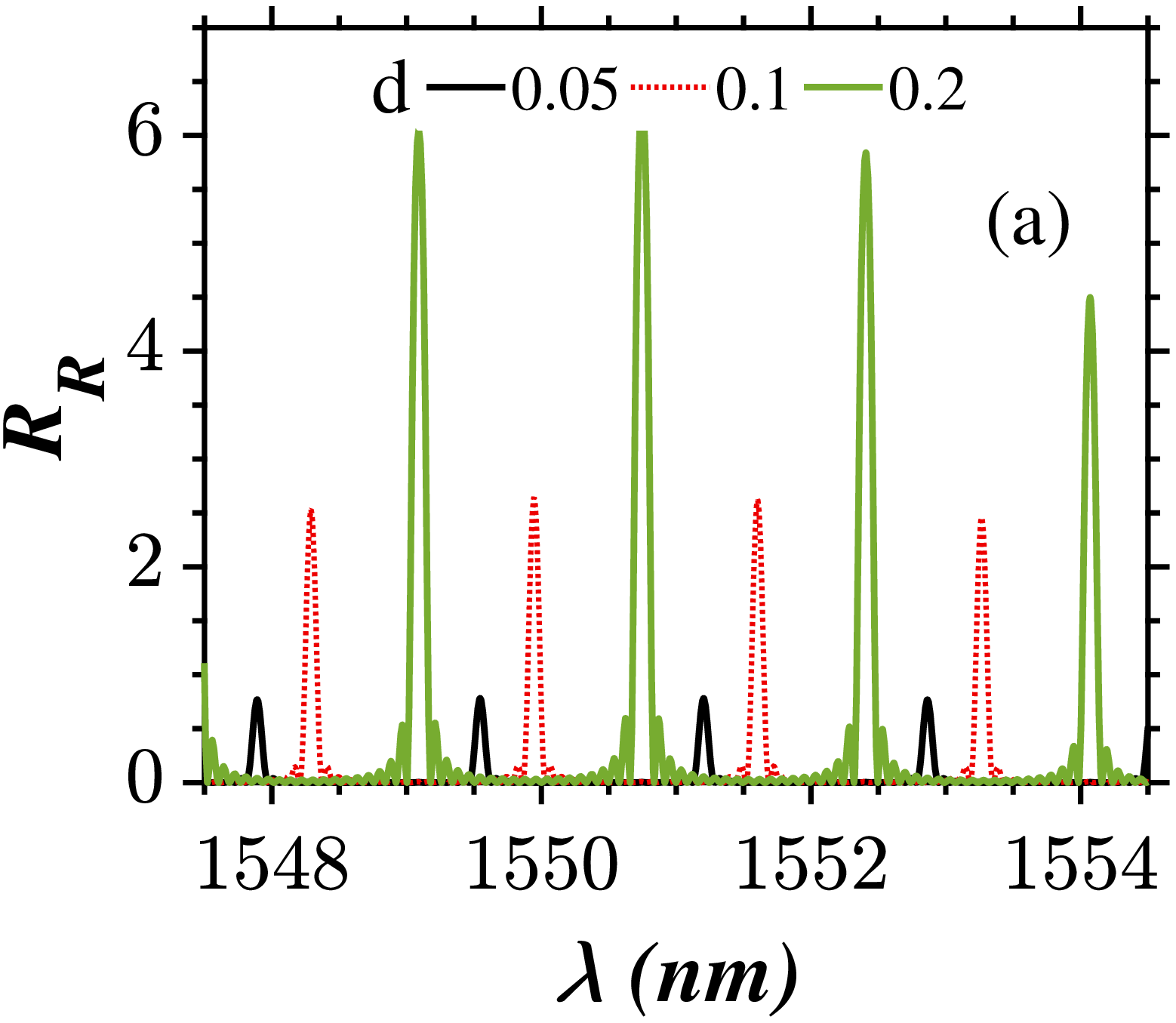}\includegraphics[width=0.5\linewidth]{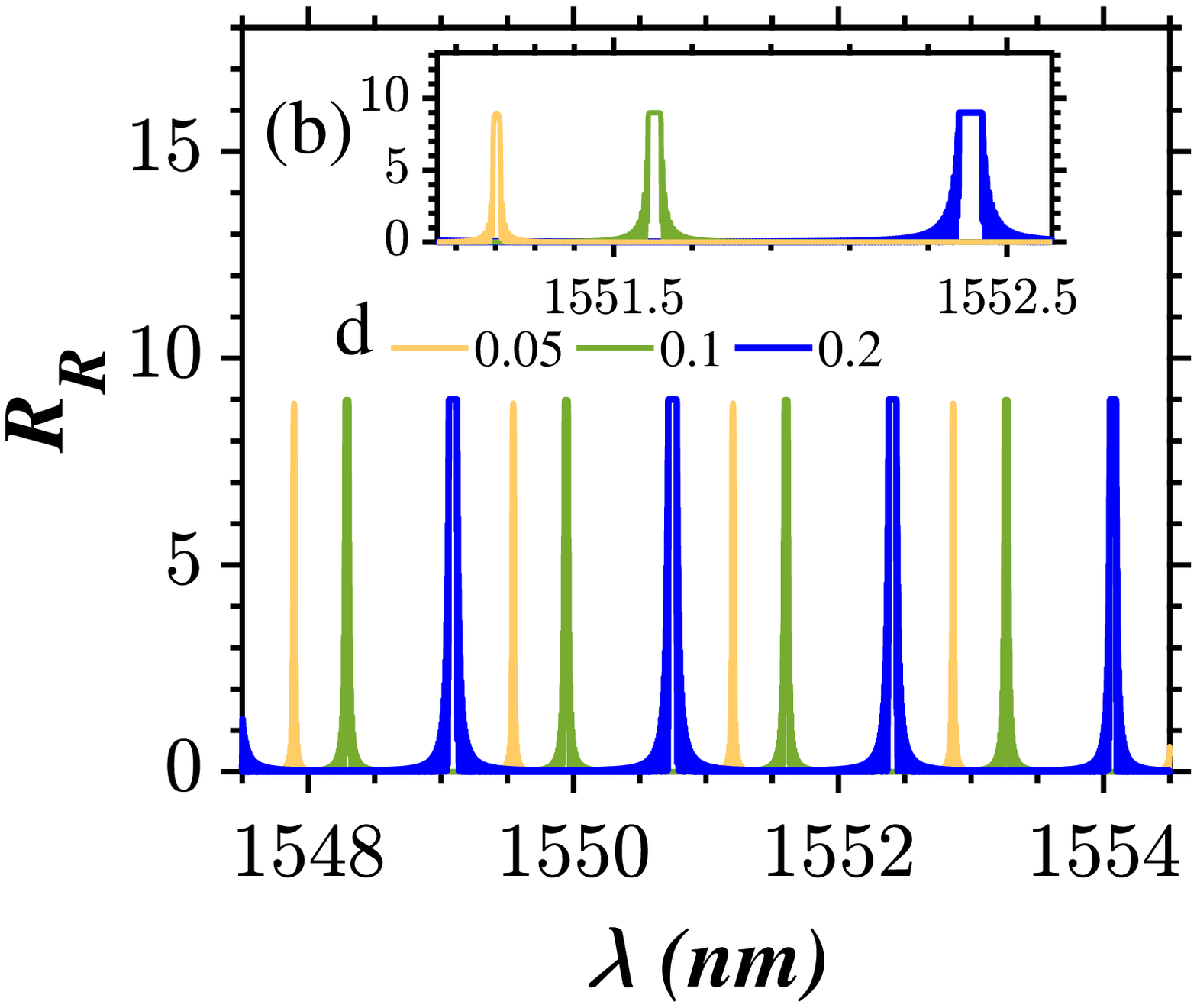}\\\includegraphics[width=0.5\linewidth]{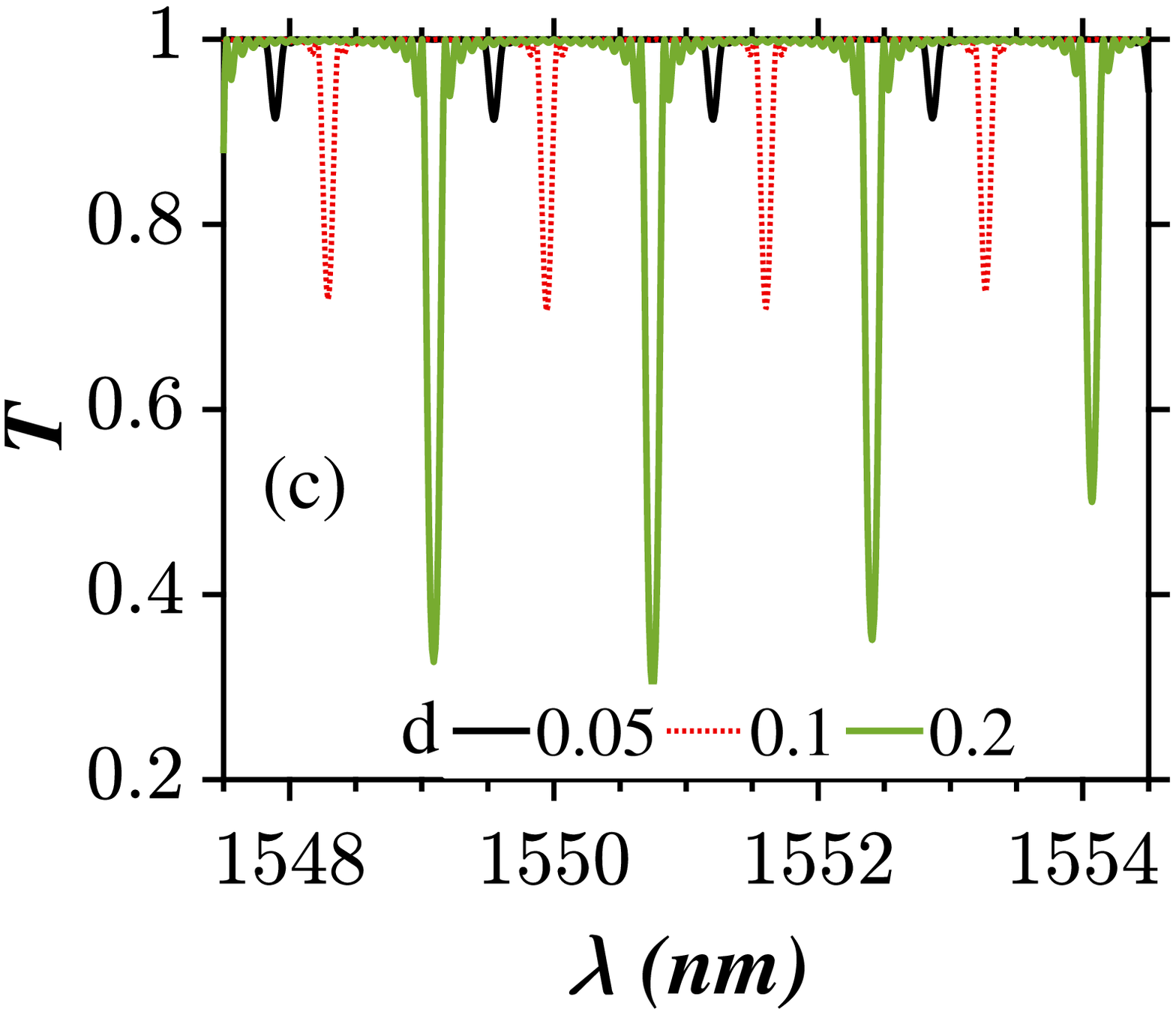}\includegraphics[width=0.5\linewidth]{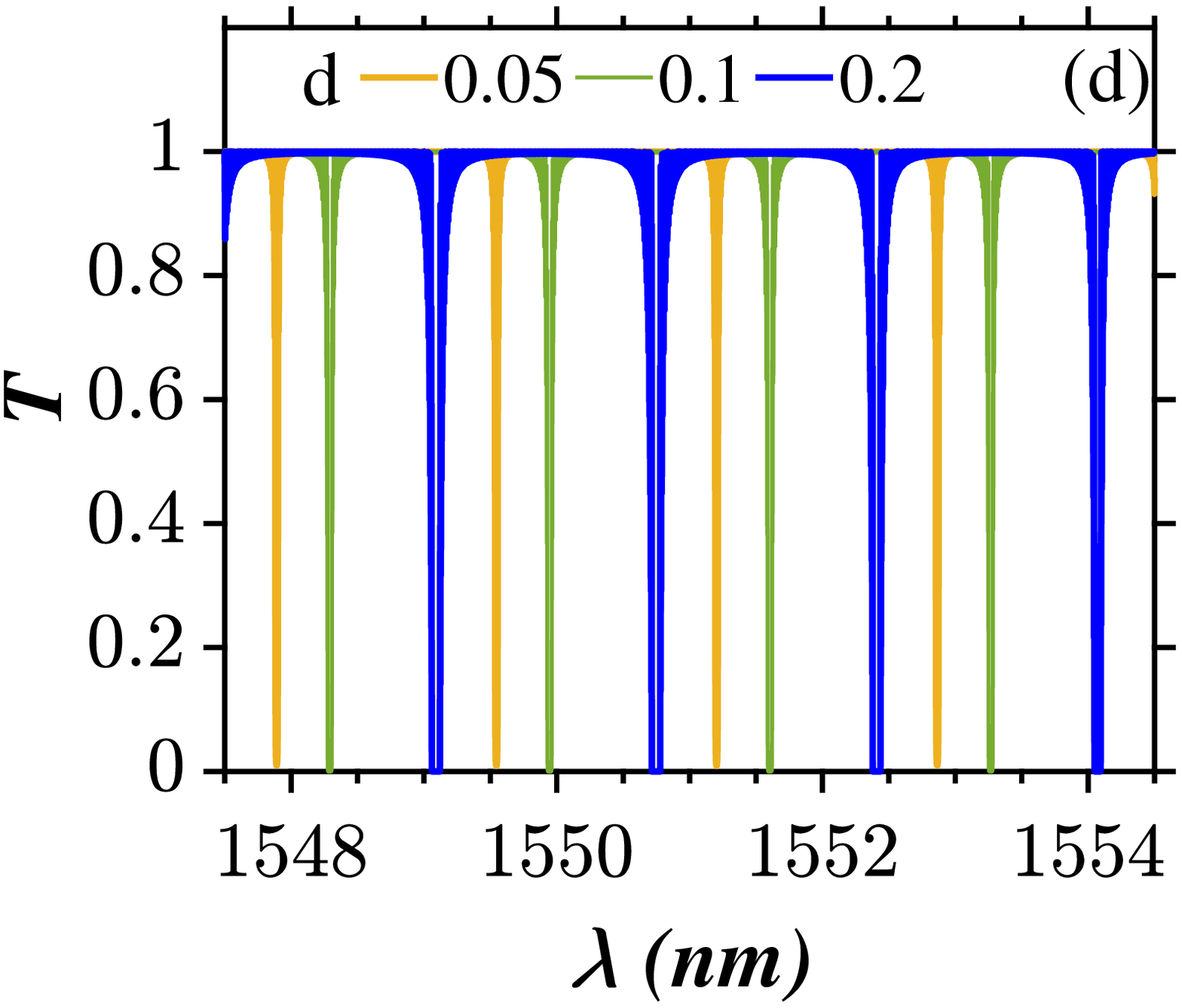}\\\includegraphics[width=0.5\linewidth]{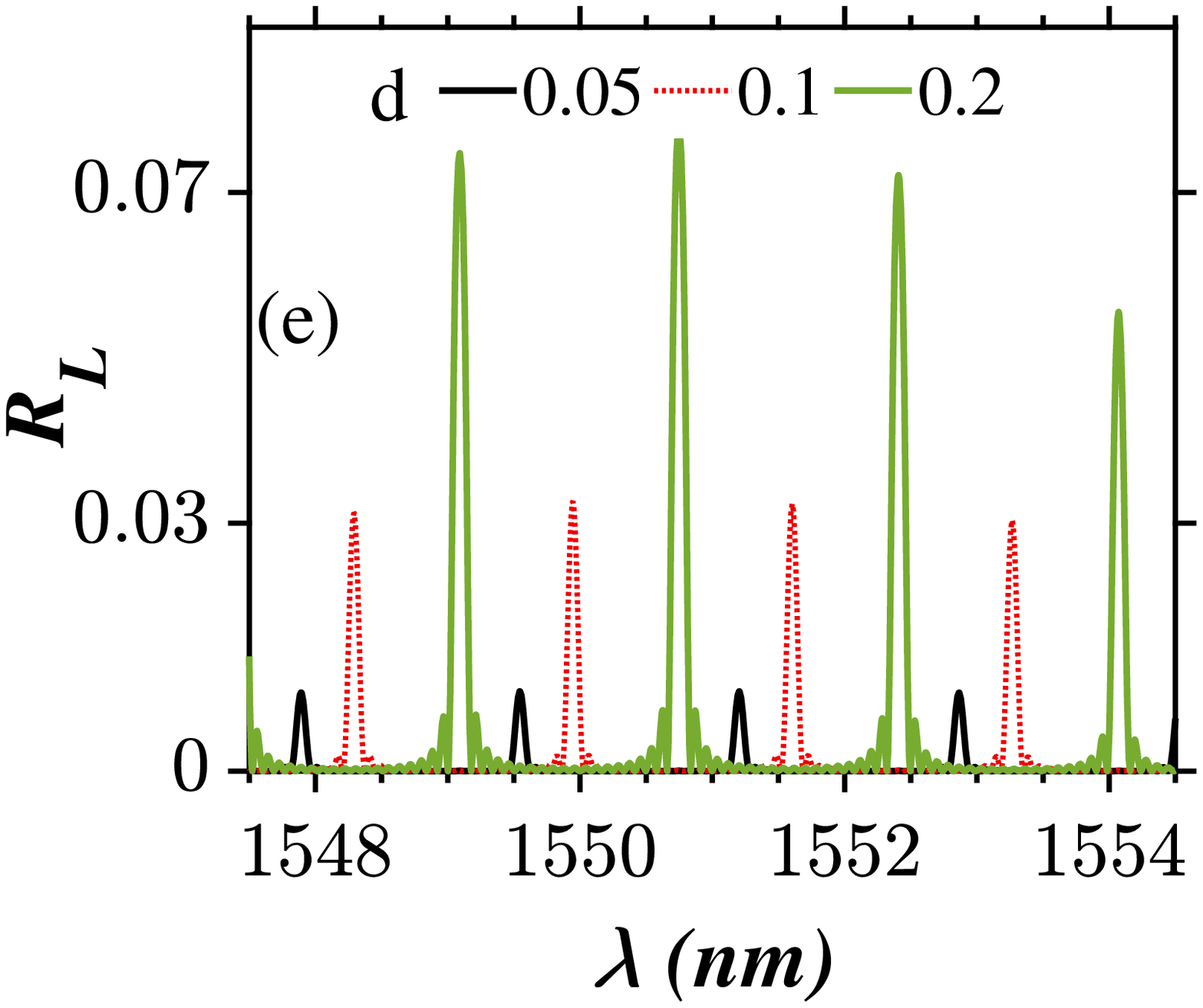}\includegraphics[width=0.5\linewidth]{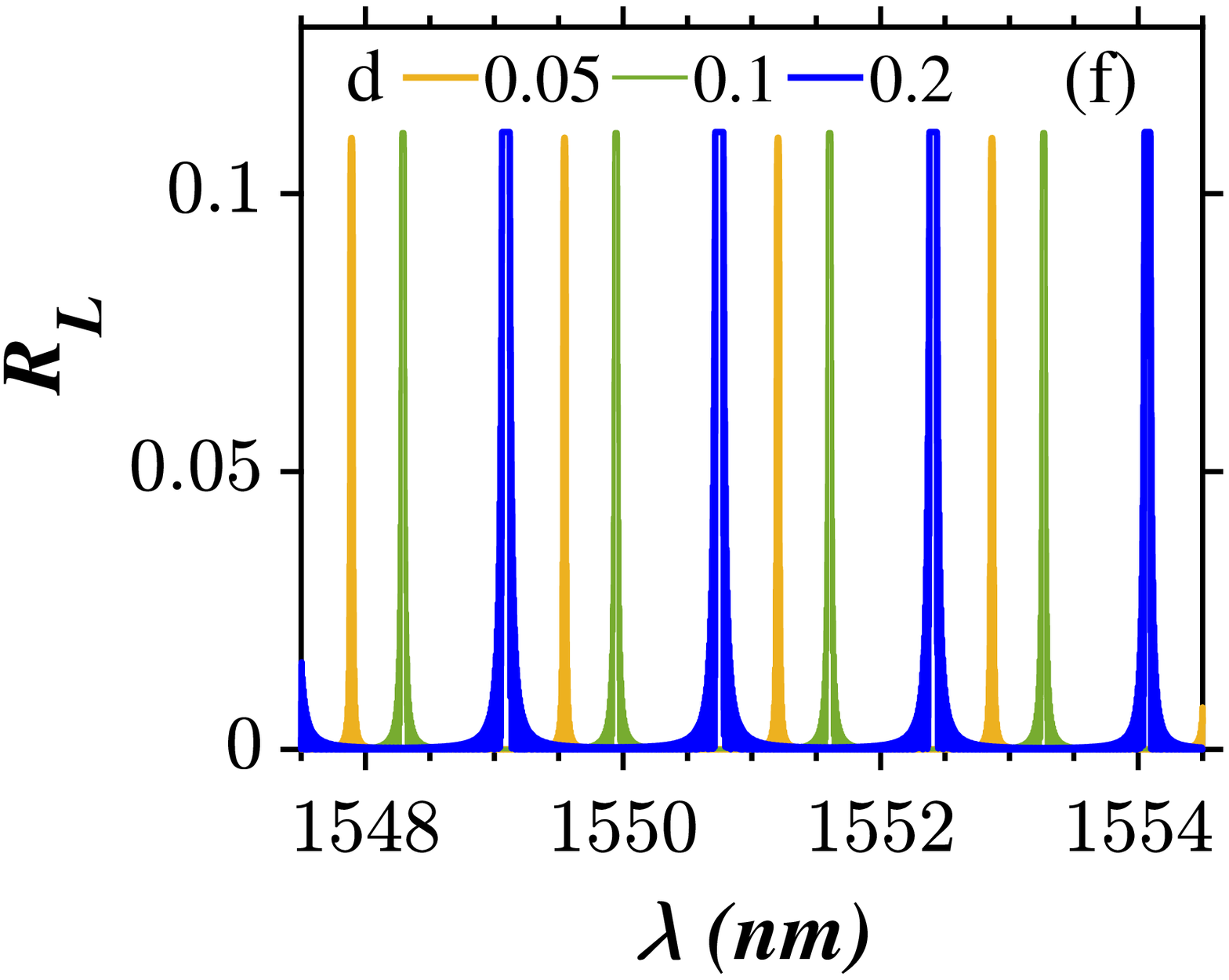}\\\includegraphics[width=0.75\linewidth]{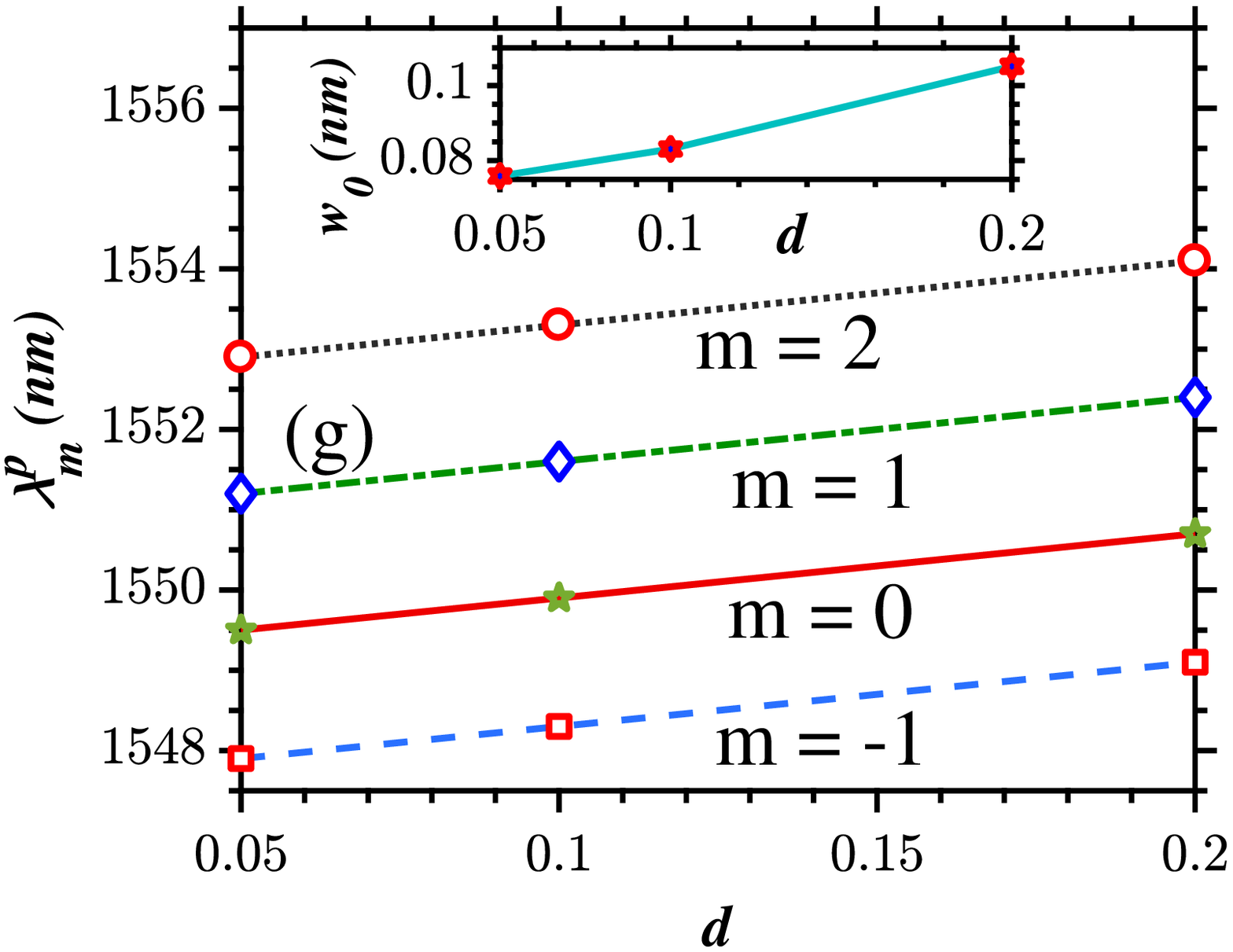}
	\caption{(a) -- (f) Reflection and transmission spectrum of a PTSFBG under varying duty cycle $d$ with $s_\Lambda = 500$ $\mu$m and $n_{1I} = 0.0004$. The plots in the left and right panel are simulated at $L = 10$ mm and $100$ mm. (g) Variation in center wavelength ($\lambda_m^{p}$) with changes in the duty cycle ($d$). Also, the variation in the full width half maximum of the zeroth order mode ($w_0$) is shown in inset. }
	\label{fig8}
\end{figure}
To enable an understanding about the role of duty cycle on the resulting PTSFBG spectra, the sampling period is kept at $s_\Lambda = 500$ $\mu$m. The duty cycle is varied between 0.05 to 0.2 for two different lengths of the grating given by $L = 10$ and $100$ mm in the left and right panels of Fig. \ref{fig8}, respectively. From these figures and Eq. (\ref{Eq:norm2}), it is very clear that if the length of the sample ($s_L$) is increased at a fixed value of sampling period ($s_\Lambda$), it gives rise to an increase in FWHM of the individual channels as shown in Fig. \ref{fig8}(a) and inset of Fig. \ref{fig8}(e). Furthermore, it causes a shift in the spectrum towards longer wavelength sides irrespective of the sampling period and length of the device as shown in Fig. \ref{fig8}(g). Also, an increase in the duty cycle leads to an increase in the magnitude of reflectivity for both left and right incidences besides growth in the dip of the transmittivity for lower values of length of the device ($L < 60$ mm) as depicted in Figs. \ref{fig8}(a), \ref{fig8}(c) and \ref{fig8}(e). Hence, it is proven that the reduction in reflectivity with a number of channels can be judiciously controlled (at lower lengths) in multiple ways, via variations in gain and loss, proper choice of duty cycle, and the device length. However, for larger values of physical length of the device such as $L = 100$ mm,  it helps only in shifting the center wavelength of the spectrum and increasing the full width half maximum ($w_m$) as shown in Figs. \ref{fig8}(b), \ref{fig8}(d) and \ref{fig8}(f).

\subsection{Influence of modulation strength  ($n_{1R}$)  on FWHM}
\begin{figure}
	\centering	\includegraphics[width=0.5\linewidth]{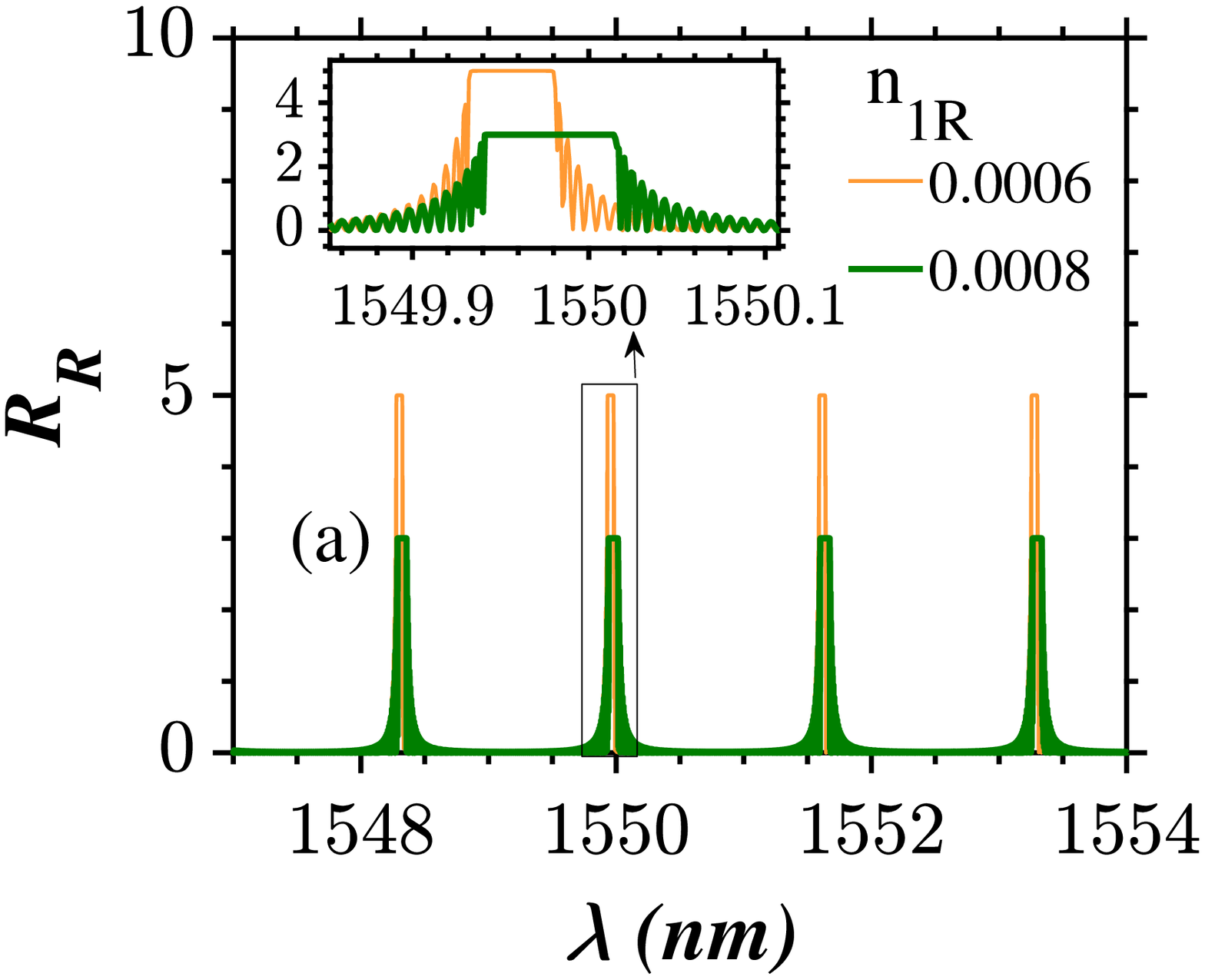}\centering	\includegraphics[width=0.5\linewidth]{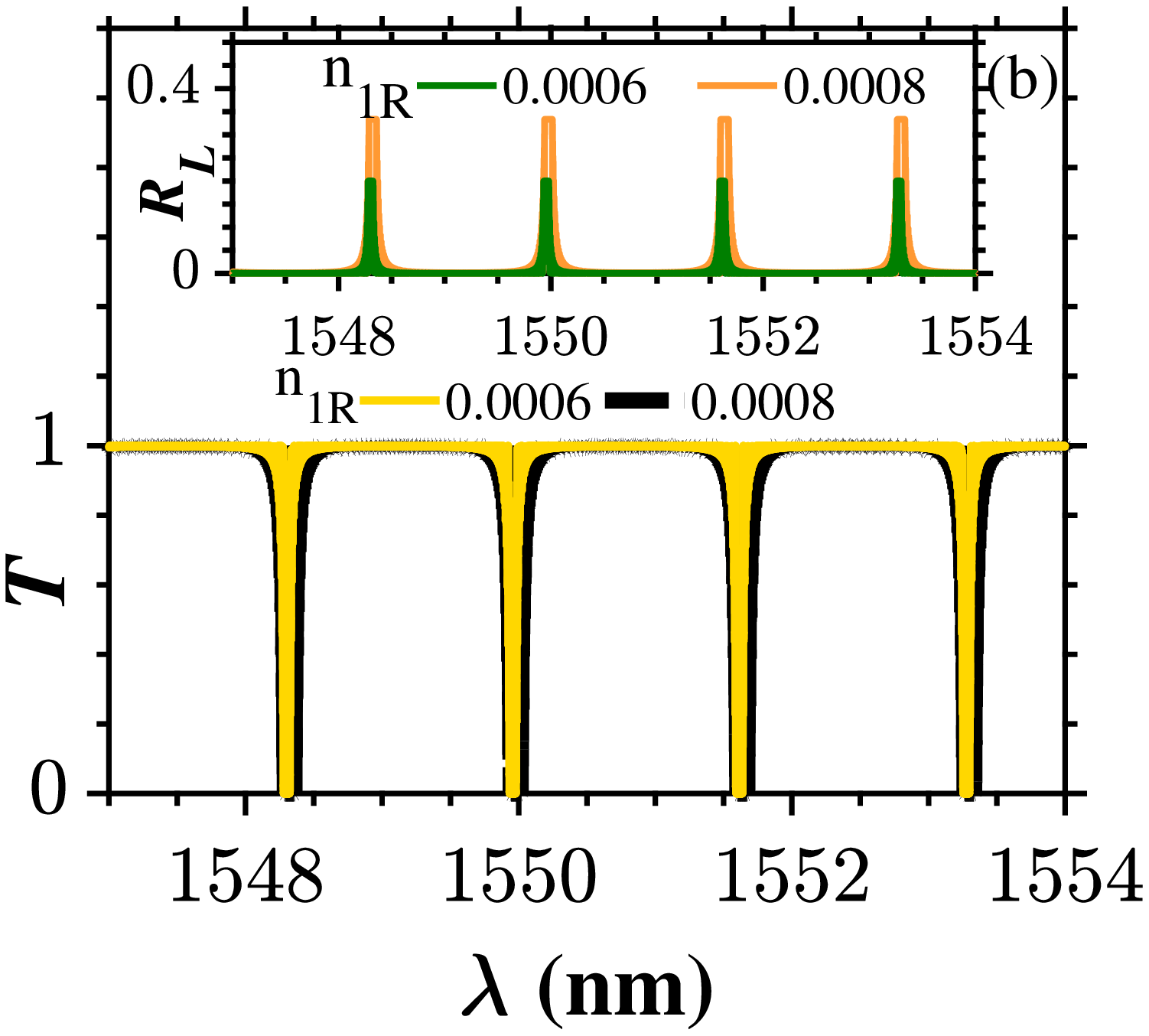}
	\caption{Variations in FWHM of PTSFBG spectrum with changes in $n_{1R}$ at $n_{1I} = 0.0004$, $d = 0.1$, and $s_\Lambda = 500$ $\mu$m.}
	\label{fig9}
\end{figure}
Finally, the role of the coupling parameter ($\kappa = \pi n_{1R}/\lambda$) on the PTSFBG spectrum is illustrated in Fig. \ref{fig9}. The shape of the stopband of individual channels and the corresponding FWHM are influenced by different values of the real part of the modulation strength ($n_{1R}$). If the individual channels need to posses a flat stopband and broader width, one can opt for larger modulation strengths as shown in the inset of Fig. \ref{fig9}(a). On the other hand, the stopband is tapered at weaker modulation strengths as shown in Figs. \ref{fig9}(a) and \ref{fig9}(b).  As depicted by Fig. \ref{fig9}(a), the decrease in $R_R$  is not a major issue here, since there are other degrees of freedom offered by the PTSFBG to compensate for such a reduction in the reflectivity. Nevertheless, the reflectivity of spectrum for left incidence increases with any increase in $n_{1R}$ as shown in the inset of Fig. \ref{fig9}(b).
\subsection{Application: Tunable RF traversal filter}
\begin{figure}[t]
	\centering	\includegraphics[width=1\linewidth]{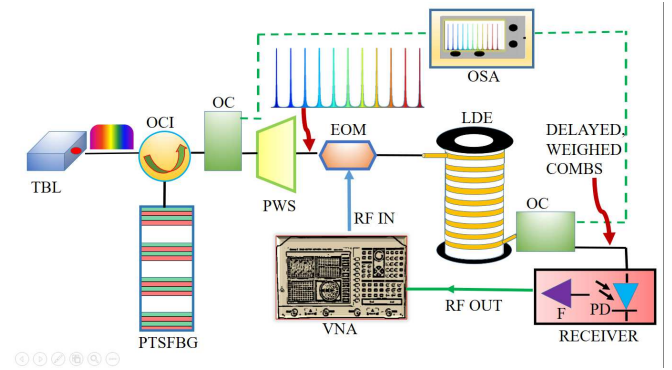}
	\caption{Schematic showing the generation of optical combs with a PTSFBG and its application to a tunable RF traversal filter \cite{xu2019advanced, pastor2001broad, leng2004optimization, davies1984fibre, hamidi2010tunable,mora2003tunable}. TBL: tunable broadband laser, OCI: optical circulator, OC: optical coupler, PWS: programmable wave shaper, EOM: electro optic modulator, LDE: linear dispersive element (single mode optical fiber spool), OSA: optical spectrum analyzer, PD: photodiode, F: filter, VNA: vector network analyzer, RF: radio frequency.}
	\label{fig10}
\end{figure}
An optical field from a tunable broadband laser source can be directed as the input to the PTSFBG via an optical ciruclator (OCI). The resulting comb spectrum is modulated by an electro optic modulator (EOM) \cite{leng2004optimization}.  To impose the desired tap weights, the intensity of each channel 
can be judiciously controlled with the aid of programmable wave shaping devices \cite{xu2019advanced,davies1984fibre,mora2003tunable,mora2002automatic}. The RF input to the EOM is nothing but the message signal which is modulated on a desired carrier frequency and must be introduced alongside the optical combs \cite{xu2019advanced,davies1984fibre}. The EOM generates replicas of RF inputs to optical outputs \cite{pastor2001broad}. The resulting optical signal is then passed into a single mode fiber (SMF) spool of length $L_f = $ 23 or 50 km \cite{pastor2001broad,leng2004optimization}. The SMF offers linear dispersion characteristics to the input signals and as a result each modulated comb channel is provided with a precise time delay ($\tau$). The magnitude of the delay  is  determined by the product of wavelength separation ($\Delta_\lambda$) between individual channels of the comb spectrum and the dispersion ($D$) offered by the SMF \cite{leng2004optimization}. In the end, the delayed and weighted optical taps are mixed at the receiver (photo diode and a optical filter) \cite{pastor2001broad,xu2019advanced}. The resulting RF output signal can then be sent to a vector
network analyzer (VNA). The VNA assists in recording and analyzing the RF response of the different frequency channels \cite{xu2019advanced,leng2004optimization}. This kind of RF traversal filters are potential candidates to realize any given RF transfer function with ease by tuning the appropriate tap weights \cite{mora2003tunable}.

\section{Unitary transmission point dynamics}
\label{Sec:4}
\begin{figure}
	\centering	\includegraphics[width=0.5\linewidth]{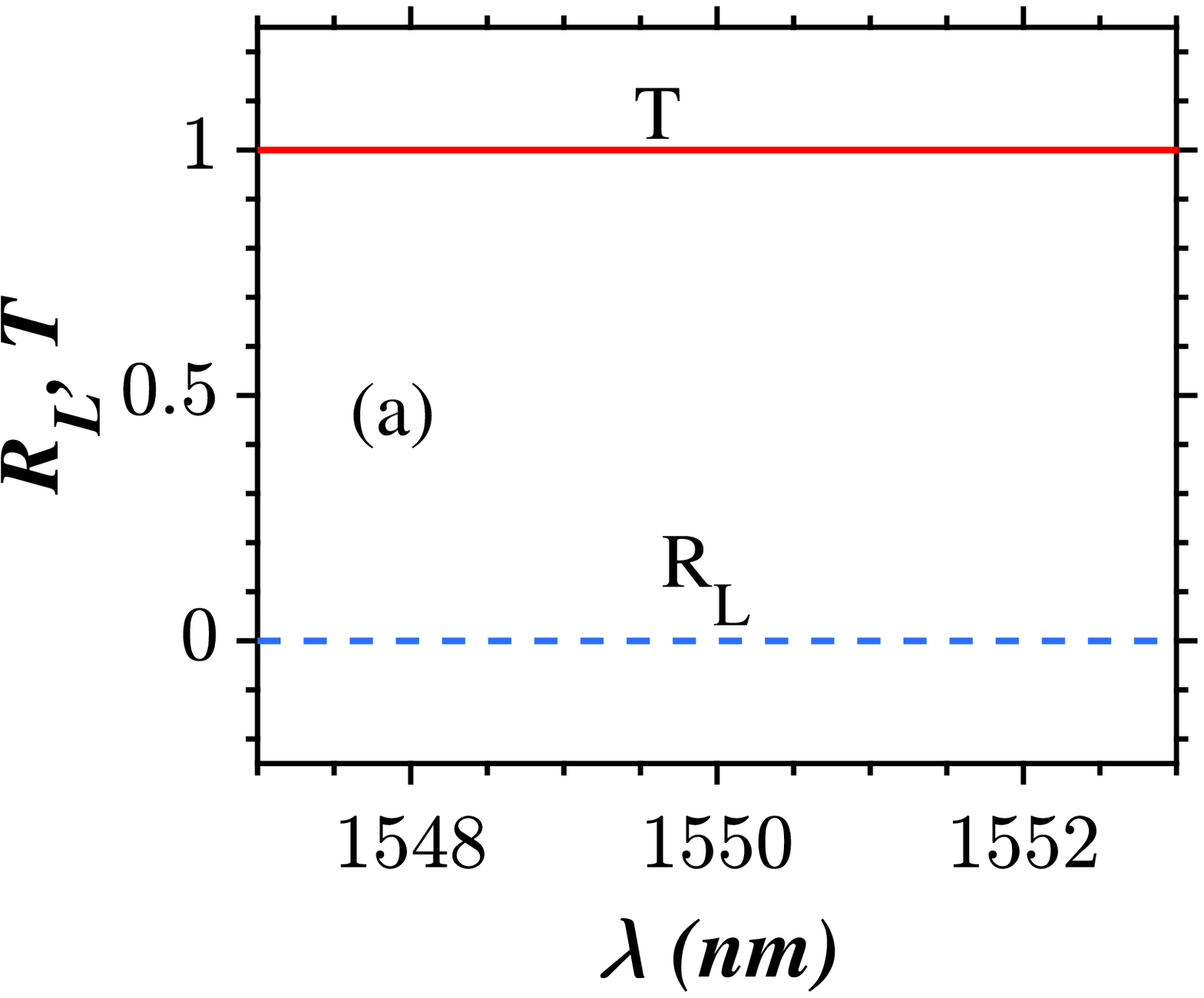}\centering	\includegraphics[width=0.5\linewidth]{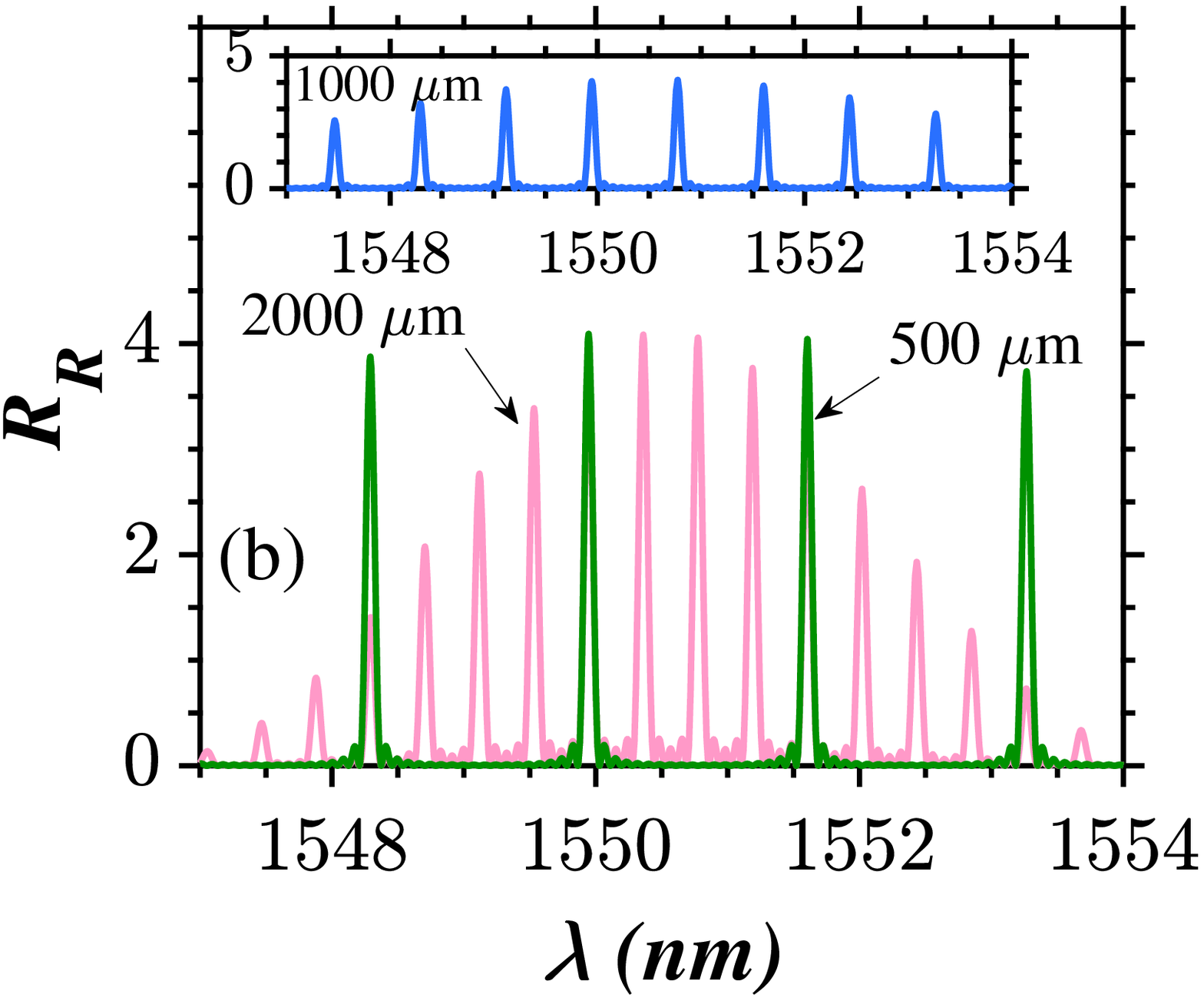}\\\centering	\includegraphics[width=0.5\linewidth]{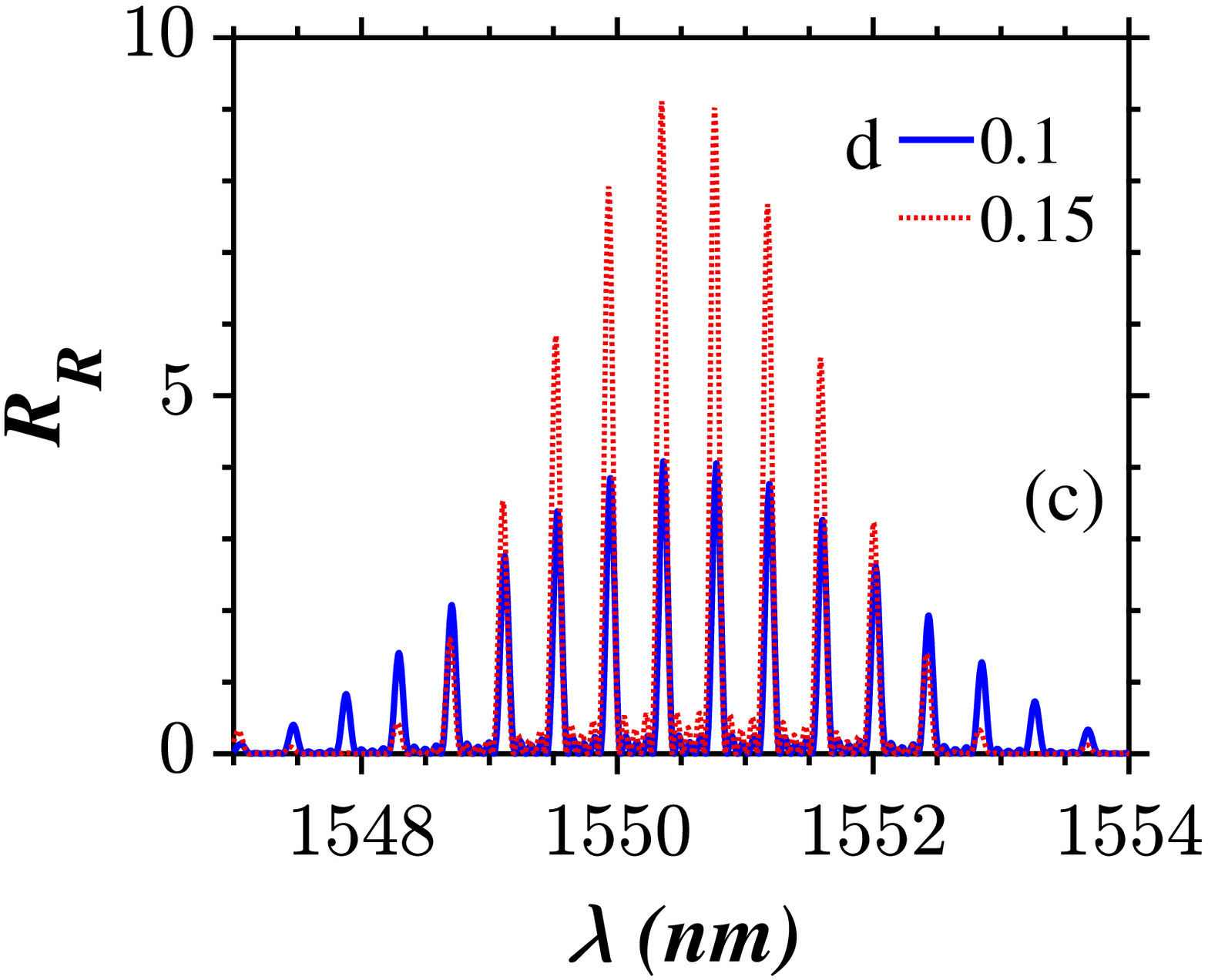}\centering	\includegraphics[width=0.5\linewidth]{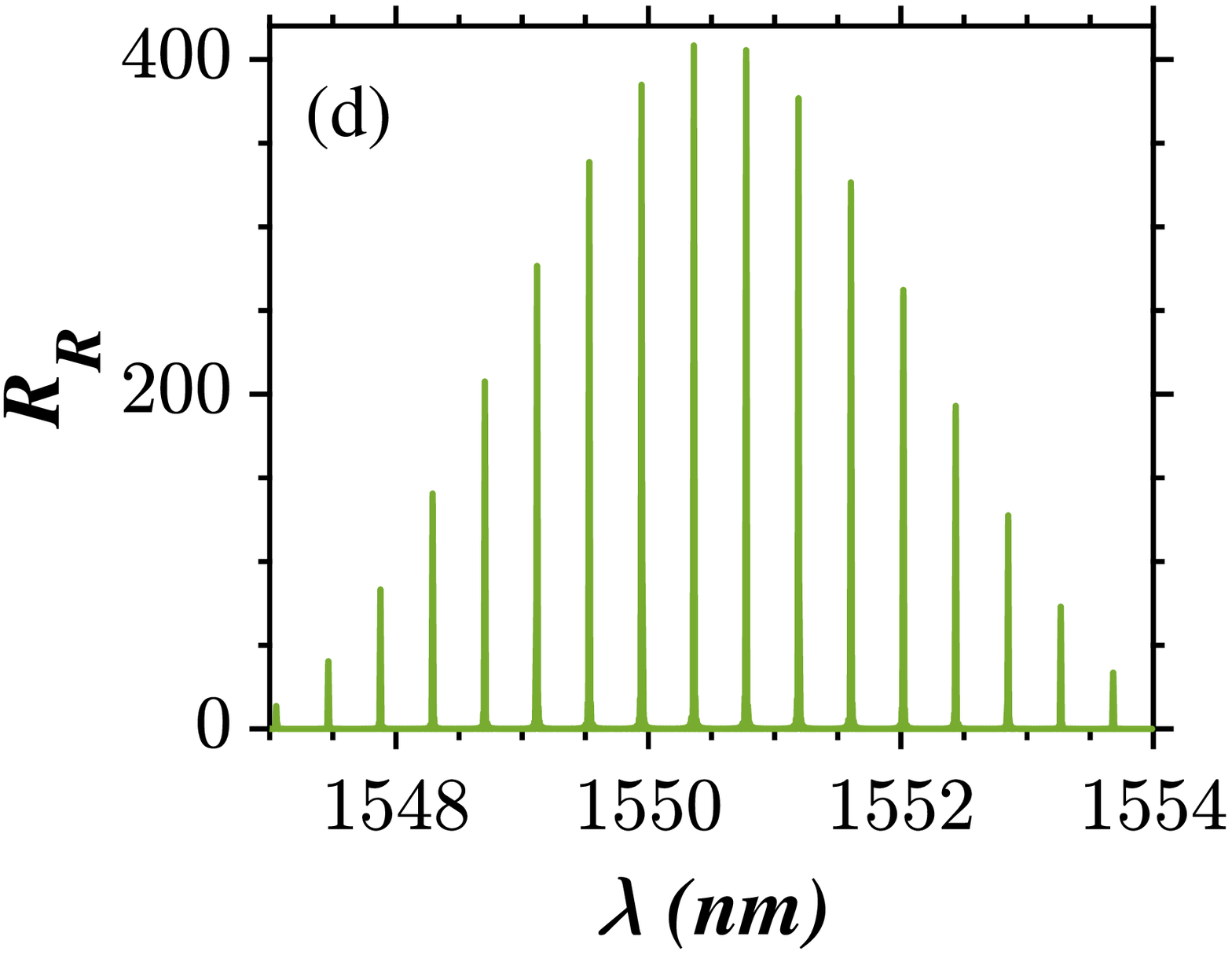}
	\caption{(a) Reflection-less wave transport phenomenon in a PTSFBG operating at the unitary transmission point ($n_{1R}$ = $n_{1I}$ = 0.0005). (b) Depicts the role of sampling parameter ($s_\Lambda$) at $L = 10$ mm and $d = 0.1$. (c) Portrays the effect of variation in the duty cycle ($d$) at $s_\Lambda$ = $2000$ $\mu$m and the length ($L$) is the same as in (b). (d) Illustrates the change in spectrum with device length $L = 100$ mm at $s_\Lambda = 2000$ $\mu$m and $d = 0.1$.  }
	\label{fig11}
\end{figure}

At this juncture, it is essential to recollect that any type of $\mathcal{PT}$-symmetric FBG device functioning at the unitary transmission point ($n_{1R} = n_{1I}$) will demonstrate ideal light transmission ($T = 1$ and $R_L = 0$), if the light is launched from the front end (left) of the device \cite{huang2014type, raja2020tailoring,raja2020phase,lin2011unidirectional,kulishov2005nonreciprocal}. From Fig. \ref{fig11}(a),
 we confirm that reflection-less light transport is exhibited by the PTSFBG irrespective of the variations in the sampling length ($s_L$), sampling period ($s_\Lambda$), duty cycle ($d$) or the length of the device ($L$). However, any increase in the sampling period ($s_\Lambda$) increases the number of channels and vice-versa for the right light incidence as shown in Fig. \ref{fig11}(b). Moreover, any increase in the duty cycle ($d$) increases the reflectivity ($R_R$) of the channels closer to $1550$ nm and marginally shifts the individual peaks as shown in Fig. \ref{fig11}(c). In contrast to Fig. \ref{fig11}(b), the reflectivity of each channel is dramatically increased in Fig. \ref{fig11}(d) when the length of the device is increased to $L = 100$ mm. It should be mentioned that these behaviors can be explored in the creation of comb filters.

\section{Broken $\mathcal{PT}$-symmetric regime}
\label{Sec:5}
Any PTFBG will exhibit lasing behavior in its spectrum under the operating conditions, $n_{1I}>n_{1R}$. Under this condition, sharp variations (exponential increase or decrease) in the reflectivity and transmittivity of the grating spectrum occur with variation in the value of $n_{1I}$. Also, the FWHM of the spectra in the broken regime is too narrow compared to the unbroken regime and thus a large amount of reflected (transmitted) power is concentrated in a narrow spectral span. For these reasons, the dynamics of the system in the broken regime is known as lasing behavior. With this note, we directly present the results pertaining to the lasing behavior exhibited by a PTSFBG in its spectrum (comb). Throughout this section, the modulation strength parameter and length are kept constant as $n_{1R} = 5\times10^{-4}$ and  $L = 10$ mm, respectively (unless specified).

\subsection{Impact of $n_{1I}$ on the lasing spectra}\label{sub:1}
\begin{figure}
	\centering	\includegraphics[width=1\linewidth]{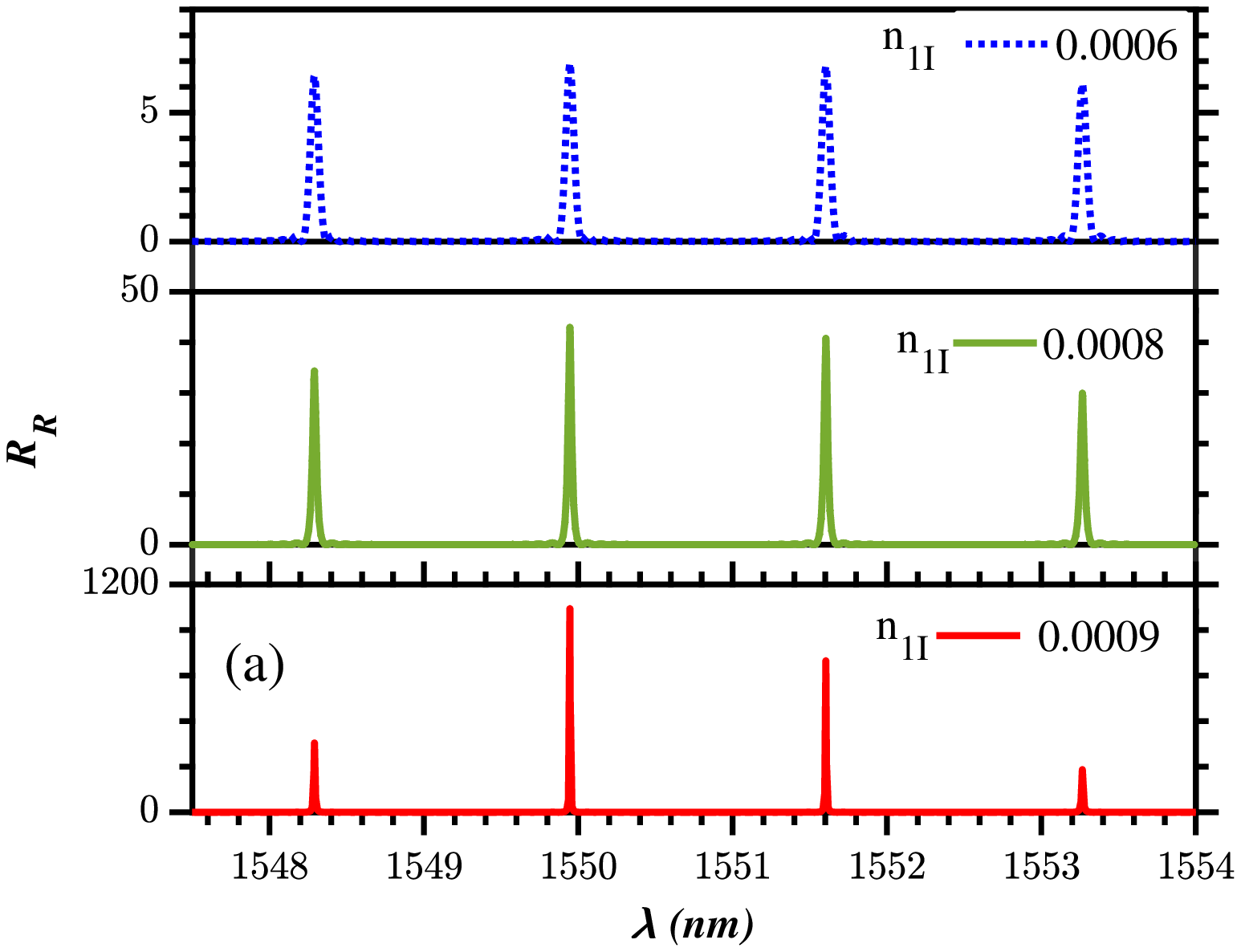}\\\centering	\includegraphics[width=1\linewidth]{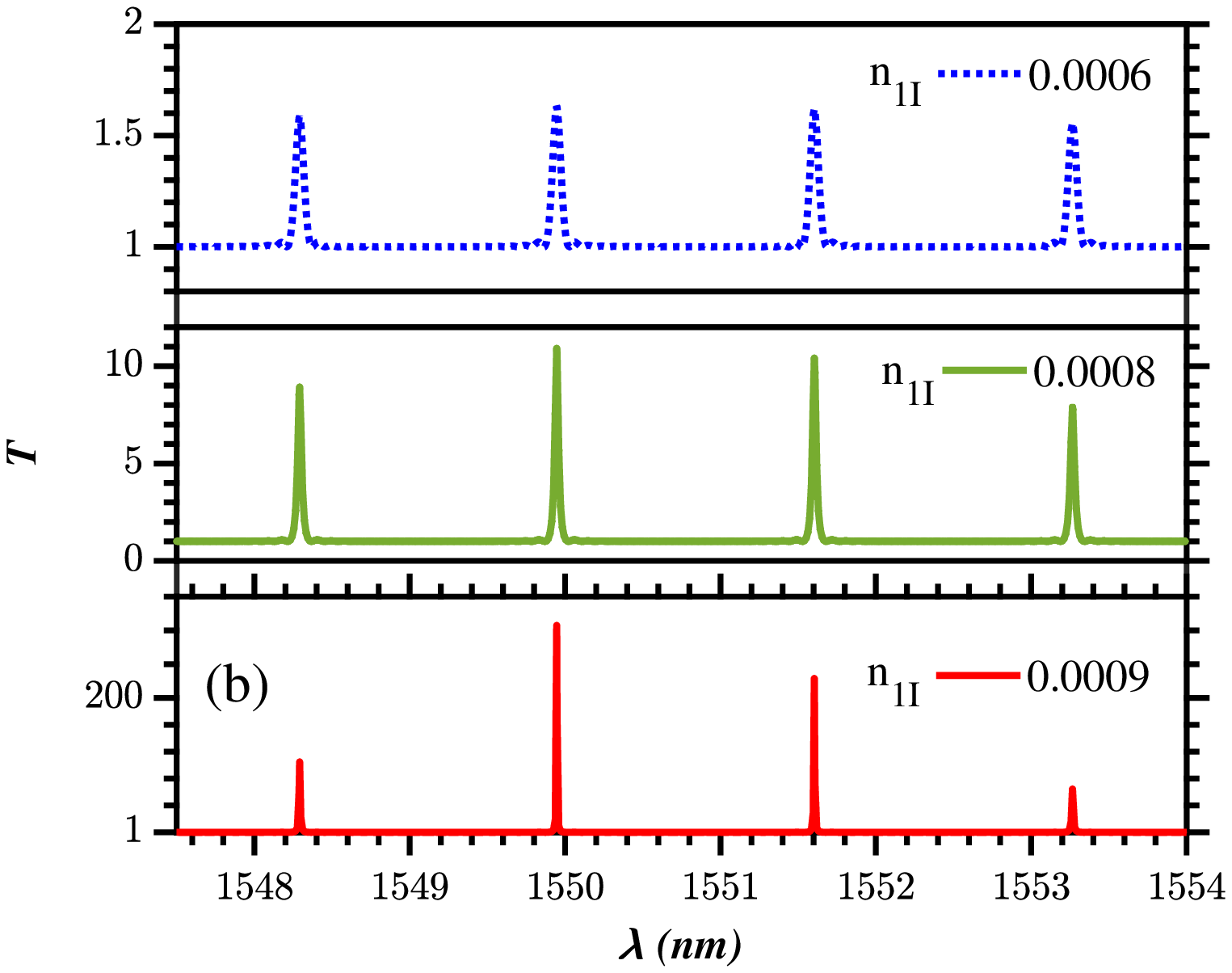}\\\includegraphics[width=1\linewidth]{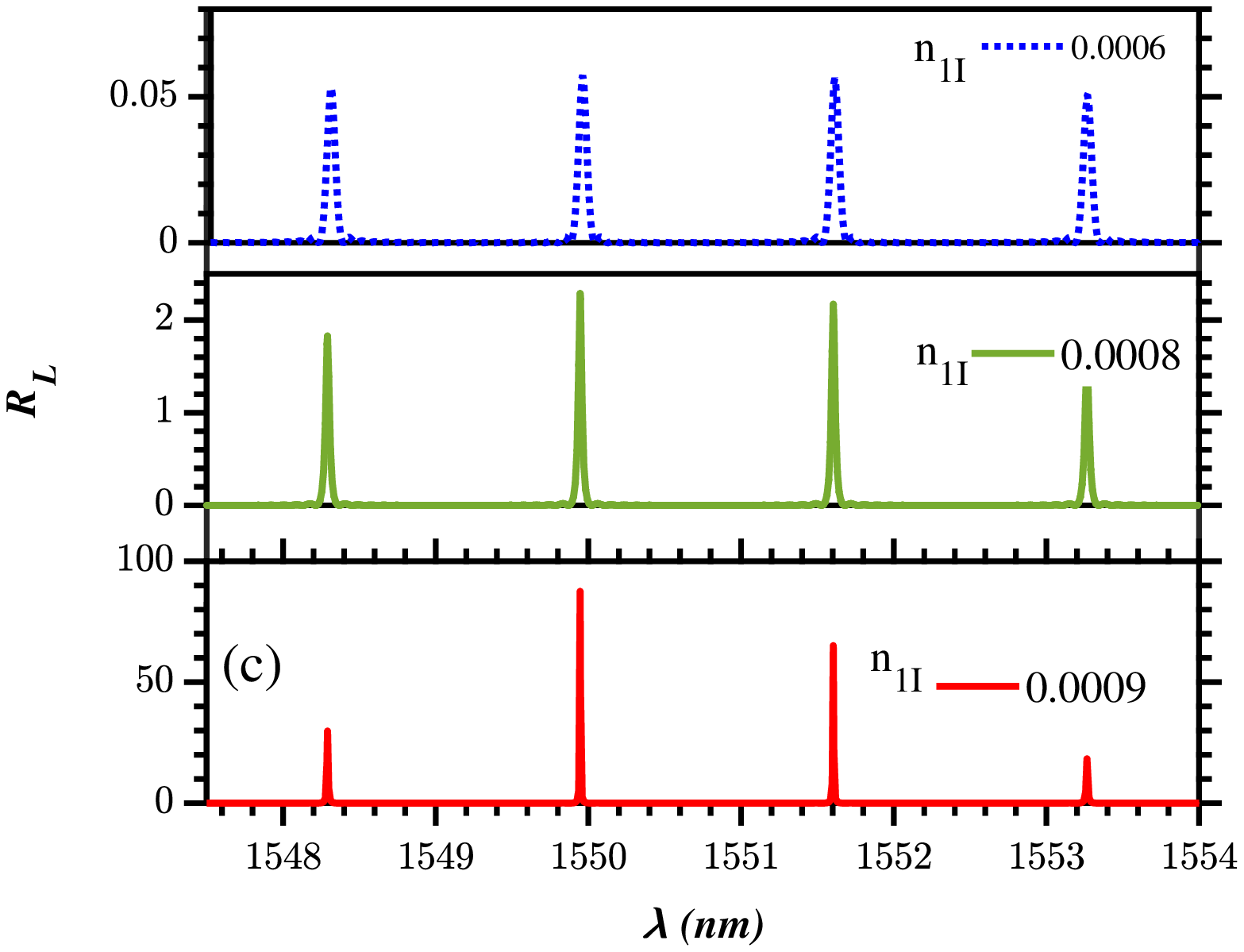}
	\caption{Lasing spectrum of a PTSFBG of length $10$ mm with a sampling period of $s_\Lambda=500$ $\mu$m and duty cycle of 0.1 under variations in $n_{1I}$. It should be noted that the minimum value of transmittivity is always greater than unity in the perspective of the system operating in the broken $\mathcal{PT}$-symmetric regime. }
	\label{fig12}
\end{figure}
Figures \ref{fig12}(a) -- \ref{fig12}(c) depict the formation of lasing combs with fewer modes. From these figures, it is very clear that the value of gain and loss affects the lasing spectrum by two means: Primarily, with an increase in the value of $n_{1I}$, the reflectivity and transmittivity of each mode get intensified. Secondly, the FWHM of the individual wavelengths of lasing spectrum gets reduced with the increase of $n_{1I}$. The mode which is closer to the Bragg wavelength (zeroth order) receives maximum amplification. On either side of the zeroth order mode, the first order lasing modes of the spectrum appear. As the order of the mode increases, both reflectivity ($R_R$ and $R_L$) and transmittivity ($T$) get decreased. The lasing modes at the edges of the spectrum (scaled to $1547.5$ to $1554$ nm here) feature lesser amplification and so the peaks corresponding to individual modes are non-identical in magnitude. Nevertheless, each mode is equally spaced ($\Delta_\lambda$) in a given wavelength range.
     
\subsection{Role of sampling period ($s_\Lambda$) on the number of the modes}\label{sub:2}
\begin{figure}
	\centering	\includegraphics[width=1\linewidth]{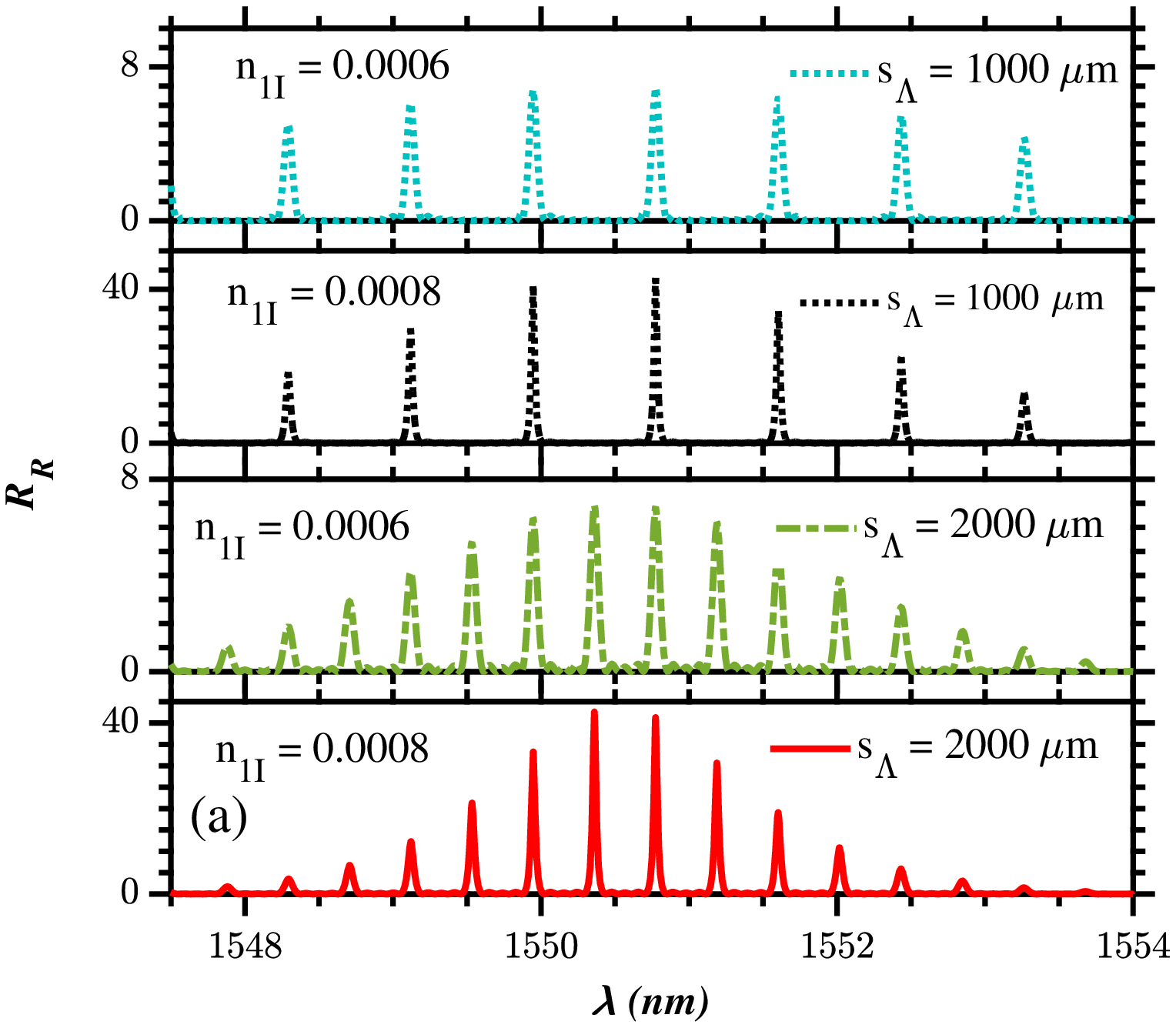}\\\centering	\includegraphics[width=1\linewidth]{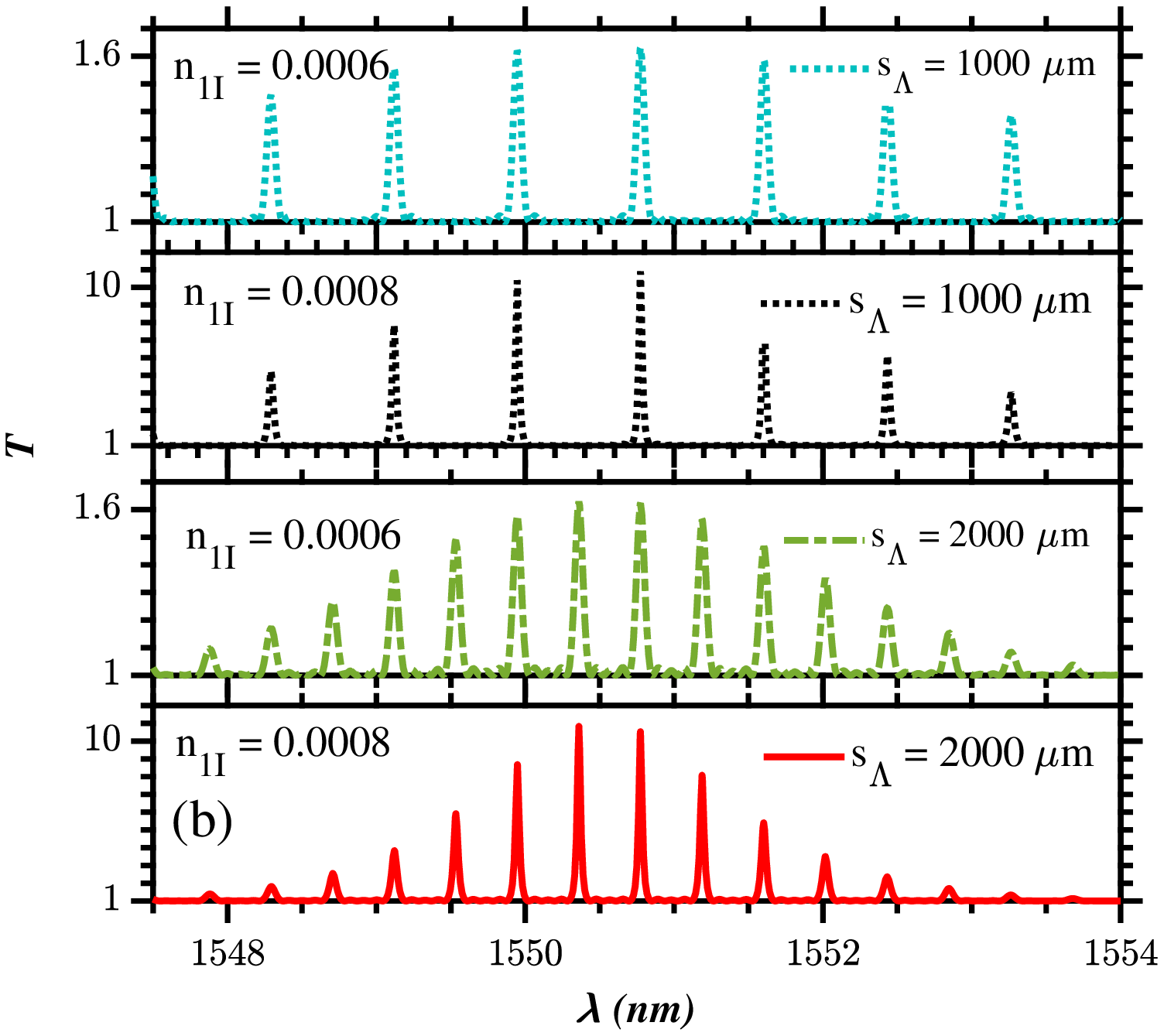}\\\includegraphics[width=1\linewidth]{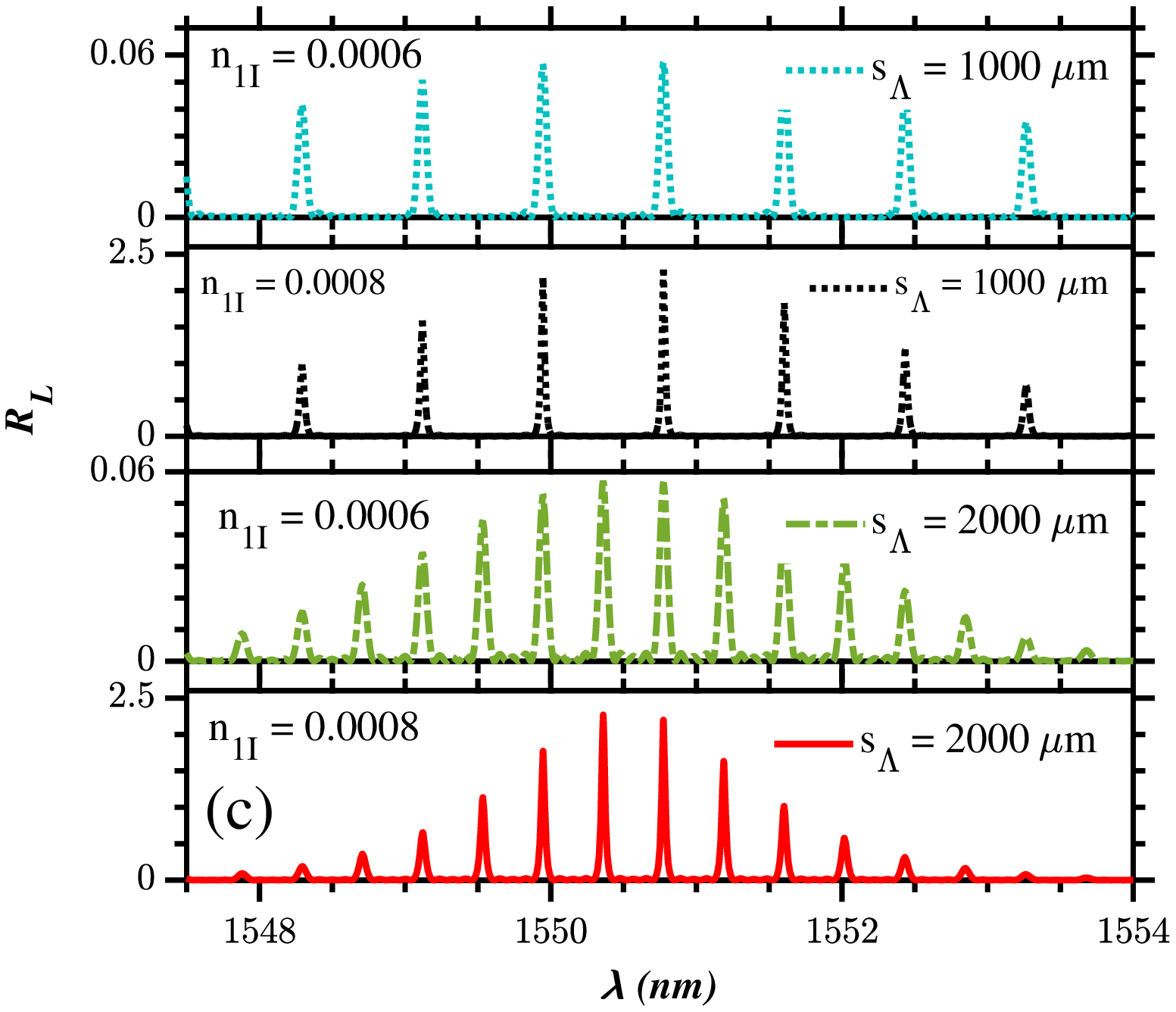}
	\caption{Variations in the number of modes in the lasing spectrum of a PTSFBG for two different sampling periods $s_\Lambda = 1000$ and $2000$ $\mu$m. The length and duty cycle parameters are same as in Fig. \ref{fig12}.}
	\label{fig13}
\end{figure}
In Fig. \ref{fig12}, the number of channels within the available range is few (four). If additional modes are desired in the lasing spectra, the obvious choice is to increase the sampling period as shown in Fig. \ref{fig13}. This also results in a decrease in the wavelength separation ($\Delta_\lambda$) between the adjacent modes of the spectra. For instance, seven channels are visible in the output spectrum for a sampling period of $s_\Lambda$ = $1000$ $\mu$m. As mentioned earlier, the reflectivity (transmittivity) of the modes far away from the Bragg wavelength is not as much as that of the zeroth order modes as depicted in Figs. \ref{fig13}(a) -- \ref{fig13}(c).  

\subsection{Variation in the duty cycle ($d$)}\label{sub:3}
\begin{figure}
	\centering	\includegraphics[width=0.5\linewidth]{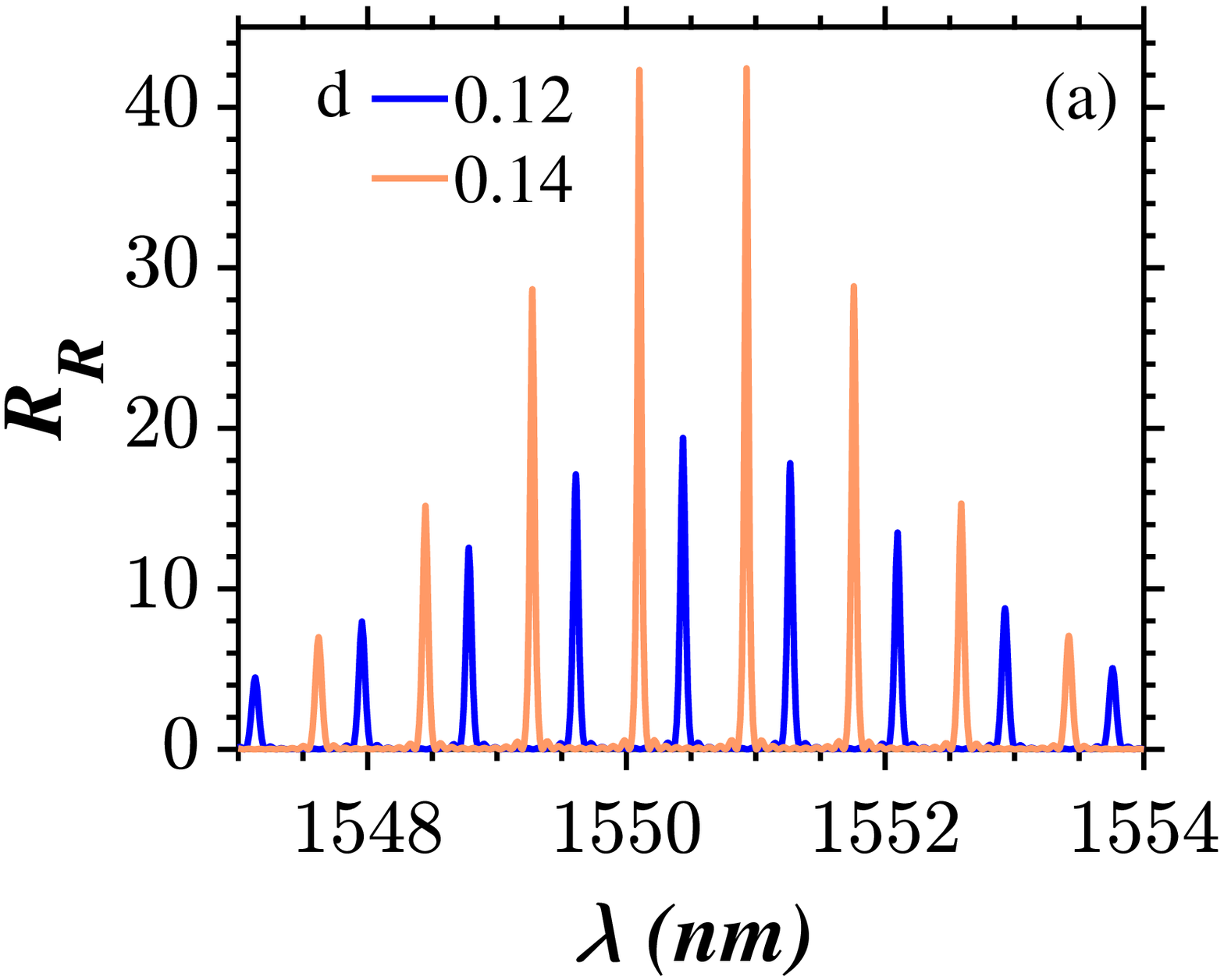}\includegraphics[width=0.5\linewidth]{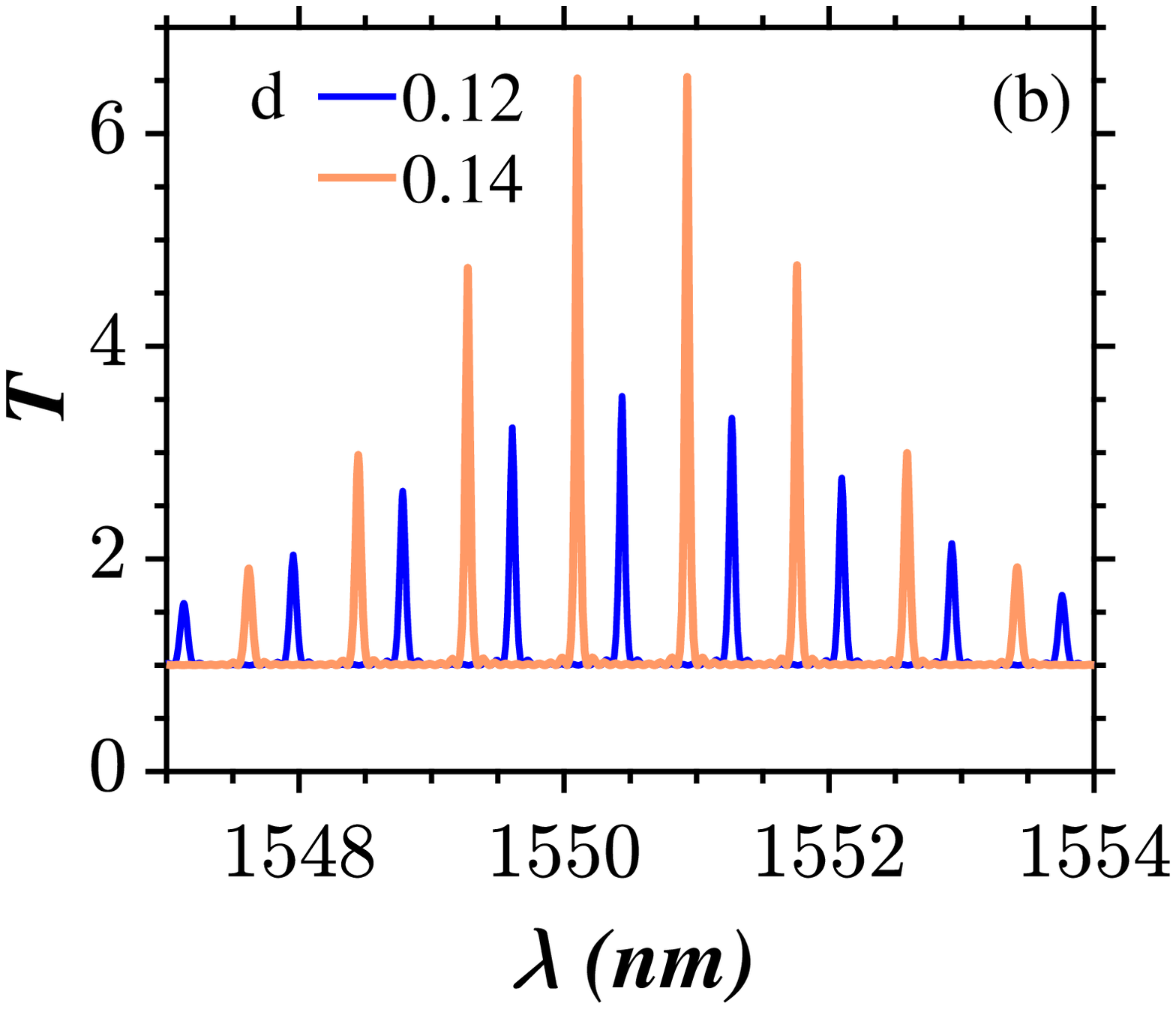}\\\includegraphics[width=0.5\linewidth]{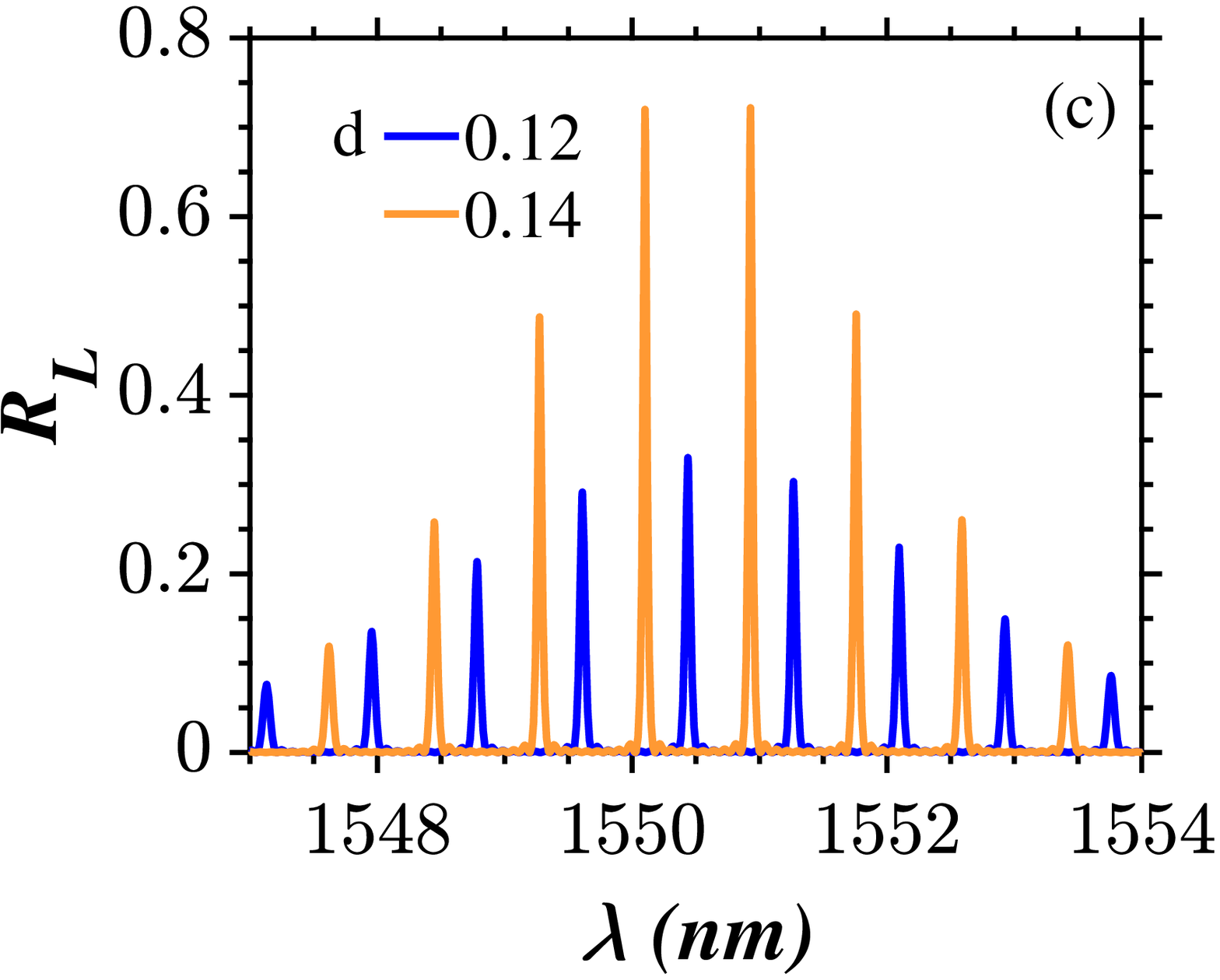}\includegraphics[width=0.5\linewidth]{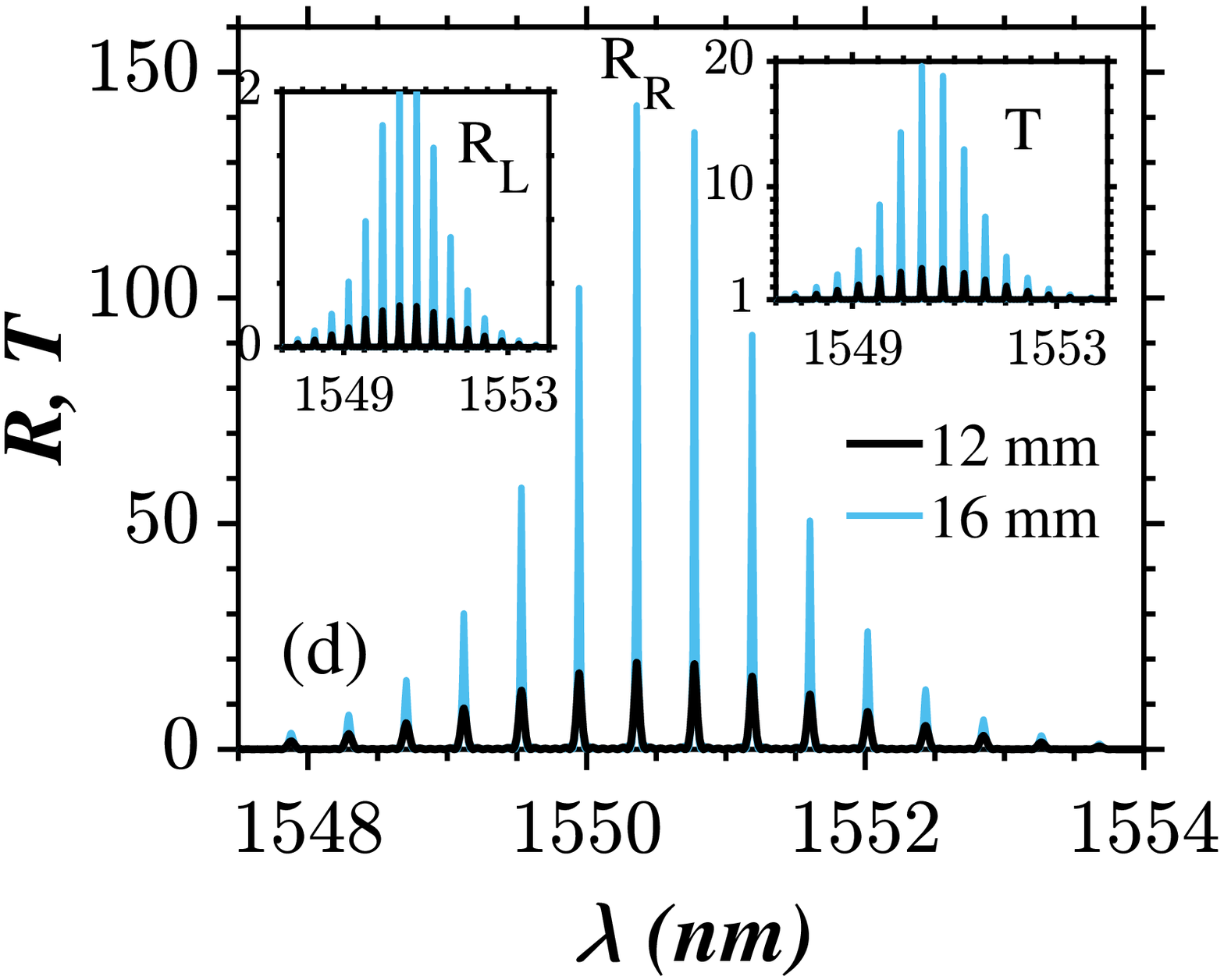}\\\includegraphics[width=0.5\linewidth]{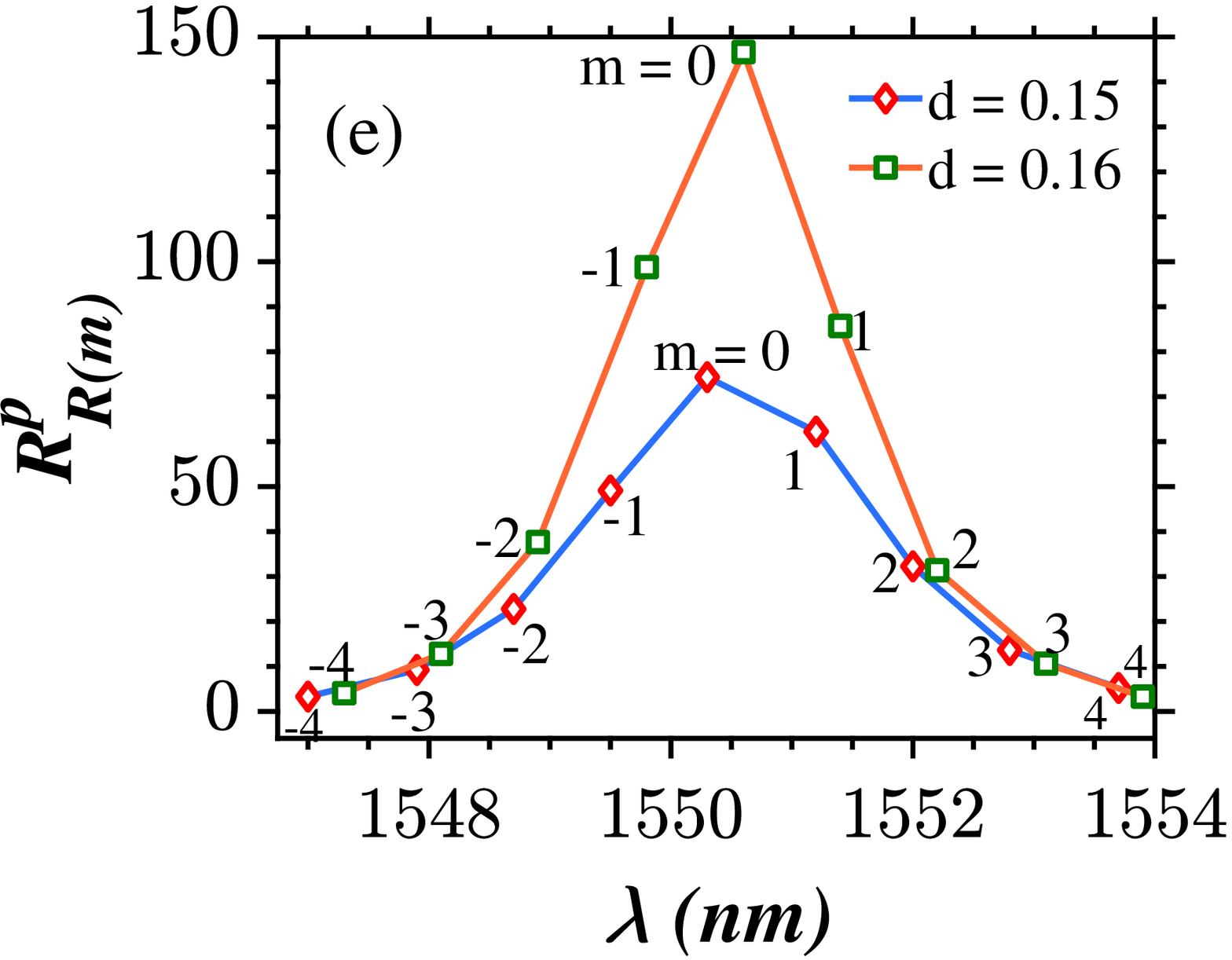}\includegraphics[width=0.5\linewidth]{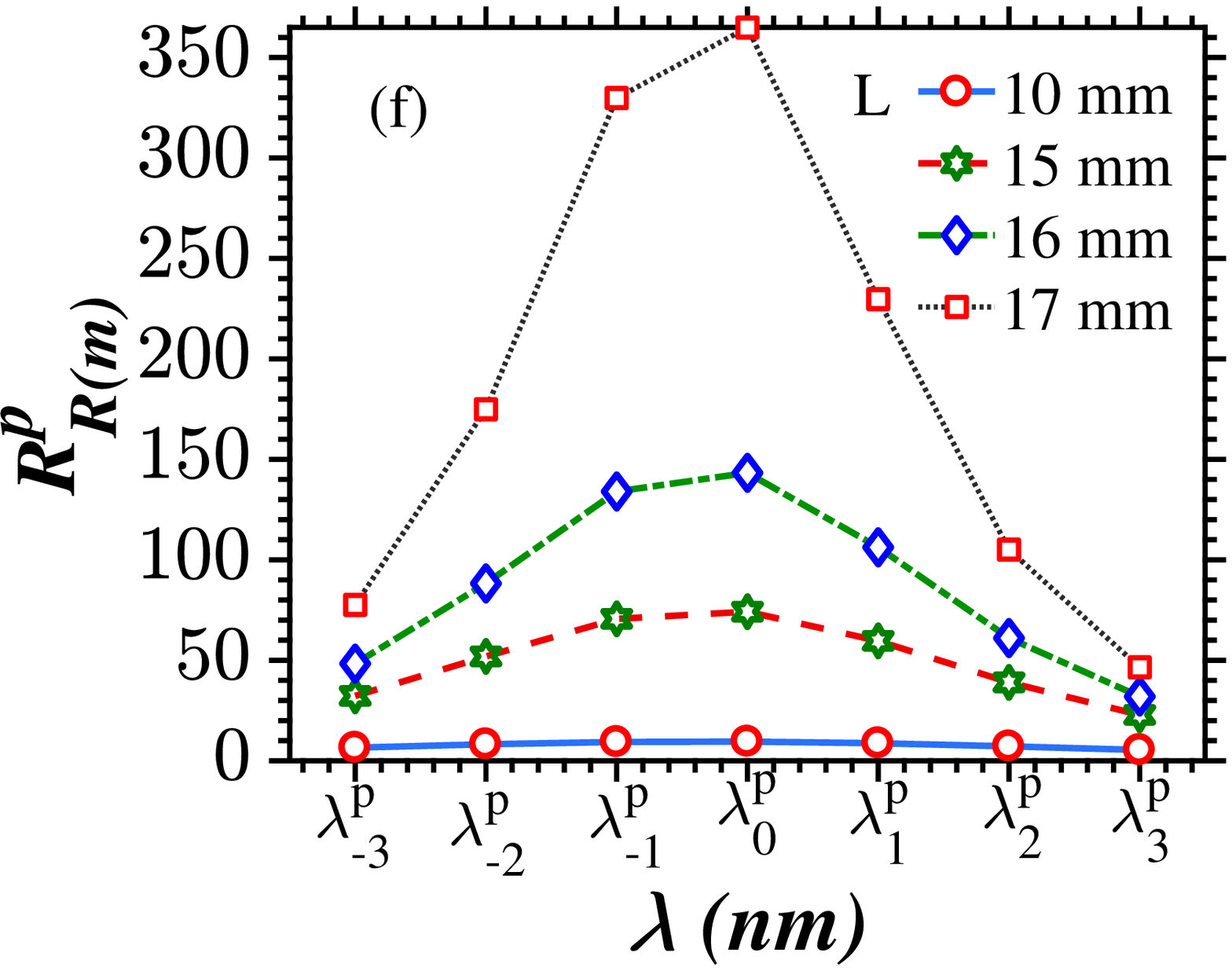}
	\caption{(a) --(c) Simultaneous shifting and increase in reflectivity (transmittivity) with changes in the duty cycle ($d$) at $n_{1I} = 6.5\times10^{-4}$ and $s_\Lambda = 1000$ $\mu$m. Also, the effect of variation in length on the lasing spectrum is shown in (d) at duty cycle value of $d = 0.1$ and the value of $g$ is same as in (a). (e) and (f) Variation in the peak reflectivity $[R_{R(m)}^p]$ with changes in the duty cycle ($d$) and length ($L$), respectively. The values of $\lambda_m^p$ in (f) are the same as mentioned in Fig. \ref{fig6}.}
	\label{fig14}
\end{figure}

The duty cycle parameter offers two important functionalities over the control of the PTSFBG lasing spectra, namely the magnitude control and location of the spectrum on the wavelength axis. As $d$ gets larger, the comb lasing spectrum shows growth in reflectivity as well as in transmittivity. Also, the combs are shifted towards longer wavelengths as shown in Figs. \ref{fig14}(a) -- \ref{fig14}(c). Under  duty cycle variations, the nonuniformity in the magnitude of the comb exists which is mainly due to the extreme amplification of the zeroth order mode as shown in Fig. \ref{fig14}(e). The degree of dissimilarity among the reflectivity of these spectral modes further builds up with any increase in the length of the structure as shown in Figs. \ref{fig14}(d) and \ref{fig14}(f). 

\subsection{Comb lasing spectrum with uniform $R$ and $T$}
\label{subsec:d}
\begin{figure}
\includegraphics[width=0.5\linewidth]{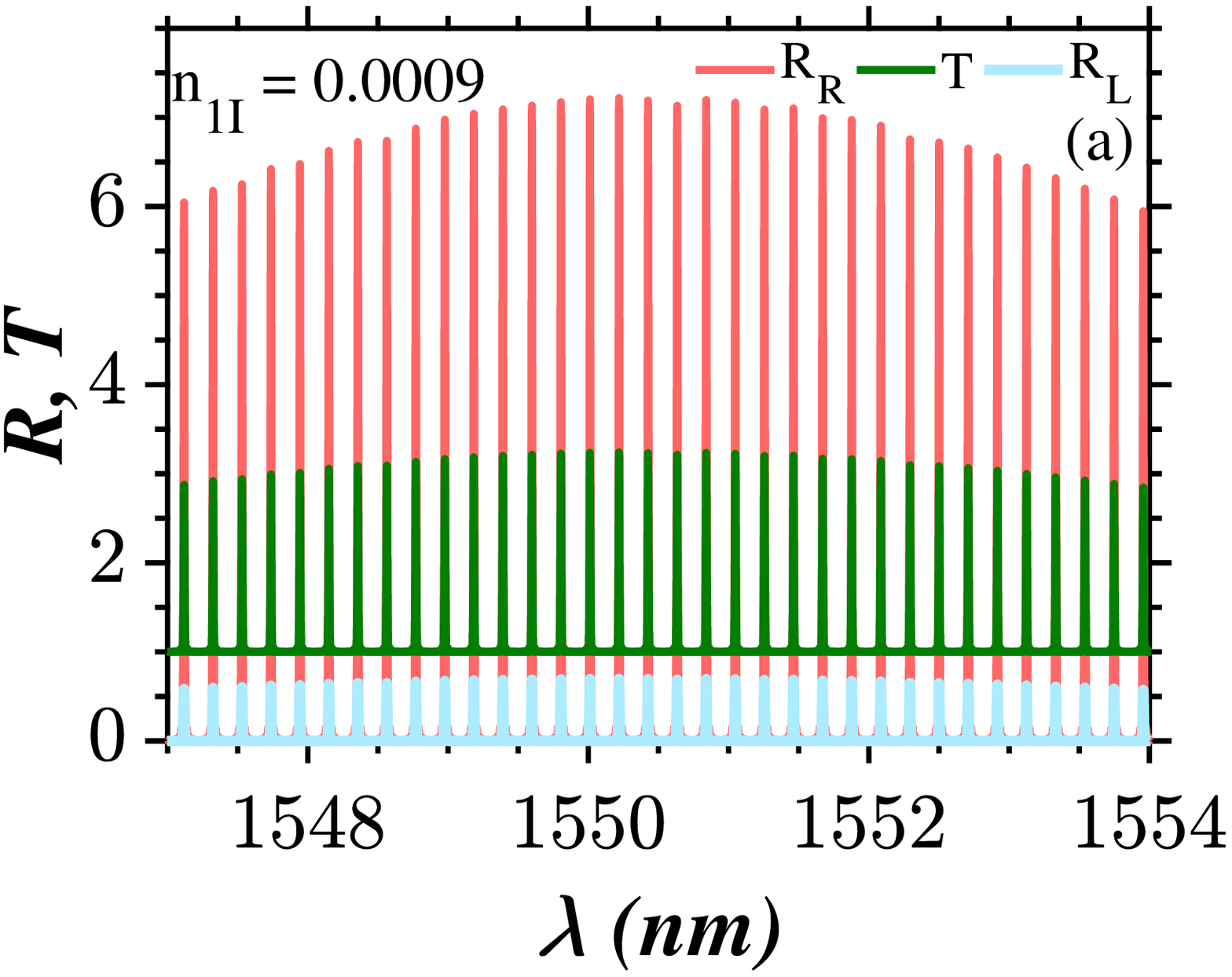}\includegraphics[width=0.5\linewidth]{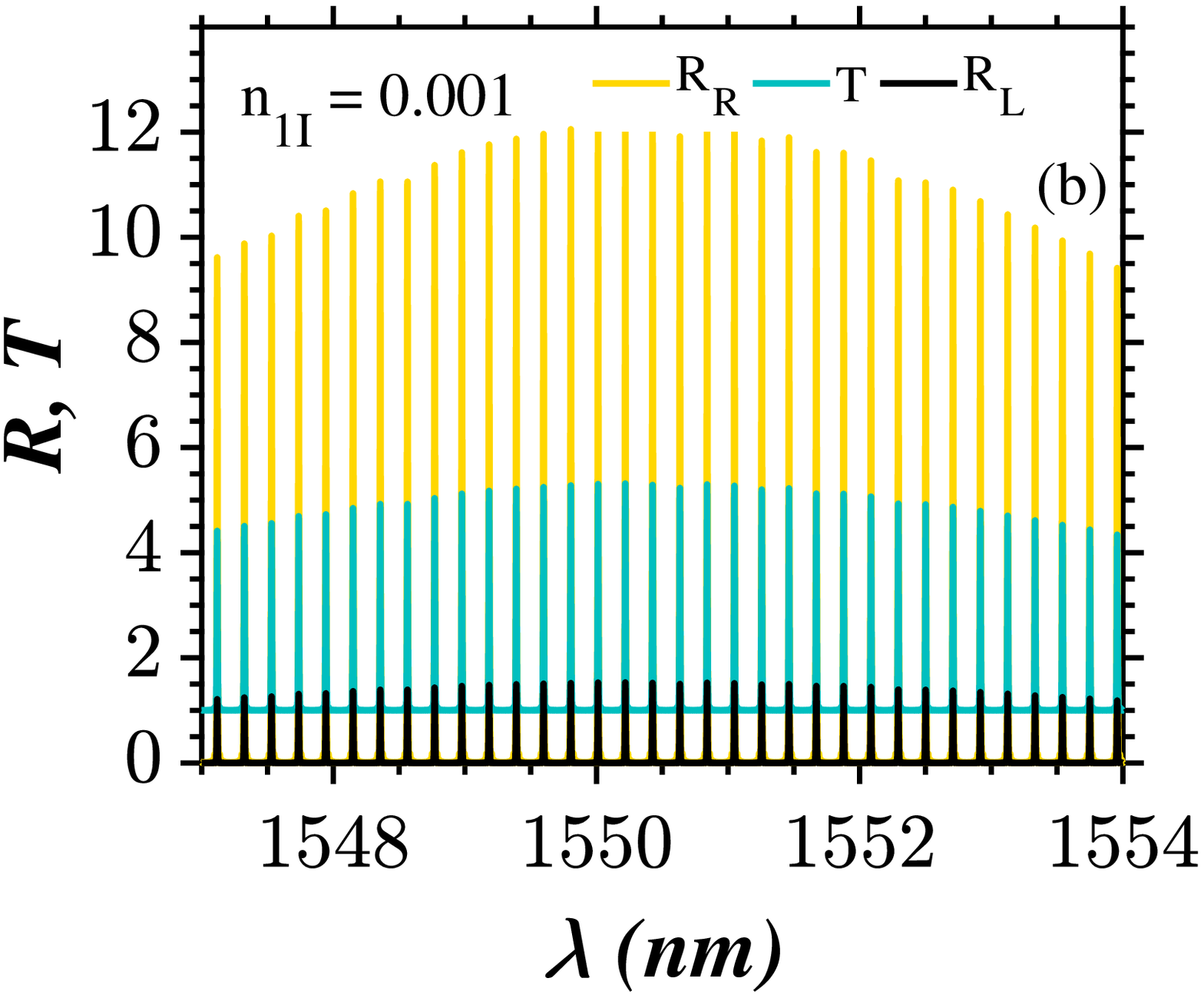}\\\includegraphics[width=0.5\linewidth]{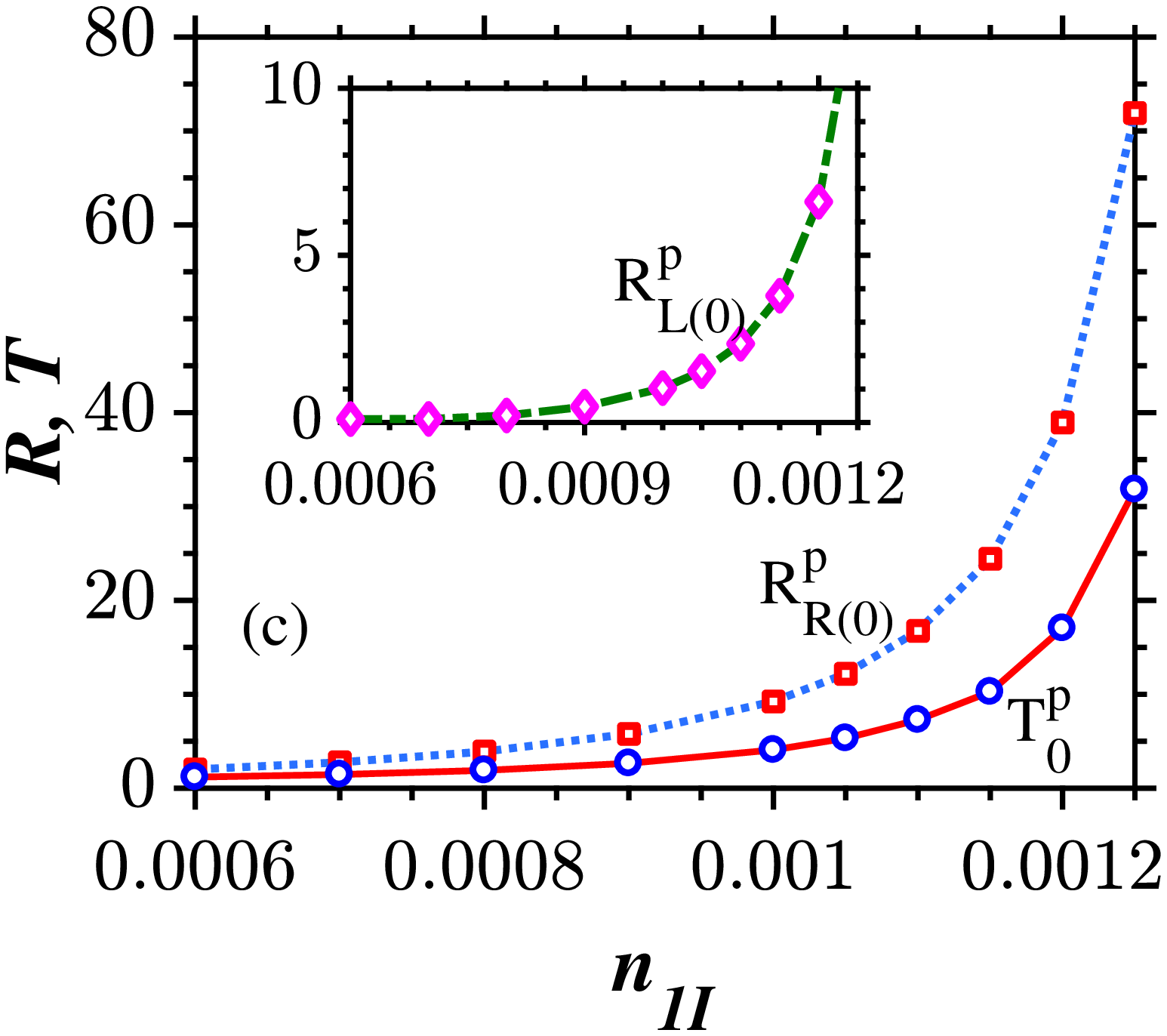}\includegraphics[width=0.5\linewidth]{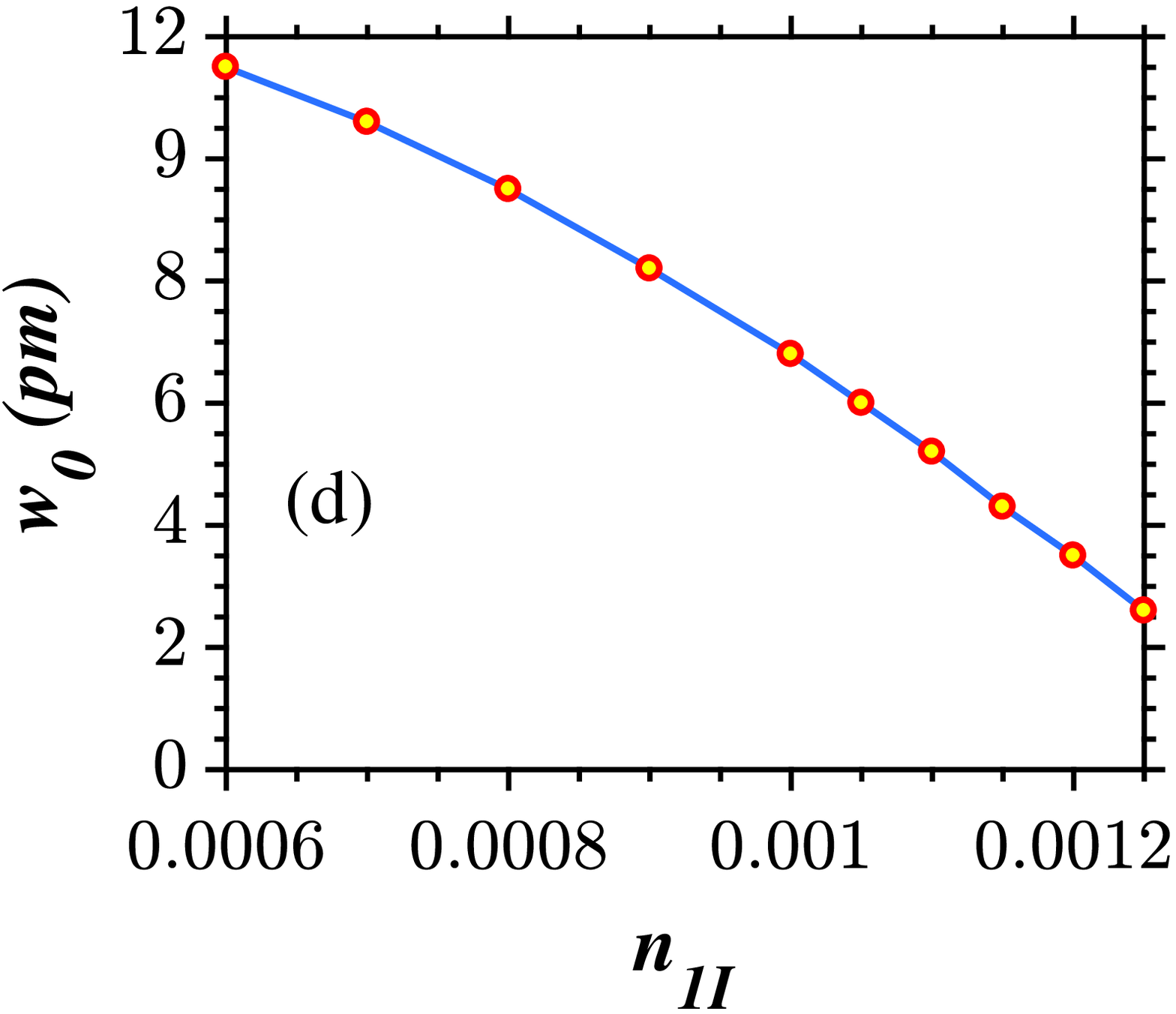}
	\caption{ (a) and (b) Gain and loss induced comb lasing spectrum of a PTSFBG in the broken $\mathcal{PT}$-symmetric regime with nearly uniform  reflectivity (transmittivity) for all the channels at a sampling period of $s_\Lambda = 4000$ $\mu$m,  duty cycle of $d = 0.01$, and $L = 60$ mm. (c) Continuous variation of reflectivity [$R_{R(m)}^p$, $R_{L(m)}^p$] and transmittivity peaks [($T^p_m$) corresponding to $m = 0$] against gain and loss ($n_{1I}$). In contrast to the unbroken $\mathcal{PT}$-symmetric regime, the transmission spectrum possesses sharp peaks rather than showing dips and hence they are denoted by $T^p_m$ rather than $T^d_m$. (d) Variations in the FWHM of the zeroth order mode with changes in the value of $n_{1I}$.}
	\label{fig15}
\end{figure}
 Unlike the unbroken $\mathcal{PT}$-symmetric regime, increasing the length of the overall structure itself cannot make the lasing spectrum uniform (nearly) as shown in Fig. \ref{fig14}(f) because of the constrain that extreme amplification of the zeroth order mode in the broken regime strongly depends on the sampling length. In other words, the inherent nature of the sample to favor stronger amplification of a particular mode must be cut down. Without managing this effect, the choice of longer device length will further build up the unevenness among the  reflectivity of the individual channels. From a theoretical perspective, different duty cycle values were tested and a value of $d = 0.01$ is found to be optimum for generation of comb lasing spectrum with nearly uniform reflectivity and transmittivity at a device length of $L = 60$ mm. It is noteworthy to mention that such a decrement in the duty cycle is not feasible in the context of conventional SFBG structures due to the fact that reduction in the duty cycle adversely decreases the reflectivity of the device. The role of the sampling period is very much the same as illustrated previously in Fig. \ref{fig13} except that the channels are nearly uniform in amplitude in all the cases. In other words, if $s_\Delta$ increases at a fixed value of sampling length $s_L$ ($s_\Lambda = s_L + s_\Delta$), the system can accommodate more channels with a reduced inter channel separation and vice-versa. The variation in gain and loss brings an increase in $R$ and $T$ as shown in Figs. \ref{fig15}(a), \ref{fig15}(b) and \ref{fig15}(c). Also, the FWHM ($w_m$) decreases with an increase in $n_{1I}$ as seen in Fig. \ref{fig15}(d). This confirms that the magnitude of the reflection spectrum is inversely proportional to the FWHM of the modes. However, larger values of $n_{1I}$ brings about nonuniformity in the values of reflectivity and transmittivity. 
 \subsection{Comb lasing spectrum with an inverted envelope}
 \begin{figure}
 	\centering	\includegraphics[width=0.5\linewidth]{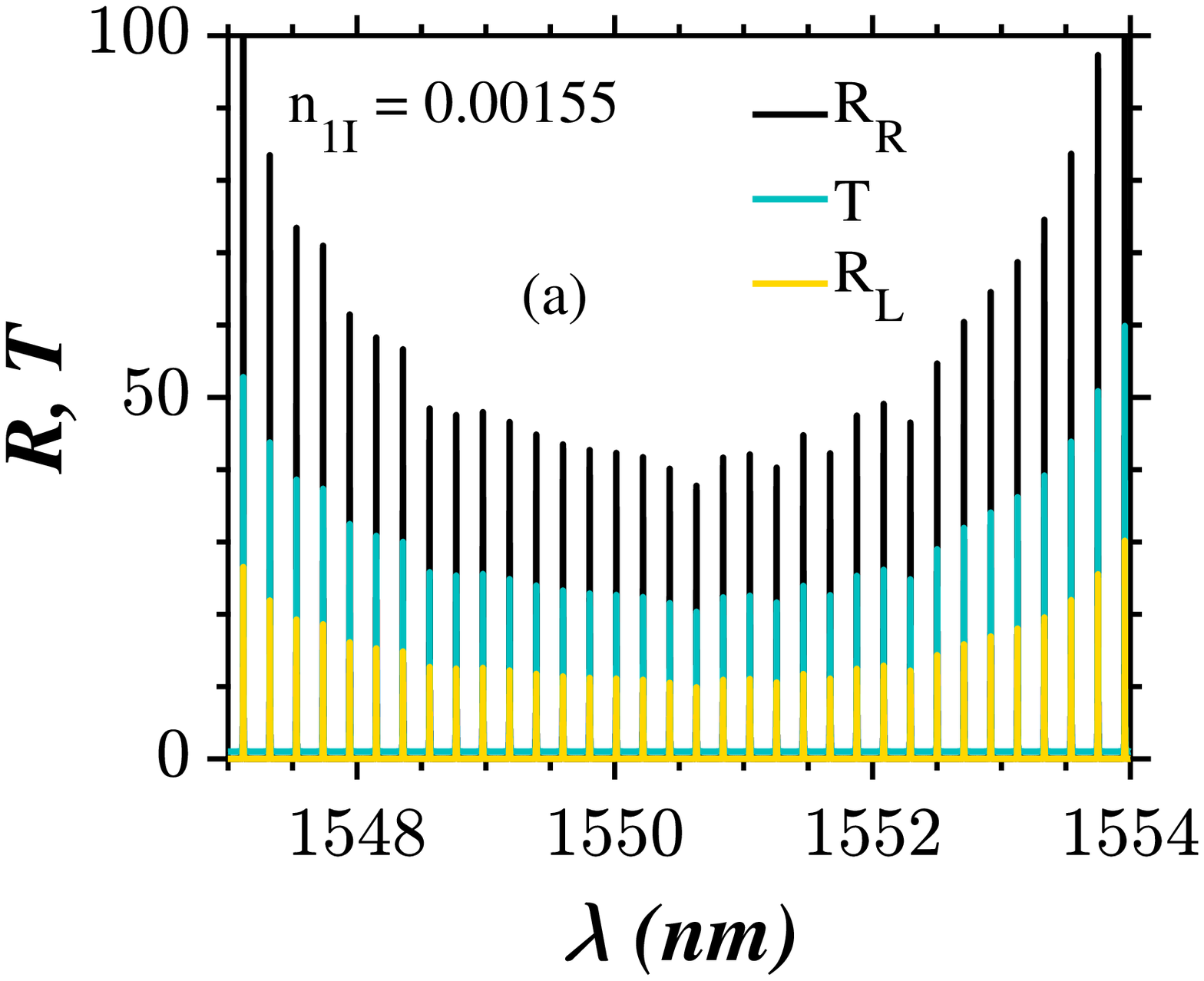}\includegraphics[width=0.5\linewidth]{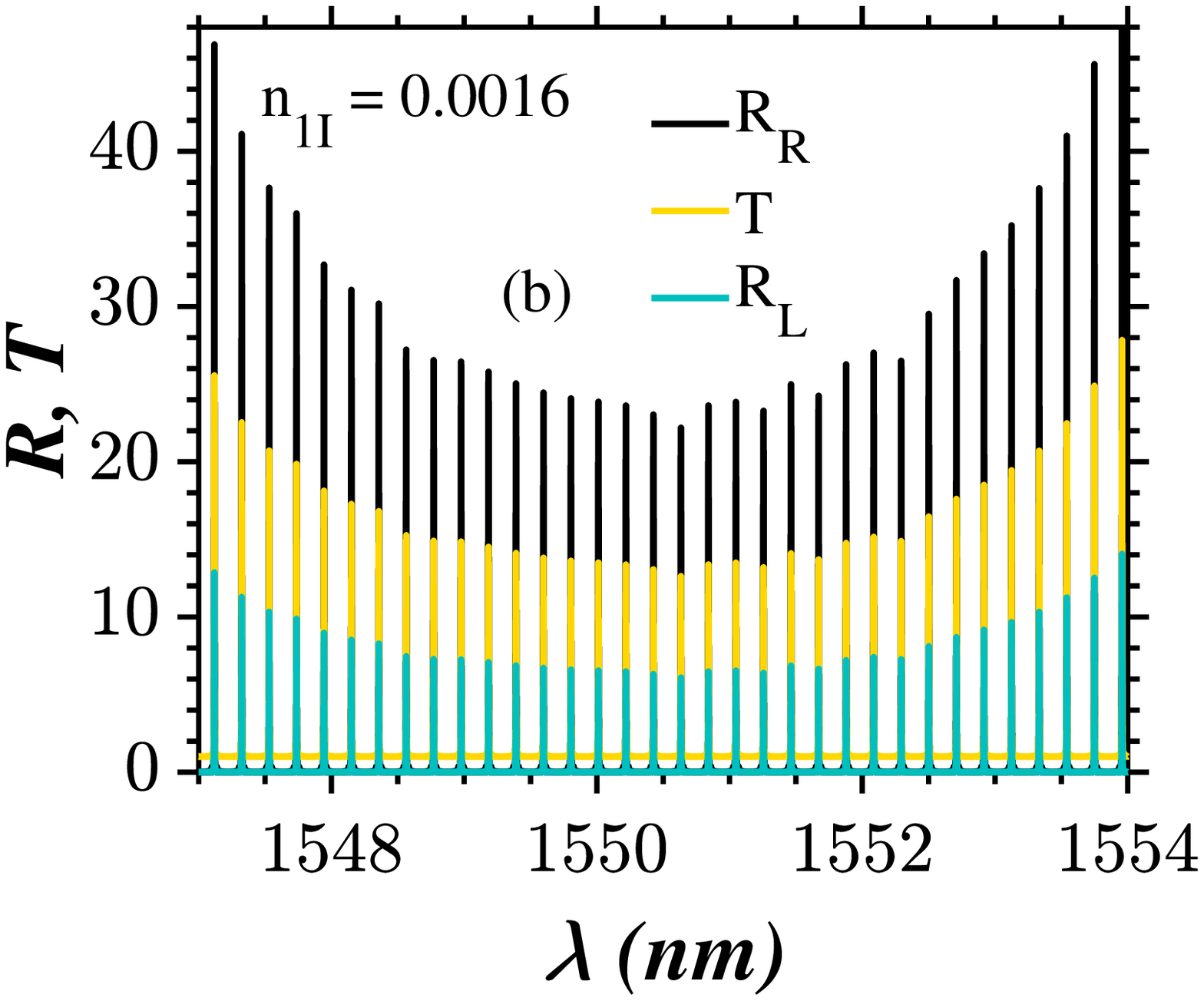}\\\includegraphics[width=0.5\linewidth]{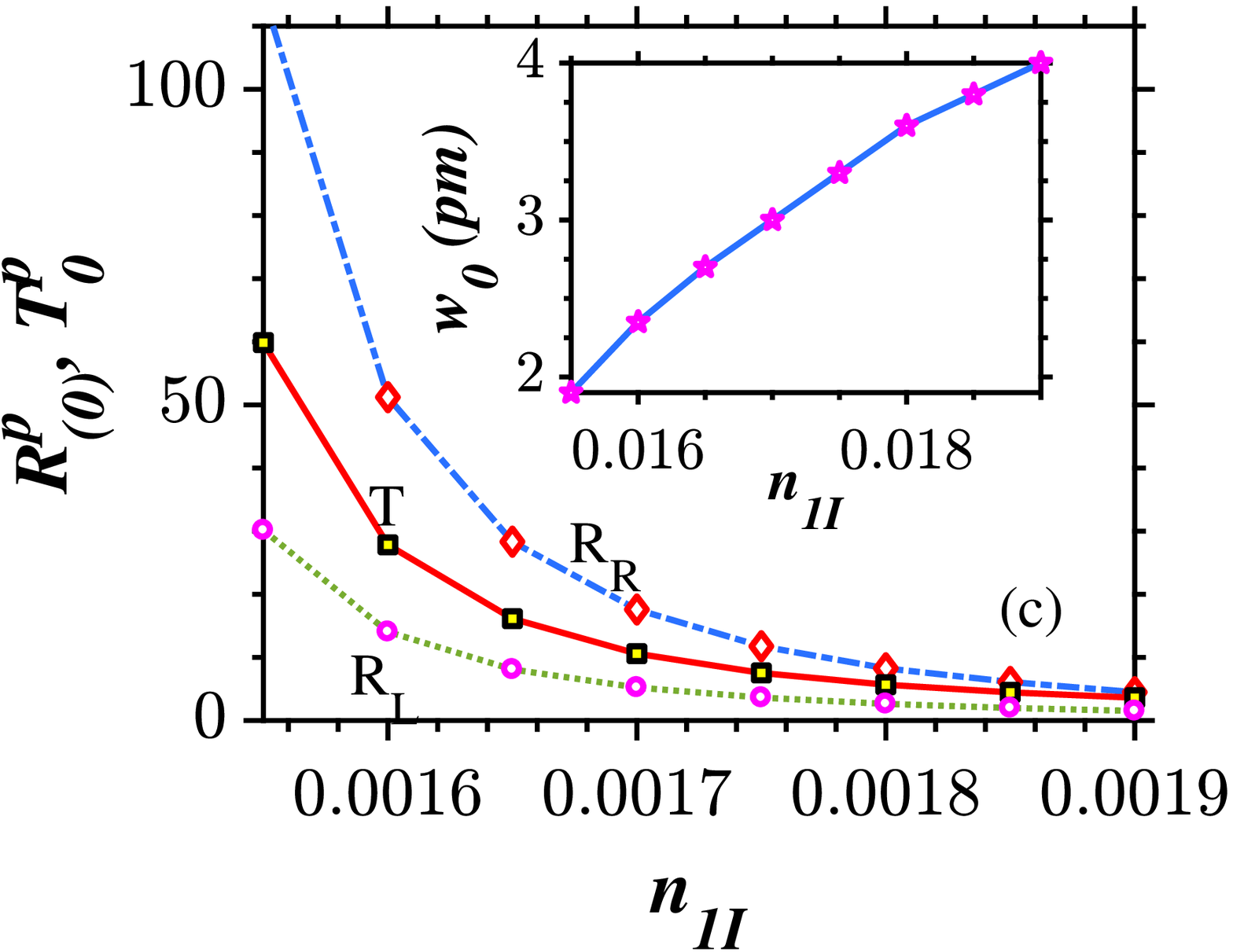}\includegraphics[width=0.5\linewidth]{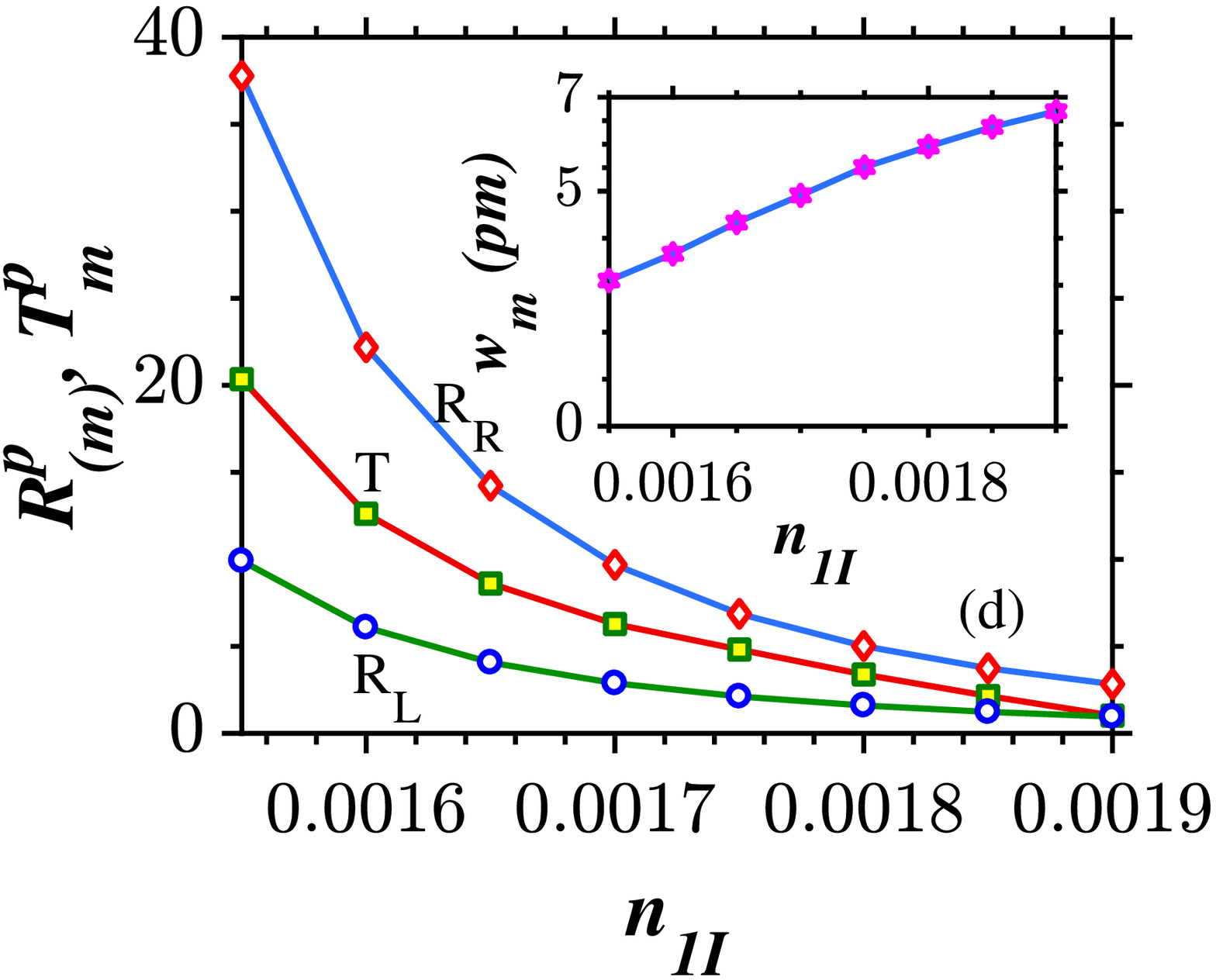}
 	\caption{(a) and (b) Comb lasing spectrum of a PTSFBG with an inverted envelope induced by variations in gain and loss. The sampling period and duty cycle values are same as in Fig. \ref{fig15}.  (c) Continuous variation of reflectivity [$R_{R(m)}^p$, $R_{L(m)}^p$] and transmittivity peaks ($T^p_m$)  against gain and loss ($n_{1I}$) for the zeroth order mode ($m = 0$). (d) Depicts the same dynamics with same system parameters as in (c) except the order of the mode is taken to be $m = -17$ [mode on the left edge of the spectra in (a) and (b)].}
 	\label{fig16}
 \end{figure}
Another distinct attribute of the gain and loss parameter is that it gives rise to a \emph{comb lasing spectrum with an uncommon envelope shape} as shown in Figs. \ref{fig16}(a) -- \ref{fig16}(b). Explicitly, the system facilitates larger amplifications for the $m^{th}$ order modes appearing at the edges of the given wavelength span. For the subsequent order modes, the reflectivity and transmittivity are lesser than the previously occurring  modes and along these lines, the zeroth order modes exhibit lowest amplification as shown in Figs. \ref{fig16}(c) and \ref{fig16}(d). This is the exact counterpart of the lasing spectrum discussed in the first three subsections of the broken $\mathcal{PT}$-symmetric regime which is characterized by an intense amplification at the center (zeroth order) and lowest amplification at the edges ($\pm m^{th}$ order). As the value of $n_{1R}$ is increased between the range $1.5 \times 10^{-3}$ and $2 \times 10^{-3}$, this behavior in the lasing spectrum is observed in a fashion that the stronger amplification at the higher order modes is inhibited progressively with an increase in $n_{1I}$. But the overall shape of the inverted envelope is maintained throughout this range of gain and loss.

\subsection{Comb lasing spectrum with dual mode lasing channels}
\begin{figure}
	\centering	\includegraphics[width=0.5\linewidth]{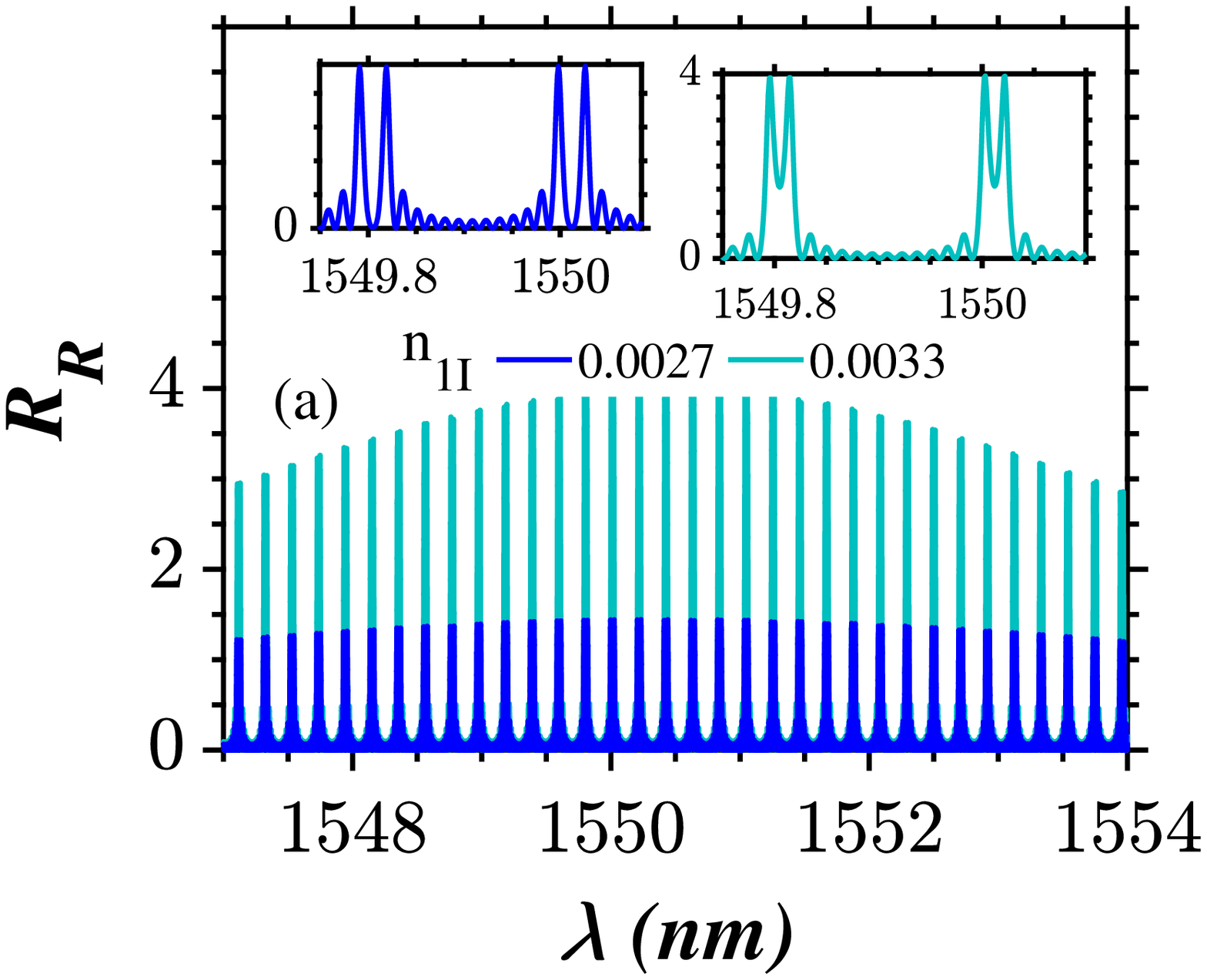}\includegraphics[width=0.5\linewidth]{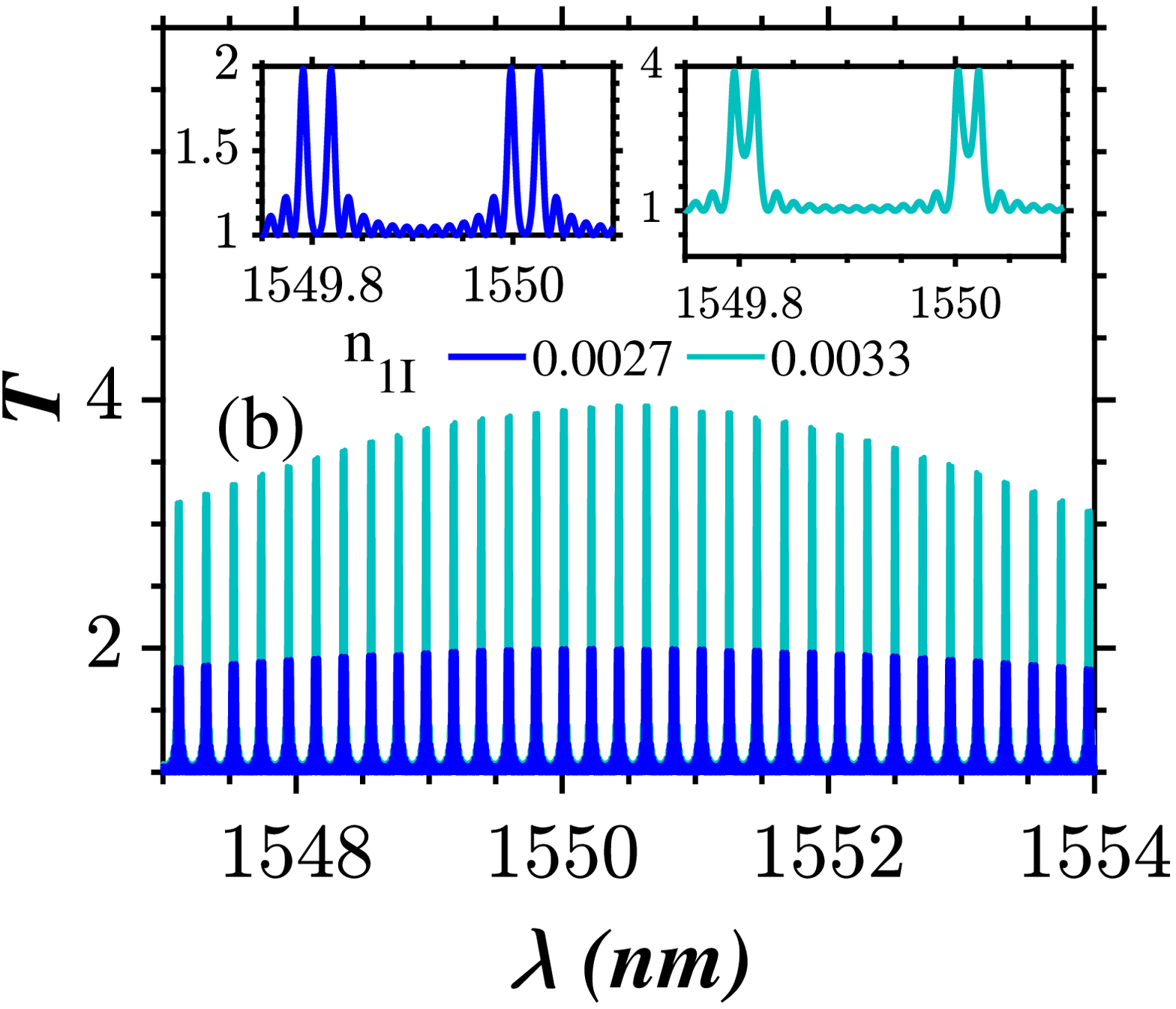}\\\includegraphics[width=0.5\linewidth]{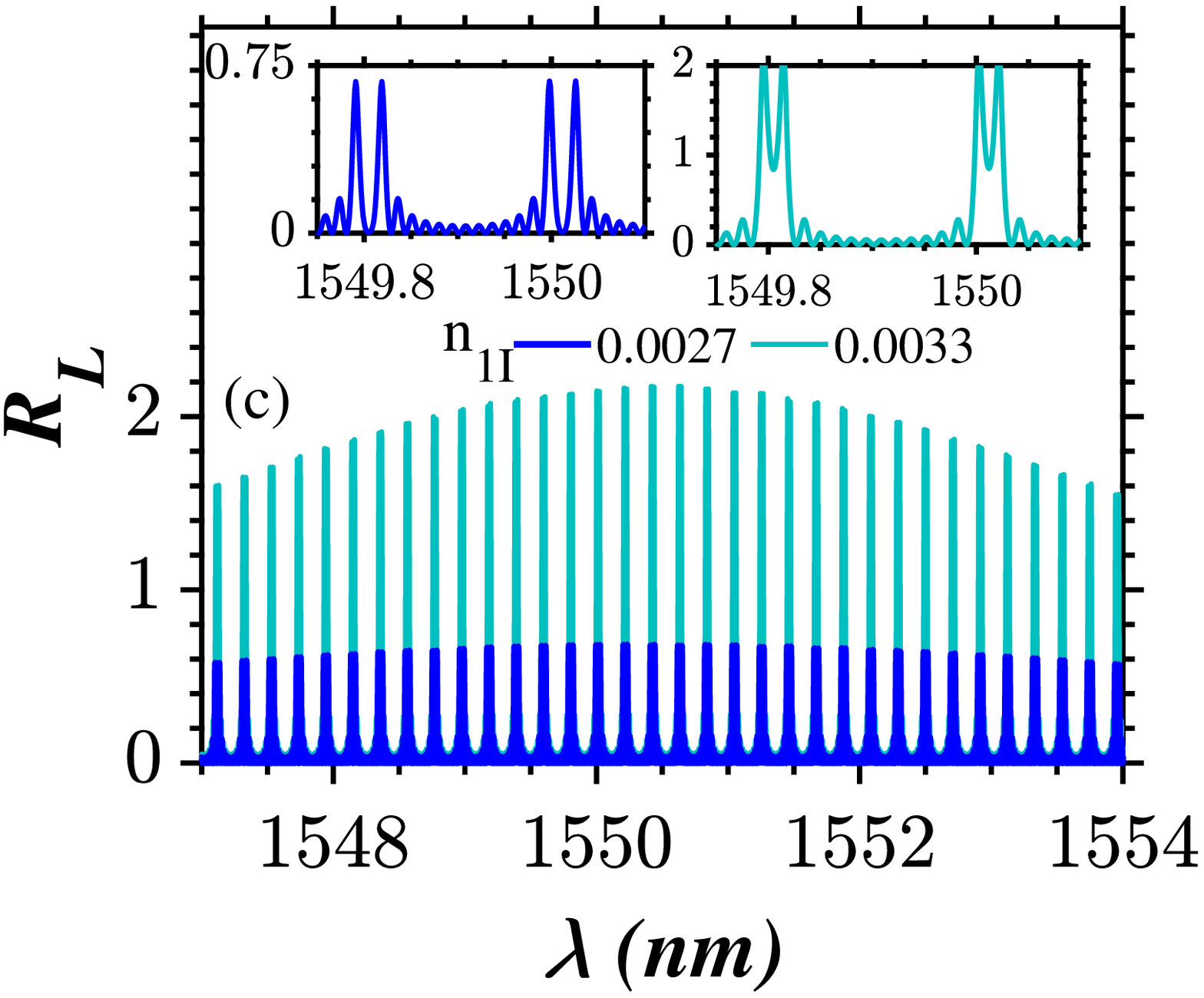}
	\caption{Comb lasing spectrum of a PTSFBG with dual mode lasing channels generated by tuning the value of gain and loss. The sampling period and duty cycle values are same as in Fig. \ref{fig15}.}
	\label{fig17}
\end{figure}
When $n_{1I} > 2 \times 10^{-3}$, we observe the usual lasing spectrum with a conventional envelope shape except that each individual channel demonstrates a dual mode lasing behavior. A dip is visible in between two peaks of the individual modes as shown in Figs. \ref{fig17}(a) -- \ref{fig17}(c). This dip exactly occurs at $R_{min}$ and $T_{min}$ (or closer to those values) when $n_{1I}$  is less.  As the value of the imaginary part of the modulation strength is gradually increased, $R$ and $T$ values of individual modes are also enhanced whereas the depth of penetration of this dip drops off and finally the dip vanishes for $n_{1I} > 3.7 \times 10^{-3}$. Beyond this, again comb spectrum with single mode lasing channels appears. 
\subsection{Application of PTSFBG in the broken $\mathcal{PT}$-symmetric regime: Tunable Laser}
\begin{figure}[t]
	\centering	\includegraphics[width=1\linewidth]{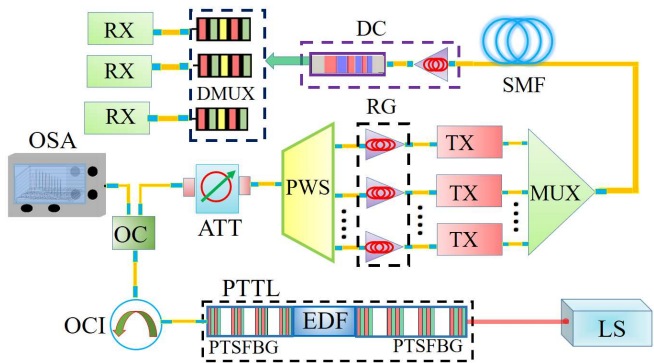}
	\caption{Schematic showing the generation of tunable multiwavelength laser source and its application to telecommunication system. LS: laser source, EDF: Erbium doped fiber, PTTL: $\mathcal{PT}$-symmetric tunable laser, OCI: optical circulator, OC: optical coupler, OSA: optical spectrum analyzer, ATT: attenuator (variable), PWS: programmable wave shaper, RG: regenerator, TX: transmitter, MUX: multiplexer, SMF: single mode fiber, BPF: band pass filter, DC: dispersion compensation, DMUX: demultiplexer, RX: receiver. PPTL consists of two PTSFBGs of different sampling periods and a gain medium (EDF). Similarly, EDF amplifiers (EDFA) are used as regenerators. SMFs are generally used as transport fibers. Uniform PTFBGS can be used as BPFs and gain flattening filters. DC consists of EDFA and chirped PTFBGS \cite{raja2020tailoring} for attenuation and dispersion compensation, respectively. For demultiplexing, phase shifted PTFBGs can be employed \cite{raja2020phase}}
	\label{fig18}
\end{figure}

 From Fig. \ref{fig15}, we infer that it possible to generate uniformly spaced lasing modes with uniform intensities and narrow FWHM with the aid of the proposed system. Since each mode in the output spectrum represents a distinct wavelength and the separation between these modes is very narrow, the device can be effectively used as tunable multi-wavelength laser source which can simultaneously be fed as inputs for multiple transmitters. To construct such a tunable laser source, two PTSFBGs with a gain element should be engineered. It should be noted that wavelength tuning of the $\mathcal{PT}$-symmetric tunable laser (PTTL) requires the two PTSFBGs to have different sampling periods \cite{jayaraman1993theory,bidaux2015extended}. The tuning is based on the principle of Vernier effect (in the reflection spectra) which occurs as a consequence of variation in the channel spacing of two dissimilar SFBGs (having different sampling periods) \cite{xu2005chirped,bidaux2015extended}. Vernier effect states that when one of these superstructured gratings is tuned, constructive interference occurs between the pair of modes which are common to both the gratings, thereby leading to lasing at these wavelengths and suppression of the other lasing modes whose center wavelength does not coincide. For instance, consider two PTSFBGs having different sampling periods $s_\Lambda = 1000$ and 2000 $\mu$m. From Fig. \ref{fig5}(e), we infer that $\lambda_m^{p} = 1550.8$ nm is one among the center wavelengths which is common to both the resulting comb spectra ($s_\Lambda = 1000$ and 2000 $\mu$m), whereas   $\lambda_m^{p} = 1550.4$ nm is not so. The resulting spectra from the PTTL will feature a comb mode at $\lambda = 1550.8$ nm and its reflectivity will be the product of reflectivities of the individual PTSFBGs. In a similar fashion, other overlapping modes are selected and amplified by the PTTL. On the other hand, the lasing at the non-overlapping wavelengths are totally suppressed, for instance $\lambda = 1550.4$ nm. 
 
  It is desirable to acquire replicas of laser output fields with the same intensities for all the channels \cite{xu2005chirped}. Therefore, PTSFBGS exhibiting uniform reflectivity across all the wavelengths (in the given range) must be employed. To flatten the envelope of the output spectra, many SFBG structures were investigated in the literature, including chirped SFBGs \cite{Dong2006}, and sinc sampled FBGs \cite{Loh1999}. In the chirped SFBGs, either the grating period or sampling length or sampling period or combinations of the above can be chirped \cite{Lee2003,Dong2006,Lee2004,Lee2004a,Loh1999,azana2005spectral}. But, the resolution of the spectrum strongly depends on the lithographic process used for fabrication of these gratings \cite{bidaux2015extended}. Alternatively, phase shifted SFBGS were proposed in which optimization of the number of phase shift regions required is tedious \cite{Lee2003,Li2008,Navruz2008,Shi2018,Zhang2019}.

Here, we demonstrate that it is possible to generate the same lasing spectrum  (with nearly uniform reflectivity) with $\mathcal{PT}$-symmetric uniform grating samples. The envelope formed by reflectivity peaks of different channels widens with larger duty cycle values in conventional SFBGS \cite{hansmann1995variation}. Reducing the duty cycle to smaller values is also not an ideal way in the perspective of conventional SFBGS as it adversely decreases the reflectivity \cite{jayaraman1993theory,Lee2003}. One could visualize from Figs. \ref{fig15} and \ref{fig17} that  PTSFBGs have two important features: First, the output spectrum features enhanced reflectivity. Second, the envelope is much flatter as a result of smaller duty cycle values (for example, $d = 0.01$). Achieving these two features simultaneously is an exceptional feature of PTSFBG, thanks to the concepts of gain and loss. It should be recalled that these structures can be fabricated using an Argon laser with the standard scan-writing technology \cite{xu2019advanced}. The modern translation stages which hold the phase mask exhibit an infinitely small moving precision of the order of 5 nm \cite{xu2005chirped} and above and thus fabricating a small length of sample should not be a tough task to our knowledge. We also visualize that much of the PTSFBG structure is unused as a result of the larger sampling period at very low duty cycles. These unused regions can be used to fabricate interleaved samples having different Bragg wavelengths \cite{Loh1999,Lee2004a}. The period ($\Lambda$) of each interleaved PTFBG should be different. Otherwise, the comb spectrum from two different PTSFBGs may impose on one another. In simpler words, the concept of interleaving refers to the fabrication of several PTSFBGs on the same fiber with different periods ($\Lambda$) in such a way that the physical concept leads to interleaving effect on the spectrum as well \cite{Loh1999}. As theoretical physicists, we believe some of the challenges remains now for the experimental realization of these $\mathcal{PT}$-symmetric devices are to find a suitable dopant material like Er$^{3+}$ and Cr$^{3+}$ for the fabrication of the PTSFBG with gain and loss regions, respectively \cite{ozdemir2019parity}. Even though it looks very simple to achieve phase transitions such as unitary transmission point and broken symmetry in $\mathcal{PT}$-symmetric systems by simply tuning the value of $n_{1I}$, certain practical challenges still remain to be addressed by the experimental physicist. One can think of tuning the value of permittivity of both the gain medium as well as the lossy medium, simultaneously by the help of external pumping source. 
	Another way  for achieving these types of phase transitions is by varying the frequency of the incident optical field, appropriately. This may lead to the violation of causality principle and $\mathcal{PT}$-symmetric operation is bound to occur only at isolated frequencies and not on a continuous interval of frequencies \cite{Ge2012}.  This type of violation occurs as a result of  Kramers-Kronig relations  which describe the fact that the variation in the imaginary part of the index can also lead to a change in its real part  \cite{Zyablovsky2014}. Nevertheless, it is to be noted that a continuous tuning of gain and loss through pumping is still feasible at the resonant frequency \cite{Nguyen2016}. In such a case,  it should be remembered that the amount of pumping should be lesser in the lossy regions and higher in the gain regions to achieve $\mathcal{PT}$-symmetry \cite{Phang2018,Phang2015res}. As theoretical  physicists, we expect the experimentalists to put forward suitable strategies to address  these challenges to make $\mathcal{PT}$-symmetric combs more realistic.

Having briefly discussed on the principle of operation, existing structures, experimental feasibility and the advantages of the proposed scheme, we look at the operation of the experimental set up illustrated by Fig. \ref{fig18}. The input optical signal to the PTTL can be pumped from a CW laser. It can be directly fed to the PTTL with an OCI as shown in the schematic or the input signal may be amplified with a EDFA and be passed through a uniform FBG  to filter out the EDFA noise \cite{chembo2016kerr}. The PTTL can be constructed with two PTSFBGs with comb like reflection spectrum as discussed above.  The EDF in the optical cavity serves as the gain medium alongside the tuning PTSFBGs and this generates a laser source over its gain profile \cite{jayaraman1993theory, xu2005chirped, othonos2006fibre,kryukov2019laser}. When one of the non identical PTSFBGs is tuned, lasing will take place at a particular channel, if and only if the comb lines from both the PTSFBGs intersect. In other words, reflectivity multiplies at the overlapping channels and the lasing at the other wavelengths is inhibited \cite{xu2005chirped,bidaux2015extended,schneider2005sampled,jayaraman1993theory}. Thus, it is possible to obtain multiwavelength comb lasers with nearly identical intensities with controlled precision and the same can be visualized via an OSA \cite{schneider2005sampled,xu2005chirped}. The variable optical attenuator is used to flatten the envelope of the comb laser further. In the case of comb lasers with extreme amplification at the center wavelength, these attenuators can be followed by a notch filter which is helpful in attenuating the power levels of these modes to comparable levels \cite{chembo2016kerr}.

 The programmable wave shapers separate the comb lines in wavelength which serve as carrier signals \cite{xu2019advanced}. Also, it is useful to separate the interleaved channels \cite{chembo2016kerr}. These carriers are regenerated by the EDFA before they are fed to the transmitters.  In the transmitter side, multiple laser sources are required to drive each transmitter which makes the system bulky \cite{bidaux2015extended}. PTSFBG are fascinating in a sense that they serve as the alternative solution to build a compact and reconfigurable transmitter system, since they serve as the building block to fabricate a broadband laser source which is then separated in terms of wavelengths by the PWS according to the number of transmitters \cite{kryukov2019laser}. The new generation transmitters come with inbuilt phase modulators and the stream of data from all these transmitters are multiplexed and sent to the transport fiber \cite{chembo2016kerr}.  SMFs are generally used as long distance transport fibers. As the signal travels in the fiber, attenuation and broadening mechanism comes into the picture and to compensate these detrimental effects, EDFA and a chirped PTFBG, respectively, can be used in the compensation module. The advantages of chirped PTFBG is that they can compensate both normal and anamolous dispersions simply by the concept of reversal of direction of incidence \cite{raja2020tailoring}.   We also reported the construction of demultiplexers with $\mathcal{PT}$-symmetric phase shifted FBGs, in our previous work which can demultiplex all the data streams from the multiplexed input signal \cite{raja2020phase}. At the receivers, all these signals are demodulated and coherently detected \cite{chembo2016kerr}. 

 \section{Conclusions}
 \label{Sec:6}
 In this article, we have presented a complete description of the comb spectrum of a PTSFBG with uniform sampling. We first illustrated the role of various device parameters, namely the sampling period, duty cycle and gain-loss parameter on the generation of the comb spectrum in the three different $\mathcal{PT}$-symmetric regimes, namely the unbroken, Unitary transmission point and broken symmetric regimes. Special emphasis was given to the generation of uniform amplitude comb filters with narrow channel spacing in the unbroken regime.  The major highlight of this section is that it provides a framework towards the generation of a large number channels with significantly large reflectivity by increasing the sampling period of the grating. It also confirms that the decrease in the reflectivity (with increasing number of channels) can be independently controlled with the aid of gain and loss parameter. An architecture which can possibly serve as a tunable RF traversal filter was proposed at the end of Sec. \ref{Sec:3}. We then presented a brief analysis on the dependence of comb spectrum on the different control parameters for the case of right light incidence at the unitary transmission point. Remarkably, the reflectionless wave transport phenomenon was observed under similar conditions when the direction of light incidence is reversed. This once again proves that the concept of unidirectional wave transport is a distinct feature of any PTFBG system and is unresponsive to any variation in other device parameters except for evenly balanced values of real and imaginary parts of the modulation strength.

 Further, the analysis required for figuring out the interdependence between the changes in the lasing spectrum against the variation in the grating parameters in the case of broken $\mathcal{PT}$-symmetric regime is discussed in Sec. \ref{Sec:5}. We then proposed an optimal way to generate comb lasing spectrum with uniform reflectivity across different wavelengths by decreasing the duty cycle of the grating. Such a reduction in the duty cycle is not feasible in the context of conventional PTSFBGs due to the dependence of reflectivity on the duty cycle and coupling coefficient. It was proved that it is possible to obtain comb spectrum flattened envelope as well as uniform reflectivity for different wavelengths as a result of the interplay between the reduced duty cycle and large gain-loss parameter. Surprisingly, the tuning of gain and loss parameter also leads to the generation of lasing spectrum with an unconventional inverted envelope and dual mode lasing behavior in the individual channels.  Towards the end, we showed that a single laser source from a PTSFBG  can drive multiple transmitters in a wavelength division multiplexing network. The architecture also integrates different modules like dispersion compensator and demultiplexer based on PTFBGs. The physical behavior exploited for the comb application is reported only from the perspective of FBGs without gain and loss. This is the very first time, to the best of our knowledge, these applications have been dealt from the viewpoint of $\mathcal{PT}$-symmetric superstructures. From a fundamental perspective, gain and loss provides an additional degree of freedom to control the intensity and FWHM of the comb spectra. From the application perspective, the inclusion of gain and loss in the form of $\mathcal{PT}$-symmetry opens up an alternative route to overcome some of the critical problems like regrowth challenges prevailing in the current hybrid integration optical technologies to build tunable and reconfigurable devices. With advancements in lightwave technology, these PTFBG based devices are expected to be available in the near future, credits to their improved spectral features and the number of degrees of freedom to manipulate them.

\section*{Acknowledgments}
SVR is indebted to a financial assistantship provided
by Anna University through an Anna Centenary Research Fellowship (CFR/ACRF-2018/AR1/24). AG and ML acknowledge the support by DST-SERB for providing a Distinguished Fellowship (Grant No. SB/DF/04/2017) to ML in which AG was a Visiting Scientist. AG is now supported by the University Grants Commission (UGC), Government of India, through a Dr. D. S. Kothari Postdoctoral Fellowship (Grant No. F.4-2/2006 (BSR)/PH/19-20/0025).


\begin{thebibliography}{75}
	\bibitem{Pilozzi2017}
	L.~Pilozzi and C.~Conti, {Topological cascade laser for frequency comb
		generation in $\mathcal{PT}$-symmetric structures}, Opt. Lett. \textbf{42},
	5174 (2017).
	
	\bibitem{jones2000carrier}
	D.~J. Jones, S.~A. Diddams, J.~K. Ranka, A.~Stentz, R.~S. Windeler, J.~L. Hall,
	and S.~T. Cundiff, Carrier-envelope phase control of femtosecond mode-locked
	lasers and direct optical frequency synthesis, Science \textbf{288}, 635
	(2000).
	
	\bibitem{fortier201920}
	T.~Fortier and E.~Baumann, 20 years of developments in optical frequency comb
	technology and applications, Commun. Phys. \textbf{2}, 1 (2019).
	
	\bibitem{jin2006absolute}
	J.~Jin, Y.-J. Kim, Y.~Kim, S.-W. Kim, and C.-S. Kang, Absolute length
	calibration of gauge blocks using optical comb of a femtosecond pulse laser,
	Opt. Express \textbf{14}, 5968 (2006).
	
	\bibitem{del2007optical}
	P.~Del’Haye, A.~Schliesser, O.~Arcizet, T.~Wilken, R.~Holzwarth, and T.~J.
	Kippenberg, Optical frequency comb generation from a monolithic
	microresonator, Nature \textbf{450}, 1214 (2007).
	
	\bibitem{Herr2014}
	T.~Herr, V.~Brasch, J.~D. Jost, C.~Y. Wang, N.~M. Kondratiev, M.~L. Gorodetsky,
	and T.~J. Kippenberg, {Temporal solitons in optical microresonators}, Nat.
	Photonics \textbf{8}, 145 (2014).
	
	\bibitem{Dong2006}
	X.~Dong, P.~Shum, X.~Yang, M.~F. Lim, and C.~C. Chan, {Bandwidth-tunable filter
		and spacing-tunable comb filter with chirped-fiber {Bragg} gratings}, Opt.
	Commun. \textbf{259}, 645 (2006).
	
	\bibitem{Lee2003}
	H.~Lee and G.~P. Agrawal, {Purely phase-sampled fiber {Bragg} gratings for
		broad-band dispersion and dispersion slope compensation}, IEEE Photonics
	Technol. Lett. \textbf{15}, 1091 (2003).
	
	\bibitem{Lee2004}
	H.~Lee and G.~P. Agrawal, {Bandwidth equalization of purely phase-sampled fiber
		{Bragg} gratings for broadband dispersion and dispersion slope compensation},
	Opt. Express \textbf{12}, 5595 (2004).
	
	\bibitem{Lee2004a}
	H.~Lee and G.~P. Agrawal, {Add-Drop Multiplexers and Interleavers With
		Broad-Band Chromatic Dispersion Compensation Based on Purely Phase-Sampled
		Fiber Gratings}, IEEE Photonics Technol. Lett. \textbf{16}, 635 (2004).
	
	\bibitem{Li2003}
	H.~Li, Y.~Sheng, Y.~Li, and J.~E. Rothenberg, {Phase-Only Sampled Fiber {Bragg}
		Gratings for High-Channel-Count Chromatic Dispersion Compensation}, J. Light.
	Technol. \textbf{21}, 2074 (2003).
	
	\bibitem{Loh1999}
	W.~H. Loh, F.~Q. Zhou, and J.~J. Pan, {Sampled fiber grating based-dispersion
		slope compensator}, IEEE Photonics Technol. Lett. \textbf{11}, 1280 (1999).
	
	\bibitem{Navruz2008}
	I.~Navruz and N.~F. Guler, {A novel technique for optical dense comb filters
		using sampled fiber {Bragg} gratings}, Opt. Fiber Technol. \textbf{14}, 114
	(2008).
	
	\bibitem{Zhang2019}
	L.~Zhang, F.~Yan, W.~Han, Y.~Bai, Z.~Bai, D.~Cheng, H.~Zhou, Y.~Suo, and
	T.~Feng, {Transmission characteristics of sampled fiber {Bragg} grating and
		phase-shifted sampled fiber {Bragg} grating in the 2$\mu$m band}, Opt. Fiber
	Technol. \textbf{50}, 263 (2019).
	
	\bibitem{Zou2006}
	X.~H. Zou, W.~Pan, B.~Luo, Z.~M. Qin, M.~Y. Wang, and W.~L. Zhang,
	{Periodically chirped sampled fiber {Bragg} gratings for multichannel comb
		filters}, IEEE Photonics Technol. Lett. \textbf{18}, 1371 (2006).
	
	\bibitem{Gohle2005}
	C.~Gohle, T.~Udem, M.~Herrmann, J.~Rauschenberger, R.~Holzwarth, H.~A.
	Schuessler, F.~Krausz, and T.~W. H{\"{a}}nsen, {A frequency comb in the
		extreme ultraviolet}, Nature \textbf{436}, 234 (2005).
	
	\bibitem{baltuvska2003attosecond}
	A.~Baltu{\v{s}}ka, T.~Udem, M.~Uiberacker, M.~Hentschel, E.~Goulielmakis,
	C.~Gohle, R.~Holzwarth, V.~Yakovlev, A.~Scrinzi, T.~W. H{\"a}nsch
	\emph{et~al.}, Attosecond control of electronic processes by intense light
	fields, Nature \textbf{421}, 611 (2003).
	
	\bibitem{Pfister2020}
	O.~Pfister, {Continuous-variable quantum computing in the quantum optical
		frequency comb}, J. Phys. B At. Mol. Opt. Phys. \textbf{53} (2020).
	
	\bibitem{udem2002optical}
	T.~Udem, R.~Holzwarth, and T.~W. H{\"a}nsch, Optical frequency metrology,
	Nature \textbf{416}, 233 (2002).
	
	\bibitem{jayaraman1993theory}
	V.~Jayaraman, Z.-M. Chuang, and L.~A. Coldren, Theory, design, and performance
	of extended tuning range semiconductor lasers with sampled gratings, IEEE J.
	Quantum Electron. \textbf{29}, 1824 (1993).
	
	\bibitem{holzwarth2000optical}
	R.~Holzwarth, T.~Udem, T.~W. H{\"a}nsch, J.~Knight, W.~Wadsworth, and P.~S.~J.
	Russell, Optical frequency synthesizer for precision spectroscopy, Phys. Rev.
	Lett. \textbf{85}, 2264 (2000).
	
	\bibitem{hall2006nobel}
	J.~L. Hall, Nobel lecture: Defining and measuring optical frequencies, Rev.
	Mod. Phys. \textbf{78}, 1279 (2006).
	
	\bibitem{hansch2006nobel}
	T.~W. H{\"a}nsch, Nobel lecture: passion for precision, Rev. Mod. Phys.
	\textbf{78}, 1297 (2006).
	
	\bibitem{ishii1993multiple}
	H.~Ishii, Y.~Tohmori, Y.~Yoshikuni, T.~Tamamura, and Y.~Kondo, Multiple-phase
	shift super structure grating {DBR} lasers for broad wavelength tuning, IEEE
	Photon. Technol. Lett. \textbf{5}, 613 (1993).
	
	\bibitem{Sourani2019}
	Y.~Sourani, A.~Bekker, B.~Levit, and B.~Fischer, {Tuning, selecting and
		switching wavelengths in lasers with chirped and sampled fiber {Bragg}
		gratings by high-order mode-locking}, Opt. Commun. \textbf{431}, 151 (2019).
	
	\bibitem{Shu2003}
	X.~Shu, K.~Chisholm, I.~Felmeri, K.~Sugden, A.~Gillooly, L.~Zhang, and
	I.~Bennion, {Highly sensitive transverse load sensing with reversible sampled
		fiber {Bragg} gratings}, Appl. Phys. Lett. \textbf{83}, 3003 (2003).
	
	\bibitem{Li2008}
	M.~Li, H.~Li, and Y.~Painchaud, {Multi-channel notch filter based on a
		phase-shift phase-only sampled fiber {Bragg} grating}, Opt. Express
	\textbf{16}, 19388 (2008).
	
	\bibitem{Eggleton1994}
	B.~J. Eggleton, P.~A. Krug, L.~Poladian, and F.~Ouellette, {Long periodic
		superstructure {Bragg} gratings in optical fibres}, Electron. Lett.
	\textbf{30}, 1620 (1994).
	
	\bibitem{DeSterke1997}
	C.~M. {De Sterke}, B.~J. Eggleton, and P.~A. Krug, {High-intensity pulse
		propagation in uniform gratings and grating superstructures}, J. Light.
	Technol. \textbf{15}, 1494 (1997).
	
	\bibitem{Eggleton1996}
	B.~J. Eggleton, R.~E. Slusher, and C.~M. de~Sterke, {Nonlinear propagation in
		superstructure {Bragg} gratings}, Opt. Lett. \textbf{21}, 1223 (1996).
	
	\bibitem{DeSterke1996}
	C.~M. de~Sterke, D.~G. Salinas, and J.~E. Sipe, {Coupled-mode theory for light
		propagation through deep nonlinear gratings}, Phys. Rev. E \textbf{54}, 1969
	(1996).
	
	\bibitem{DeSterke1995}
	C.~M. de~Sterke and N.~G.~R. Broderick, {Coupled-mode equations for periodic
		superstructure {Bragg} gratings}, Opt. Lett. \textbf{20}, 2039 (1995).
	
	\bibitem{erdogan1997fiber}
	T.~Erdogan, Fiber grating spectra, J. Light. Technol. \textbf{15}, 1277 (1997).
	
	\bibitem{ibsen1998sinc}
	M.~Ibsen, M.~K. Durkin, M.~J. Cole, and R.~I. Laming, Sinc-sampled fiber
	{Bragg} gratings for identical multiple wavelength operation, IEEE Photon.
	Technol. Lett. \textbf{10}, 842 (1998).
	
	\bibitem{kottos2010optical}
	T.~Kottos, Optical physics: Broken symmetry makes light work, Nat. Phys.
	\textbf{6}, 166 (2010).
	
	\bibitem{el2007theory}
	R.~El-Ganainy, K.~Makris, D.~Christodoulides, and Z.~H. Musslimani, Theory of
	coupled optical $\mathcal{PT}$-symmetric structures, Opt. Lett. \textbf{32},
	2632 (2007).
	
	\bibitem{ruter2010observation}
	C.~E. R{\"u}ter, K.~G. Makris, R.~El-Ganainy, D.~N. Christodoulides, M.~Segev,
	and D.~Kip, Observation of parity--time symmetry in optics, Nat. phys.
	\textbf{6}, 192 (2010).
	
	\bibitem{sarma2014modulation}
	A.~K. Sarma, Modulation instability in nonlinear complex parity-time symmetric
	periodic structures, J. Opt. Soc. Am. B \textbf{31}, 1861 (2014).
	
	\bibitem{lin2011unidirectional}
	Z.~Lin, H.~Ramezani, T.~Eichelkraut, T.~Kottos, H.~Cao, and D.~N.
	Christodoulides, Unidirectional invisibility induced by
	$\mathcal{PT}$-symmetric periodic structures, Phys. Rev. Lett. \textbf{106},
	213901 (2011).
	
	\bibitem{phang2013ultrafast}
	S.~Phang, A.~Vukovic, H.~Susanto, T.~M. Benson, and P.~Sewell, Ultrafast
	optical switching using parity--time symmetric {Bragg} gratings, J. Opt. Soc.
	Am. B \textbf{30}, 2984 (2013).
	
	\bibitem{miri2012bragg}
	M.-A. Miri, A.~B. Aceves, T.~Kottos, V.~Kovanis, and D.~N. Christodoulides,
	Bragg solitons in nonlinear $\mathcal{PT}$-symmetric periodic potentials,
	Phys. Rev. A \textbf{86}, 033801 (2012).
	
	\bibitem{huang2014type}
	C.~Huang, R.~Zhang, J.~Han, J.~Zheng, and J.~Xu, Type-{II} perfect absorption
	and amplification modes with controllable bandwidth in combined
	$\mathcal{PT}$-symmetric and conventional {Bragg}-grating structures, Phys.
	Rev. A \textbf{89}, 023842 (2014).
	
	\bibitem{govindarajan2018tailoring}
	A.~Govindarajan, A.~K. Sarma, and M.~Lakshmanan, Tailoring
	$\mathcal{PT}$-symmetric soliton switch, Opt. Lett. \textbf{44}, 663 (2019).
	
	\bibitem{kulishov2005nonreciprocal}
	M.~Kulishov, J.~M. Laniel, N.~B{\'e}langer, J.~Aza{\~n}a, and D.~V. Plant,
	Nonreciprocal waveguide {Bragg} gratings, Opt. Express. \textbf{13}, 3068
	(2005).
	
	\bibitem{lupu2016tailoring}
	A.~T. Lupu, H.~Benisty, and A.~V. Lavrinenko, Tailoring spectral properties of
	binary $\mathcal{PT}$-symmetric gratings by duty-cycle methods, IEEE J. Sel.
	Top. Quantum Electron. \textbf{22}, 35 (2016).
	
	\bibitem{raja2020tailoring}
	S.~V. Raja, A.~Govindarajan, A.~Mahalingam, and M.~Lakshmanan, Tailoring
	inhomogeneous $\mathcal{PT}$-symmetric fiber-{Bragg}-grating spectra, Phys.
	Rev. A \textbf{101}, 033814 (2020).
	
	\bibitem{raja2020phase}
	S.~V. Raja, A.~Govindarajan, A.~Mahalingam, and M.~Lakshmanan, Phase-shifted
	$\mathcal{PT}$-symmetric periodic structures, Phys. Rev. A \textbf{102},
	013515 (2020).
	
	\bibitem{longhi2015supersymmetric}
	S.~Longhi, Supersymmetric {Bragg} gratings, J. Opt. \textbf{17}, 045803 (2015).
	
	\bibitem{longhi2010optical}
	S.~Longhi, Optical realization of relativistic non-{Hermitian} quantum
	mechanics, Phys. Rev. Lett. \textbf{105}, 013903 (2010).
	
	\bibitem{phang2015versatile}
	S.~Phang, A.~Vukovic, T.~M. Benson, H.~Susanto, and P.~Sewell, A versatile
	all-optical parity-time signal processing device using a {Bragg} grating
	induced using positive and negative {Kerr}-nonlinearity, Opt. Quant.
	Electron. \textbf{47}, 37 (2015).
	
	\bibitem{phang2014impact}
	S.~Phang, A.~Vukovic, H.~Susanto, T.~M. Benson, and P.~Sewell, Impact of
	dispersive and saturable gain/loss on bistability of nonlinear parity--time
	{Bragg} gratings, Opt. Lett. \textbf{39}, 2603 (2014).
	
	\bibitem{longhi2010pt}
	S.~Longhi, $\mathcal{PT}$-symmetric laser absorber, Phys. Rev. A \textbf{82},
	031801(R) (2010).
	
	\bibitem{kashyap2009fiber}
	R.~Kashyap, \emph{Fiber {Bragg} Gratings}  (Academic {P}ress 2009).
	
	\bibitem{xu2019advanced}
	X.~Xu, M.~Tan, J.~Wu, T.~G. Nguyen, S.~T. Chu, B.~E. Little, R.~Morandotti,
	A.~Mitchell, and D.~J. Moss, Advanced adaptive photonic {RF} filters with 80
	taps based on an integrated optical micro-comb source, J. Light. Technol.
	\textbf{37}, 1288 (2019).
	
	\bibitem{pastor2001broad}
	D.~Pastor, J.~Capmany, and B.~Ortega, Broad-band tunable microwave transversal
	notch filter based on tunable uniform fiber {Bragg} gratings as slicing
	filters, IEEE Photonics Technol. Lett. \textbf{13}, 726 (2001).
	
	\bibitem{leng2004optimization}
	J.~Leng, W.~Zhang, and J.~A. Williams, Optimization of superstructured fiber
	{Bragg} gratings for microwave photonic filters response, IEEE Photonics
	Technol. Lett. \textbf{16}, 1736 (2004).
	
	\bibitem{davies1984fibre}
	D.~Davies and G.~James, Fibre-optic tapped delay line filter employing coherent
	optical processing, Electron. Lett. \textbf{20}, 95 (1984).
	
	\bibitem{hamidi2010tunable}
	E.~Hamidi, D.~E. Leaird, and A.~M. Weiner, Tunable programmable microwave
	photonic filters based on an optical frequency comb, IEEE Trans. Microwave
	Theory and Tech. \textbf{58}, 3269 (2010).
	
	\bibitem{mora2003tunable}
	J.~Mora, M.~Andres, J.~Cruz, B.~Ortega, J.~Capmany, D.~Pastor, and S.~Sales,
	Tunable all-optical negative multitap microwave filters based on uniform
	fiber {Bragg} gratings, Opt. Lett. \textbf{28}, 1308 (2003).
	
	\bibitem{mora2002automatic}
	J.~Mora, B.~Ortega, J.~Capmany, J.~Cruz, M.~Andres, D.~Pastor, and S.~Sales,
	Automatic tunable and reconfigurable fiber-optic microwave filters based on a
	broadband optical source sliced by uniform fiber {Bragg} gratings, Opt.
	Express \textbf{10}, 1291 (2002).
	
	\bibitem{bidaux2015extended}
	Y.~Bidaux, A.~Bismuto, C.~Tardy, R.~Terazzi, T.~Gresch, S.~Blaser, A.~Muller,
	and J.~Faist, Extended and quasi-continuous tuning of quantum cascade lasers
	using superstructure gratings and integrated heaters, Appl. Phys. Lett.
	\textbf{107}, 221108 (2015).
	
	\bibitem{xu2005chirped}
	X.~Xu, Y.~Dai, X.~Chen, D.~Jiang, and S.~Xie, Chirped and phase-sampled fiber
	{Bragg} grating for tunable {DBR} fiber laser, Opt. Express \textbf{13}, 3877
	(2005).
	
	\bibitem{azana2005spectral}
	J.~Aza{\~n}a, C.~Wang, and L.~R. Chen, Spectral self-imaging phenomena in
	sampled {Bragg} gratings, J. Opt. Soc. Am. B \textbf{22}, 1829 (2005).
	
	\bibitem{Shi2018}
	N.~Shi, T.~Hao, W.~Li, N.~Zhu, and M.~Li, {A reconfigurable microwave photonic
		filter with flexible tunability using a multi-wavelength laser and a
		multi-channel phase-shifted fiber {Bragg} grating}, Opt. Commun.
	\textbf{407}, 27 (2018).
	
	\bibitem{hansmann1995variation}
	S.~Hansmann, H.~Hillmer, H.~Walter, H.~Burkhard, B.~Hubner, and E.~Kuphal,
	Variation of coupling coefficients by sampled gratings in complex coupled
	distributed-feedback lasers, IEEE J. Sel. Top. Quantum Electron. \textbf{1},
	341 (1995).
	
	\bibitem{ozdemir2019parity}
	{\c{S}}.~{\"O}zdemir, S.~Rotter, F.~Nori, and L.~Yang, Parity--time symmetry
	and exceptional points in photonics, Nat. Mater. \textbf{18}, 783 (2019).
	
	\bibitem{chembo2016kerr}
	Y.~K. Chembo, Kerr optical frequency combs: theory, applications and
	perspectives, Nanophotonics \textbf{5}, 214 (2016).
	
	\bibitem{othonos2006fibre}
	A.~Othonos, K.~Kalli, D.~Pureur, and A.~Mugnier, Fibre {Bragg} gratings, in
	\emph{Wavelength Filters in Fibre Optics}  (Springer 2006), pp. 189--269.
	
	\bibitem{kryukov2019laser}
	P.~G. Kryukov, Laser optical frequency combs and their applications in optical
	fibre communication systems and astrophysics, Quantum Electron. \textbf{49},
	895 (2019).
	
	\bibitem{schneider2005sampled}
	L.~Schneider, M.~Pfeiffer, J.~Piprek, A.~Witzig, and B.~Witzigmann,
	Sampled-grating {DBR} lasers: calibrated 3d simulation of tuning
	characteristics, in \emph{Opto-Ireland 2005: Optoelectronics, Photonic
		Devices, and Optical Networks}  (International Society for Optics and
	Photonics 2005), volume 5825, pp. 95--106.
	
\bibitem{Zyablovsky2014}
	A.~A. Zyablovsky,  A.~ P.Vinogradov,  A.~ V.Dorofeenko, A.~ A. Pukhov, and A.~ A. Lisyansky, Causality and phase transitions in
	$\mathcal{PT}$-symmetric optical systems, Phys. Rev. A, \textbf{89}, 3, 033808 (2014).
	
	\bibitem{Phang2015res}
	S.~Phang, A.~ Vukovic, S.~C.Creagh, T.~ M. Benson,
	P.~D. Sewell, and G.~Gradoni, Parity-time symmetric
	coupled microresonators with a dispersive gain/loss, Opt. Express, \textbf{23}, 9, 11493 (2015).
	
		\bibitem{Nguyen2016}
	N.~B. Nguyen, S.~A. Maier, M.~Hong, and R.~F.Oulton, Recovering parity-time symmetry in highly dispersive coupled
	optical waveguides, New J. Phys, \textbf{18}, 12, 125012 (2016).
	
	\bibitem{Phang2018}
	S.~Phang, T.~M. Benson, H.~Susanto, S.~C. Creagh, G.~Gradoni,
	P.~D. Sewell, and A.~Vukovic, Theory and Numerical Modelling of
	Parity-Time Symmetric Structures in Photonics: Introduction and
	Grating Structures in One Dimension, arXiv, arXiv-1801 (2018). 
	
	
	\bibitem{Ge2012}
	L.~Ge,  Y.~D.Chong,  and A.~D.Stone, Conservation relations
	and anisotropic transmission resonances in one-dimensional
	$\mathcal{PT}$-symmetric photonic heterostructures, Phys. Rev. A, \textbf{85}, 2, 023802 (2012).
	

\end{thebibliography}

\end{document}